\documentclass[11pt]{article}
\usepackage{fullpage}
\usepackage[ruled,vlined]{algorithm2e}
\usepackage{enumerate}
\usepackage{latexsym,amsmath,amssymb}
\usepackage{graphicx,subfigure}
\usepackage{color}
\newtheorem{definition}{Definition}
\newtheorem{lemma}{Lemma}
\newtheorem{theorem}{Theorem}

\newtheorem{example}{Example}
\newtheorem{corollary}{Corollary}
\newtheorem{proposition}{Proposition}
\newtheorem{observation}{Observation}
\newtheorem{remark}{Remark}

\newenvironment{pf}{\begin{trivlist}
\item[\hspace{\labelsep}{\em\noindent Proof: }]
}{\hfill$\Box$\end{trivlist}}

\title{Structure and Recognition of 3,4-leaf Powers of Galled Phylogenetic  Networks}

\author{Michel \textsc{Habib}$^1$ and Thu-Hien  \textsc{To}$^2$\\
LIAFA - CNRS and Universit\'e Paris Diderot - Paris VII, \\
Case 7014, 75205 Paris Cedex 13, France\\
$^1$ habib@liafa.jussieu.fr \hspace*{.5cm} $^2$ toth@liafa.jussieu.fr 
}

\begin{document}
\maketitle

\begin{abstract}
A graph is a $k$-leaf power of a tree $T$ if its vertices are leaves of $T$ and two vertices are adjacent in $T$ if and only if their distance in $T$ is at most $k$. Then $T$ is a $k$-leaf root of $G$. This notion was introduced in \cite{NRT02} motivated by the search for underlying phylogenetic trees.
We study here an extension of the $k$-leaf power graph recognition problem. This extension is motivated by a new  biological question for the  evaluation of the latteral gene transfer on a population of viruses. We allow the host graph to slightly differs from a tree and allow some cycles. In fact we study phylogenetic galled networks in which cycles are pairwise vertex disjoint. We show some structural results and propose polynomial algorithms for the cases $k=3$ and $k=4$. As a consequence, squares of galled networks can also be recognized in polynomial time.
\end{abstract}

\section{Introduction}
The reconstruction of the evolutionary history of a set of species, based on quantitative biological data, is one of the most challenging problems in computational biology.

Most often the evolutionary history is modeled by an evolutionary tree called phylogeny whose leaves are labeled by species and each internal node represents a speciation event whereby an ancestral specie gives rise to two children.
Practically from the biological data one constructs a similarity undirected graph on the set of species where adjacency indicates evolutionary closeness, and then the problem is to build a phylogeny tree from this data. If we understand  the distance between species as a distance prescribed ($k$) in the phylogeny, the reconstruction problem is known as the recognition of $k$-leaf power graphs \cite{NRT02}, for a survey on leaf powers see \cite{B08}.

Our motivation comes from a question biological question from \cite{Bap07, Bap09, L09}.  They ask us how to measure the  relative importance of latteral genes transfer between viruses \cite{LVT08} compare to the normal evolution Darwinian rules. In other words can we infer a tree from the virus biological data or something else (a tree-like structure plus extra edges).

Therefore we consider the problem in which the evolutionary history is modelled by a generalization of trees, the so-called galled networks in which cycles are allowed but under the constraint that they pairwise do not intersect.

In the next sections we first consider some easy facts about phylogenetic networks and in section 4, when studying  $4$-leaf g-networks we show an important difference with $4$-leaf power of trees. Using maximal clique graphs and block decomposition we propose a polynomial algorithm to recognize these graphs.

\section{Preliminaries}
It is easy to see that $G$ is a $k$-leaf $\mathcal{N}$ power iff each of its connected component is. So, we may assume  in the following that \textbf{$G$ is connected}. The case $k=2$ is obvious since only sibling leaves can be adjacent, therefore only disjoint union of cliques can be $2$-leaf  $\mathcal N$ power graphs.

Most of the tools developed for the study of $k$-leaf tree power graphs, are also useful in the generalisation to other classes of phylogenetic networks. Therefore we will extend the notions of basic tree, visible vertex  and critical clique as can be found in \cite{BL06,BLS08}. 
 
A tree or a network is called \textbf{basic} \index{basic network} if each inner vertex is attached to at most one leaf. 
 
A vertex of $N$ is \textbf{visible} \index{visible vertex} if there is a leaf attached to it, otherwise it is \textbf{invisible}. \index{invisible vertex}
For any leaf $x$ of $N$, let $\textbf{p(x)}$ \index{$p(x)$} (for parent of $x$) be the inner vertex that $x$ is attached to. For any not leaf $u$, let $\textbf{l(u)}$ \index{$l(u)$} (for leaf of $u$) be the leaf attached to $u$ if $u$ is visible, and be empty if $u$ is invisible.

The two following results were already known for $k$-leaf tree power and can be easily extended to other classes of networks. 
\begin{proposition}
\label{propo:1}
If $G$ is a $k$-leaf  $\mathcal N$  power, then there exists a $k$-leaf $\mathcal{N}$ root $N'$ of $G$ such that there is a bijection between the critical cliques \index{critical clique} of $G$ and the set of leaves which are adjacent to the same vertex in $N'$.
\end{proposition}

\begin{pf}
Let $X$ be a critical clique of $G$ and $N \in \mathcal N$ be one of its roots. 
Let us suppose that in $N$, the leaves labelled by $X$ are adjacent to several different visible vertices $v_1, \dots , v_m$, and denote $X_i$ the set of leaves adjacent to $v_i$. By moving all $X_2, \dots , X_m$ so that they become adjacent to $v_1$, and then delete all new created leaves if there is any, we obtain a network $N'$. It is obvious that the leaf set of $N'$ and of $N$ is the same, and $N'$ is still a  $\mathcal N$ network, since we ask the class $\mathcal N$ to be hereditary and closed under addition and removing leaves.
Let $G'$ be the $k$-leaf power of $N'$. We will prove that $G'$ is equal to $G$.


For any two leaves $a,b$, if $a,b \notin X$ then their distance in $N$ and in $N'$ are the same. So, if they are connected (resp. non connected) in $G$, they are also connected (resp. non connected) in $G'$.

If $a$ is in $X$ but $b \notin X$ . Suppose that in $G$, $b$ is adjacent to $a$, so it is adjacent with all $X$ because $X$ is a critical clique of $G$. Then the distance between $b$ and the leaves of $X_1$ in $N$ are at most $k$, i.e in $N'$ the distance from $b$ to the leaves of $X$ are at most $k$. Then, $b$ is adjacent to all $X$ in the $G'$. The same, if $b$ is not adjacent to $X$ in $G$, then $b$ is not adjacent to any leaf of $X$ in $G'$.

If $a, b$ are both in $X$. Since $X$ is a clique of $G$, $ab$ is connected in $G$. They are obviously connected in $G'$ because their distance in $N'$ is exactly $2$ (they are attached to the same node $v_1$).
So, $G'$ is equal to $G$.

By using this construction for any critical clique of $G$, then for each critical clique $X$ of $G$, there is a vertex $v_X$ in $N'$ such that all vertices in $X$ are attached to $v_X$ of $N'$. Suppose that there are two critical cliques $X,Y$ such that $v_X=v_Y$, then it is easy to see that in the $k$-leaf power of $N'$, $X \cup Y$ is a critical clique. However, the $k$-leaf power of $N'$ is $G$, it is a contradiction because $X,Y$ are critical cliques in $G$ so $X \cup Y$ can not be a critical clique of $G$. This achieves the proof.

\end{pf}  

\begin{proposition} \label{CC(G)}
\label{lem:basic}
$G$ is a $k$-leaf $\mathcal N$ power if and only if $CC(G)$ \index{$CC(G)$} is a $k$-leaf basic $\mathcal N$  power.
\end{proposition}

\begin{pf}
Suppose that $G$ is a $k$-leaf $\mathcal N$ power and $N$ is one of its roots. By the above proposition, we can choose $N$ such that the leaves of any critical clique of $G$ are attached to the same visible vertex. By contracting each set of leaves of $N$ which adjacent to the same node, we obtain a basic network $N' \in \mathcal N$. It is easy to see that $N'$ is a $k$-leaf power of $CC(G)$.

Conversely, if $CC(G)$ is a $k$-leaf basic $\mathcal N$ network power and $N'$ is one of its basic roots. By replacing each leaf of $N'$ by the set of leaves of the corresponding critical cliques, we obtain a network $N$ which is a $k$-leaf root of $G$.
\end{pf}

\section{$3$-leaf power of $\mathcal N$ networks.}
It is known that a connected graph $G$ is a $3$-leaf tree power if and only if its critical clique graph is a tree \cite{R06,BL06,DGHN06}. Our next result generalizes this result  to $\mathcal N$ network powers.

\begin{theorem}
A graph $G$ is a $3$-leaf  $\mathcal N$ power if and only if its critical clique graph belongs to $\mathcal N$.
\end{theorem}

\begin{pf}
Using previous Proposition \ref{CC(G)}, let us suppose that $CC(G)=(V,E)$ is a 3-leaf $\mathcal N$ power graph and let $N$ be one of its roots. By definition it exists a bijective mapping $\theta$ from $V$ to the leaves of $N$. 

For every $t, t'$  leaves of $N$   $ dist(t, t') \leq 3$ iff  $p(t) p(t')$ is an edge of $N$. Therefore  $\theta^{-1}(t) \theta^{-1}(t') \in E$ iff  $p(t) p(t')$ is an edge of $N$. So $CC(G)$ is isomorphic to a subgraph of $N$. We have forced the class $\mathcal N$ to be
 hereditary, so $CC(G)$ belongs to $\mathcal N$.

Conversely, if  $CC(G)$ belongs to $\mathcal N$, let us build a phylogenetic network $N$ from $CC(G)$ by adding to every $x \in V$ a leaf $l(x)$ attached to $x$.  $\mathcal N$ is supposed to be closed by adding leaves, so $N \in \mathcal N$. Furthermore we immediately have $d_{N}(l(x), l(y)) =3$  iff $xy \in E$.

\end{pf}
\begin{corollary}
 The recognition of 3-leaf $\mathcal N$ power is equivalent (same complexity) of the recognition problem for $\mathcal N$.
\end{corollary}
\begin{corollary}
 $3$-leaf g-network powers can be recognized in linear time.
\end{corollary}

\begin{pf}
Since it is well-known that computing twins (false or true) can be done in linear time see  \cite{HPV98},  therefore calculating $CC(G)$ can be done in linear time. Recognition of g-networks can also be done in linear time simply by using a depth-first search traversal and computing the block (2-connected components) decomposition of the graph.
To conclude we just have to check if every block contains at most one cycle and if these cycles intersect.

\end{pf}

\section{$4$-leaf basic g-network powers}
\label{sec:4-leaf}
For $k=4$ the problem becomes harder to solve and therefore for technical purpose we choose to consider first a slight generalization of trees namely g-networks. \index{g-network}

\begin{definition}
Let $\mathcal{L}$ be a set of $n$ species.
A g-network on $\mathcal{L}$ is a connected undirected graph containing $n$ vertices of degree $1$ which are bijectively labelled by $\mathcal{L}$ and in which induced cycles are pairwise vertex-disjoint.
\label{def:g-network}
\end{definition}

Moreover, by computing $CC(G)$, one can easily transform a $4$-leaf g-network power to a $4$-leaf basic g-network power. So, we study the problem only on basic models.

\subsection{Preliminaries}
As a first remark we notice that if $G$ is a $4$-leaf basic g-network power then it has a root $N$ which does not contain any two adjacent invisible vertices. This is simply because if there were an edge between two invisibles vertices, this edge cannot be used to connect $2$ leaves via a path of length $\leq 4$. Therefore this edge can be deleted in $N$. So we can suppose that our networks do not contain two adjacent invisible vertices. This assumption implies that for any two vertices $x,y$ of $G$, there is a path from $x$ to $y$ in $G$ iff there is a path from $\theta(x)$ to $\theta(y)$ in $N$.

\begin{definition}
\label{def:N_M}
For a subset of vertices $M$ of $G$ let us define  for a root $N$ of $G$, $N[M]$ \index{$N[M]$} to be the minimal connected induced subgraph of $N$ whose leaves are exactly $\theta(M)$ and for which $\forall x, y \in M, d_{N[M]}(\theta(x),\theta(y)) =d_{N}(\theta(x),\theta(y))$.
\end{definition}

When $N$ is known, for every subset of vertices $M$, $N[M]$ is uniquely defined and easy to compute. 

\begin{figure}[ht]
\begin{center}
\subfigure[An invisible star\label{fig:invisible_star}]{\includegraphics[scale=0.7]{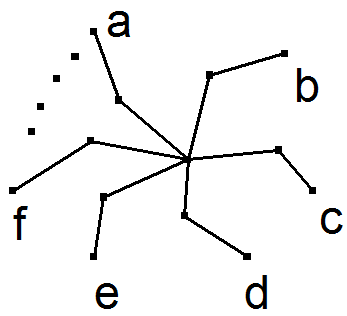}}\;\;\;
\subfigure[An  invisible quasi star\label{fig:invisible_nearly_star}]{\includegraphics[scale=0.7]{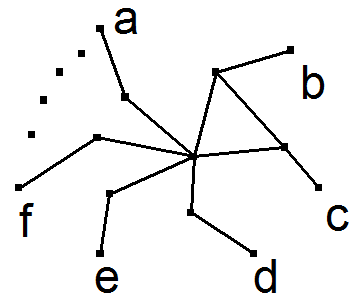}}\;\;\;
\subfigure[A visible star\label{fig:visible_star}]{\includegraphics[scale=0.7]{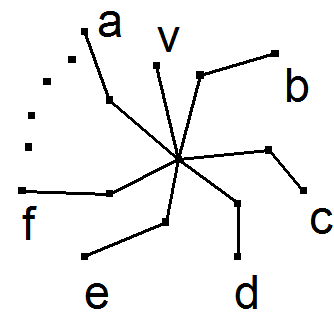}}\;\;\;
\subfigure[A   visible quasi star\label{fig:visible_nearly_star}]{\includegraphics[scale=0.7]{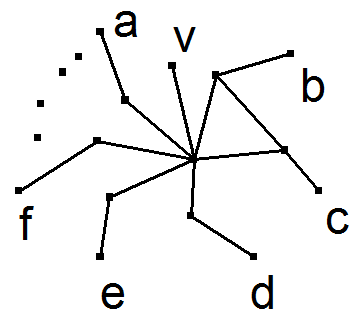}}\;\;\;
\subfigure[$N_4$ \label{fig:4-cycle}]{\includegraphics[scale=0.7]{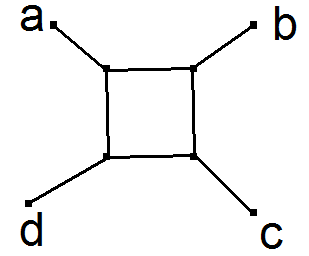}}\;\;\;
\subfigure[$N_5$\label{fig:5-cycle}]{\includegraphics[scale=0.7]{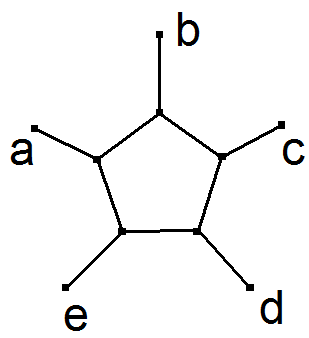}}\;\;\;
\subfigure[$N'_5$\label{fig:5-cycle_1}]{\includegraphics[scale=0.7]{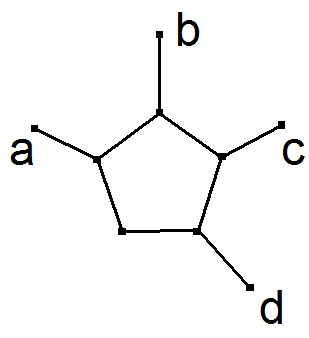}}\;\;\;
\subfigure[$N''_5$\label{fig:5-cycle_2}]{\includegraphics[scale=0.7]{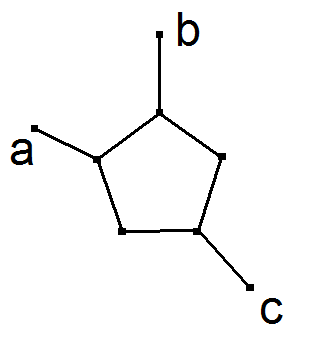}}\;\;\;
\subfigure[$N_6$ \label{fig:6-cycle}]{\includegraphics[scale=0.7]{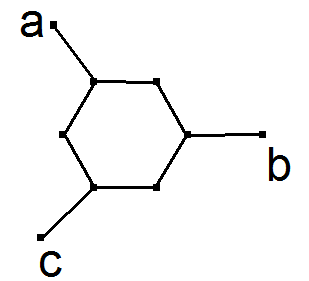}}\;\;\;
\caption{$\mathcal F$ the family of roots of maximal clique for  $4$-leaf basic g-networks power}
\label{fig:maximal_clique}
\end{center}
\end{figure}

The family $\mathcal F$ \index{family $\mathcal F$} of g-networks described in Figure \ref{fig:maximal_clique} is made up with an infinite family of visible, invisible (quasi) stars together with $5$ networks $N_4$, $N_5$, $N'_5$, $N''_5$, $N_6$ based on cycles of length $4$ to $6$. 

\begin{observation}
Suppose that $G$ has a $4$-leaf basic g-network root $N$. Then $M$ is a maximal clique of $G$ iff $N[M]$ is isomorphic to one of the subgraphs in $\mathcal F$.
\label{ob:1} 
\end{observation}

We denote \textit{i-cycle} \index{i-cycle} for a chordless cycle of length $i$. So, $N_4$ \index{$N_4$} is a  $4$-cycle without invisible vertex,  $N_5$ \index{$N_5$} is a $5$-cycle without invisible vertex, $N'_5$ \index{$N'_5$} is a $5$-cycle with one invisible vertex,  $N''_5$ \index{$N''_5$} is a $5$-cycle with two invisible vertices and  $N_6$ \index{$N_6$} is a $6$-cycle with $3$ visible vertices of pairwise distance $2$.

An \textit{invisible star} \index{invisible star} (Figure \ref{fig:invisible_star}) consists of an invisible vertex $u$, called the \textit{middle vertex} \index{middle vertex} of it, and all visible vertices adjacent to $u$ as well as the leaves attached to these visible vertices. Two visible vertices of an invisible star can be adjacent, in this case we have an \textit{invisible quasi star} \index{invisible quasi star} (Figure \ref{fig:invisible_nearly_star}). Because there is no intersecting cycle in the constructing networks, there are at most two adjacent visible vertices in each invisible quasi star.

A \textit{visible star} \index{visible star} (Figure \ref{fig:visible_star}) consists of a visible vertex $u$, called the \textit{middle vertex} \index{middle vertex} of it, the leaf of $u$ and all visible vertices adjacent to $u$ as well as the leaves attached to these visible vertices. If there are two adjacent visible vertex $u,u'$ such that they are not adjacent to any other visible vertex, then the visible star of $u$ is exactly the visible star of $u'$, we can consider either $u$ or $u'$ as the middle vertex of this visible star.
Similarly we define also \textit{visible quasi star} \index{visible  quasi star} (Figure \ref{fig:visible_nearly_star}). 

Moreover, if a visible or invisible star is totally included in another visible or invisible star, or a network $N_4, N_5, N'_5, N''_5, N_6$ then we do not consider it as a visible or invisible star. For example, this is the case of the visible star whose middle vertex is $p(b)$ in Figure \ref{fig:4-cycle} when $p(b)$ is not adjacent to any other visible vertex different from $p(a), p(c)$; or the invisible stars having $\{a,d\}$ as the leaf set in Figure \ref{fig:5-cycle_1} when the middle vertex of this star is not adjacent to any other visible vertex different from $p(a), p(d)$.

\begin{pf}
$\Rightarrow$ Let $M =  \{x_1, \dots, x_k\}$ be a maximal clique of $G$, then in $N$ we have $d(x_i,x_j) \le 4$ for any $i,j$. 

We first prove that $N[M]$ cannot contain more than one cycle. In fact, if there are two cycles $C_1, C_2$ in $N[M]$, they are vertex disjoint and be separated by a path containing at least one edge. 

\begin{minipage}[b]{12cm}
Let $u_1,u_2$ be the extremities of this path on $C_1, C_2$. There must exist a visible vertex $v_i$ on $C_i$ different from $u_i$ for any $i=1,2$. Indeed, suppose otherwise, then in order that the distance between every pair of leaves in $M$ are at least $4$, all leaves of $M$ are attached to either $u_1$ or $u_2$. It implies by the definition of $N[M]$ that $N[M]$ consists of only the path from $u_1$ to $u_2$ but not the cycles $C_1,C_2$, a contradiction. Let $x_i$ be the leaf attached to $v_i$. Hence, $d_N(x_1,x_2) \ge 5$,  it means that they are not connected in $G$, a contradiction.
\end{minipage}\hfill
\begin{minipage}[b]{4cm}
\begin{center}
\includegraphics[scale=.8]{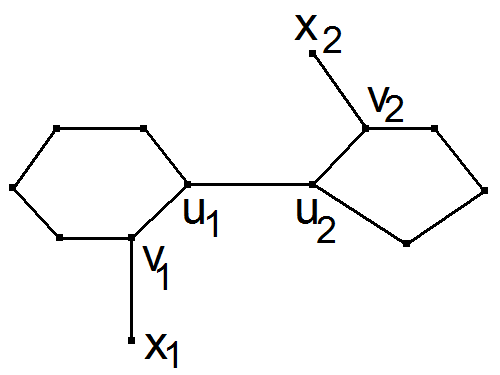}
\vspace*{.5cm}
\end{center}
\end{minipage}

If $N[M]$ contains a cycle, that cycle has at most $6$ vertices. Otherwise there are two leaves whose distance is greater than $4$. So, the Figures \ref{fig:invisible_nearly_star}, \ref{fig:visible_nearly_star}, \ref{fig:4-cycle}, \ref{fig:5-cycle}, \ref{fig:5-cycle_1}, \ref{fig:5-cycle_2}, \ref{fig:6-cycle} correspond to all the possible configurations of $N[M]$ in this case.

If $N[M]$ does not contain any cycle, then in order that $d(x_i,x_j) \le 4$ for any $i,j$, $N[M]$ must be a star (visible or invisible). 

$\Leftarrow$ It is obvious that the $4$-leaf power of any network of $\mathcal{F}$ is a clique. By definition, not any network of $\mathcal{F}$ is included in another one. So each one correspond to a maximal clique of $G$.
\end{pf}

Therefore, one can construct a $4$-leaf g-network leaf of a graph $G$ by replacing each maximal clique $M$ of $G$ by a network in the family $\mathcal{F}$ in an appropriate way. 

It should be noticed that the networks $N_4, N_5, N'_5, N''_5$ and  $N_6$ of the family $\mathcal F$ are not chordal graphs. Nevertheless maximal cliques still play a great role as will be shown in the following sections.

\begin{figure}[ht]
\begin{center}
\subfigure[$G$\label{fig:G}]{\includegraphics[scale=.8]{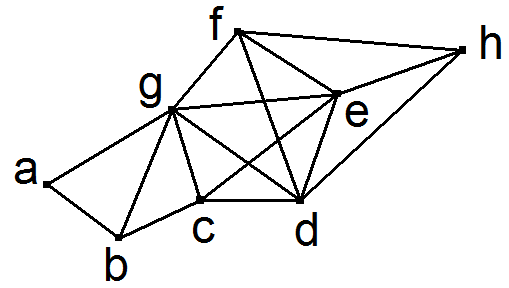}}
\subfigure[$MC(G)$\label{fig:MC(G)}]{\includegraphics[scale=.8]{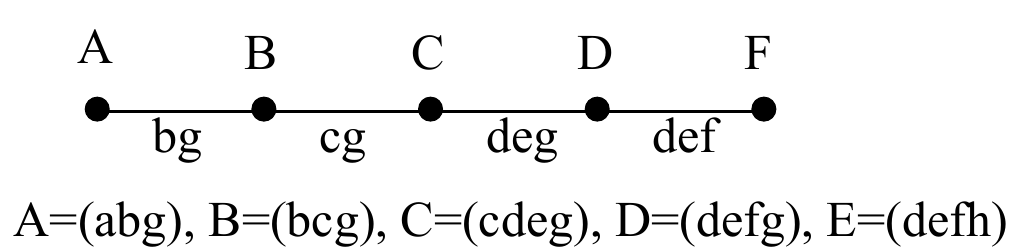}}
\caption{An example of maximal cliques graph}
\label{fig:MCG}
\end{center}
\end{figure}

\begin{definition}[Maximal cliques graph]

Let $G$ be a biconnected graph. The \textit{maximal cliques graph} \index{maximal cliques graph} of $G$, denoted by  $MC(G)$ \index{$MC(G)$} is a graph whose nodes correspond to maximal cliques of $G$. Two nodes $A, B$ are connected by an edge if and only if the corresponding maximal cliques of $A$ and $B$ have common vertices, and furthermore $A \cap B$ is maximal with respect to  $A$ or $B$.
($A \cap B$ is maximal with respect to $A$ if there is not any node $C$ such that $A \cap C$ is included in $A \cap B$). 
\end{definition}

The \textit{label of a node} $A$ of $MC(G)$, denoted by $l(A)$, is equal to the vertex set of the maximal clique $A$. The \textit{label of an edge} $AB$, denoted by $l(AB)$, is equal to the common vertex set of the two maximal cliques $A,B$. The \textit{label of a subgraph} $S$, denoted by $l(S)$, is equal to the union of the labels of the edges of $S$. The \textit{weight} of an edge of $MC(G)$ is the cardinal of its label. The \textit{size} of a node is the cardinal of its label.

Let $A$ be a node of $MC(G)$, denote by $N[A]$ the induced subgraph of $N$ as defined in Definition \ref{def:N_M}, on the maximal clique of $G$ corresponding to the node $A$.
For the sake of simplicity, when we represent $N[A]$ in figures, except the leaves of $A$ which have concrete labels, we denote the labels of the other leaves by $A$.

We say that a network $n$ is a $4$-leaf root of a subgraph $S$ of $MC(G)$ iff $MC(n^{4l})=S$. 

Let $S$ be a subgraph of $MC(G)$, then denote by $G[S]$ the induced subgraph of $G$ on the vertex set consisting of all $l(A)$ for every node $A$ of $S$. 
If a network $n$ is a root of $G[S]$, we also say that it is a root of $S$.

With this definition of $MC(G)$, we can deduce several properties of $MC(G)$ related to the family $\mathcal F$ in the observation below. 

\begin{observation}Let $N$ be a $4$-leaf basic g-network root of $G$.

\begin{minipage}[b]{12cm}
(i) If $AB$ is an edge of $MC(G)$ such that $l(AB)=\{x,y\}$ and $N[A], N[B]$ are visible (quasi) star, then $x,y$ are the two leaves attached to the middle vertices of $N[A]$ and $N[B]$. 
\end{minipage}\hfill
\begin{minipage}[b]{3cm}
\includegraphics[scale=.7]{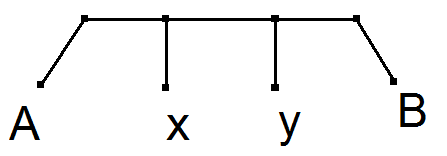}
\end{minipage}

(ii) If $AB$ is an edge of $MC(G)$ such that $|AB|=3$ then  either $N[A]$, $N[B]$ are visible quasi stars sharing a same triangle or one of them is $N_4$ or $N_5$ or $N'_5$ and the other is a visible star.

\begin{figure}[ht]
\begin{center}
\def\svgwidth{10cm}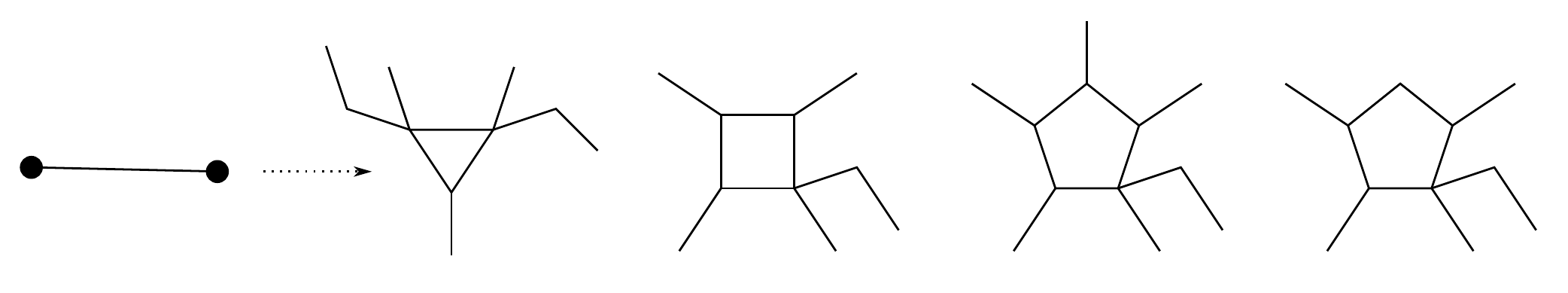
\label{fig:AB3}
\end{center}
\end{figure}

(iii) If $ABC$ is a triangle of $MC(G)$, then:

\begin{minipage}[b]{8cm}
- either $|AB|=|BC|=|AC|=2$, then one of $N[A], N[B], N[C]$ is either an invisible quasi stars or a $N''_5$.

\def\svgwidth{7cm}

\begingroup
  \makeatletter
  \providecommand\color[2][]{%
    \errmessage{(Inkscape) Color is used for the text in Inkscape, but the package 'color.sty' is not loaded}
    \renewcommand\color[2][]{}%
  }
  \providecommand\transparent[1]{%
    \errmessage{(Inkscape) Transparency is used (non-zero) for the text in Inkscape, but the package 'transparent.sty' is not loaded}
    \renewcommand\transparent[1]{}%
  }
  \providecommand\rotatebox[2]{#2}
  \ifx\svgwidth\undefined
    \setlength{\unitlength}{384pt}
  \else
    \setlength{\unitlength}{\svgwidth}
  \fi
  \global\let\svgwidth\undefined
  \makeatother
  \begin{picture}(1,0.3125)%
    \put(0,0){\includegraphics[width=\unitlength]{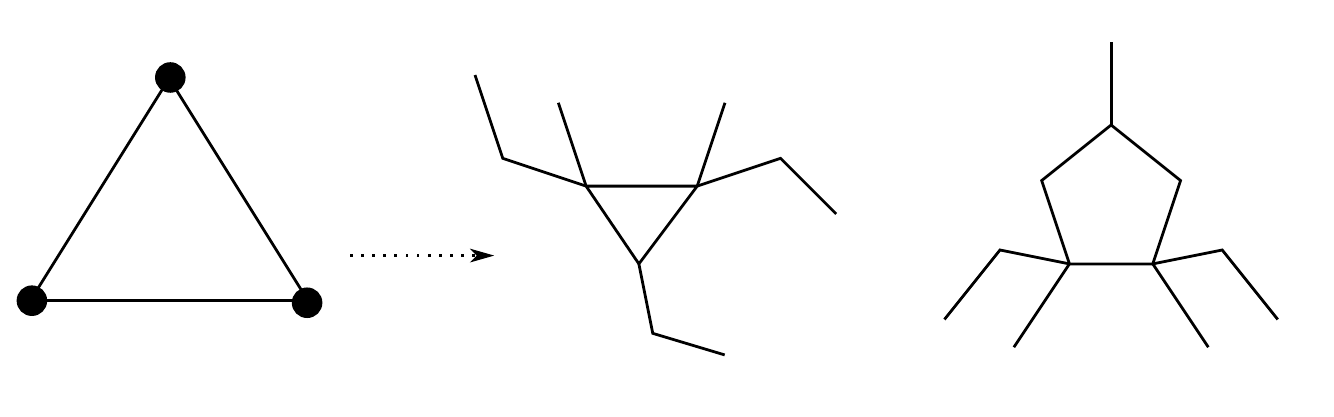}}%
    \put(0.09560436,0.04427744){\color[rgb]{0,0,0}\makebox(0,0)[lb]{\smash{$xy$}}}%
    \put(-0.00037527,0.02871602){\color[rgb]{0,0,0}\makebox(0,0)[lb]{\smash{$A$}}}%
    \put(0.21153549,0.0274414){\color[rgb]{0,0,0}\makebox(0,0)[lb]{\smash{$B$}}}%
    \put(0.33958333,0.25833333){\color[rgb]{0,0,0}\makebox(0,0)[lb]{\smash{$A$}}}%
    \put(0.61875,0.11666666){\color[rgb]{0,0,0}\makebox(0,0)[lb]{\smash{$B$}}}%
    \put(0.40833333,0.24583333){\color[rgb]{0,0,0}\makebox(0,0)[lb]{\smash{$x$}}}%
    \put(0.53958333,0.25208333){\color[rgb]{0,0,0}\makebox(0,0)[lb]{\smash{$y$}}}%
    \put(0.73818696,0.02083333){\color[rgb]{0,0,0}\makebox(0,0)[lb]{\smash{$x$}}}%
    \put(0.89189568,0.01920776){\color[rgb]{0,0,0}\makebox(0,0)[lb]{\smash{$y$}}}%
    \put(0.95369746,0.04467722){\color[rgb]{0,0,0}\makebox(0,0)[lb]{\smash{$B$}}}%
    \put(0.8375,0.27916666){\color[rgb]{0,0,0}\makebox(0,0)[lb]{\smash{$z$}}}%
    \put(0.10420191,0.27845052){\color[rgb]{0,0,0}\makebox(0,0)[lb]{\smash{$C$}}}%
    \put(0.02211355,0.19331296){\color[rgb]{0,0,0}\makebox(0,0)[lb]{\smash{$xy$}}}%
    \put(0.17166653,0.19200159){\color[rgb]{0,0,0}\makebox(0,0)[lb]{\smash{$xy$}}}%
    \put(0.53816446,0.01285476){\color[rgb]{0,0,0}\makebox(0,0)[lb]{\smash{$C$}}}%
    \put(0.67395833,0.04536412){\color[rgb]{0,0,0}\makebox(0,0)[lb]{\smash{$A$}}}%
  \end{picture}%
\endgroup

\end{minipage}\hfill
\begin{minipage}[b]{7cm}
- or $|AB|=|BC|=|AC|=3$, then $N[A],N[B],N[C]$ are $3$ visible quasi stars sharing the same triangle.

\def\svgwidth{4cm}

\begingroup
  \makeatletter
  \providecommand\color[2][]{%
    \errmessage{(Inkscape) Color is used for the text in Inkscape, but the package 'color.sty' is not loaded}
    \renewcommand\color[2][]{}%
  }
  \providecommand\transparent[1]{%
    \errmessage{(Inkscape) Transparency is used (non-zero) for the text in Inkscape, but the package 'transparent.sty' is not loaded}
    \renewcommand\transparent[1]{}%
  }
  \providecommand\rotatebox[2]{#2}
  \ifx\svgwidth\undefined
    \setlength{\unitlength}{256pt}
  \else
    \setlength{\unitlength}{\svgwidth}
  \fi
  \global\let\svgwidth\undefined
  \makeatother
  \begin{picture}(1,0.46875)%
    \put(0,0){\includegraphics[width=\unitlength]{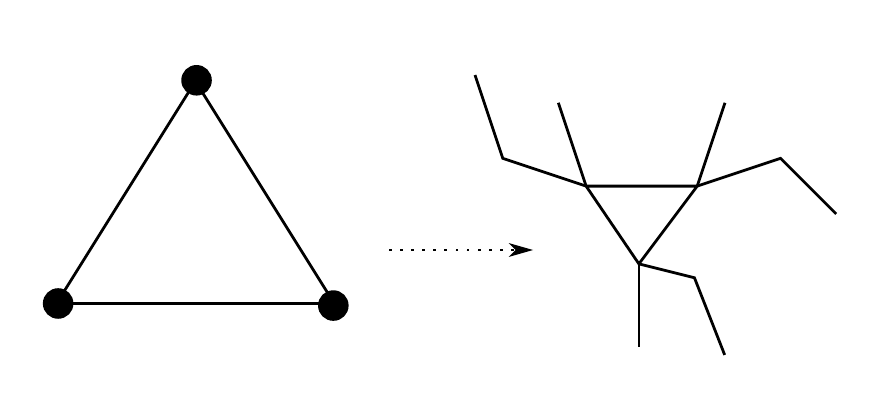}}%
    \put(0.15729845,0.06481628){\color[rgb]{0,0,0}\makebox(0,0)[lb]{\smash{$xyz$}}}%
    \put(0.02895401,0.03053664){\color[rgb]{0,0,0}\makebox(0,0)[lb]{\smash{$A$}}}%
    \put(0.35463262,0.03174972){\color[rgb]{0,0,0}\makebox(0,0)[lb]{\smash{$B$}}}%
    \put(0.503125,0.39218749){\color[rgb]{0,0,0}\makebox(0,0)[lb]{\smash{$A$}}}%
    \put(0.928125,0.17968749){\color[rgb]{0,0,0}\makebox(0,0)[lb]{\smash{$B$}}}%
    \put(0.6125,0.36874999){\color[rgb]{0,0,0}\makebox(0,0)[lb]{\smash{$x$}}}%
    \put(0.809375,0.37812499){\color[rgb]{0,0,0}\makebox(0,0)[lb]{\smash{$y$}}}%
    \put(0.69375,0.02812499){\color[rgb]{0,0,0}\makebox(0,0)[lb]{\smash{$z$}}}%
    \put(0.18581978,0.41451339){\color[rgb]{0,0,0}\makebox(0,0)[lb]{\smash{$C$}}}%
    \put(0.00799974,0.28993205){\color[rgb]{0,0,0}\makebox(0,0)[lb]{\smash{$xyz$}}}%
    \put(0.2870167,0.28484){\color[rgb]{0,0,0}\makebox(0,0)[lb]{\smash{$xyz$}}}%
    \put(0.80412159,0.00834464){\color[rgb]{0,0,0}\makebox(0,0)[lb]{\smash{$C$}}}%
  \end{picture}%
\endgroup

\vspace*{.35cm}
\end{minipage}

(iv) If $AB$ is an edge of $MC(G)$ such that $|AB|=1$ then either $N[A]$ or $N[B]$ is an invisible (quasi) star.

(v) If $MA,MB$ are two edges of $MC(G)$ such that $|MA|=|MB|=3$ and $AB$ is not an edge of $MC(G)$ then $N[M]$ is either $N_4$ or $N_5$ or $N'_5$ and $N[A]$, $N[B]$ are visible stars.
\label{ob:2}
\end{observation}

\begin{pf}
(i) We check all the possible configurations of $N[A], N[B]$ such that they have two common leaves $x,y$. 

\begin{minipage}[b]{12.5cm}
\hspace*{.5cm}Suppose these common vertices are not middle vertices of $N[A], N[B]$, and the middle vertices of $N[A], N[B]$ are respectively $a,b$.
 So, there is a $4$-cycle $N_4$ with the leaves $x,y,a,b$. Let $C$ be the maximal clique corresponding to this cycle. Then $A \cap C = \{x,a,b\}$ and $B \cap C = \{y,a,b\}$, i.e. $A \cap B$ is include in $A \cap C$ and $B \cap C$. By definition of $MC(G)$, $AB$ is not an edge of $MC(G)$, a contradiction.
\end{minipage}\hfill
\begin{minipage}[b]{3cm}
\includegraphics[scale=.7]{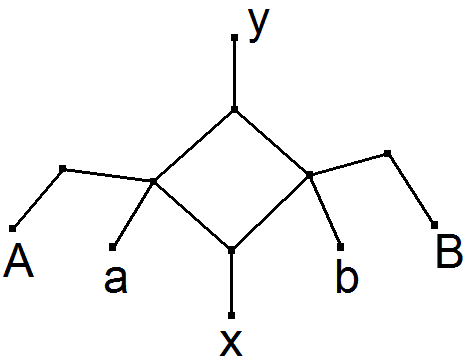}
\vspace*{.2cm}
\end{minipage}

\begin{minipage}[b]{13cm}
\hspace*{.5cm} Suppose that $x$ is the label of the middle vertex of $N[A]$, but $y$ is not the label of the middle vertex of $N[B]$. Because $x$ is contained in the star $N[B]$, so $b$ is also in the star having $p(x)$ as middle vertex. In other words, $b$ is contained in $N[A]$, i.e. $A,B$ have $3$ common vertices $x,y,b$, a contradiction.
\end{minipage}\hfill
\begin{minipage}[b]{2.5cm}
\includegraphics[scale=.7]{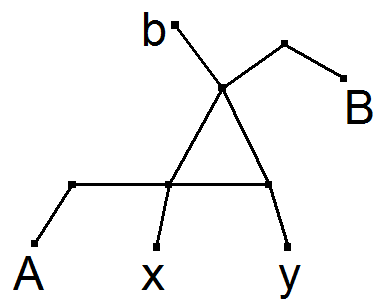}
\end{minipage}

So, $x,y$ are the labels of the middle vertices of $N[A], N[B]$.

(ii,iii) It is easy to check theses claim by considering all possible configurations of $N[A]$, $N[B]$, (and $N[C]$) with the condition of vertex-disjoint cycles in $N$.

(iv) Suppose that neither $N[A]$ nor $N[B]$ is an invisible (quasi) star, and let $x$ be the unique common leaf of $N[A]$ and $N[B]$. 
By considering all other possible configurations of $N[A]$ and $N[B]$, we can deduce that it is not possible. 

\begin{minipage}[b]{12cm}
\hspace*{.5cm}For example suppose that $N[A], N[B]$ are visible (quasi) stars. Let $x_A,x_B$ be the leaves of the middle vertex of $N[A]$ and $N[B]$. Let consider the visible (quasi) star having $p(x)$ as its middle vertex, it corresponds to the maximal clique $C$. So, $C$ contains $x_A,x_B$. We deduce that $A \cap B \subset A \cap C$ and $A \cap B \subset B \cap C$. By definition of $MC(G)$, $AB$ is not an edge, a contradiction.

\hspace*{.5cm}If $N[A]$ is visible (quasi) star, $N[B]$ is $N_4$, let $x_A$ be the leaf of the middle vertex of $N[A]$. Let consider the visible (quasi) star having $p(x)$ as its middle vertex, it corresponds to the maximal clique $C$. So, $C$ contains $x_B$ and $y,t$. We deduce that $A \cap B \subset A \cap C$ and $A \cap B \subset B \cap C$. By definition of $MC(G)$, $AB$ is not an edge.
\end{minipage}\hfill
\begin{minipage}[b]{3.5cm}
\begin{center}
\includegraphics[scale=.7]{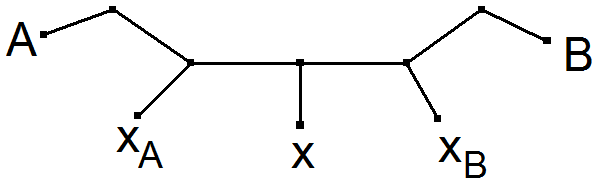}
\vspace*{.5cm}

\includegraphics[scale=.7]{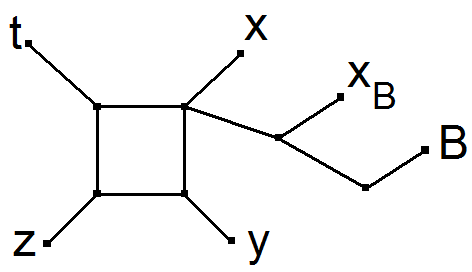}
\vspace*{.5cm}
\end{center}
\end{minipage}

Similarly, we can check that in any other case, there is always a maximal clique $C$ such that $A \cap B \subset A \cap C$ and $A \cap B \subset B \cap C$, a contradiction.

(v) We have $|MA|=3$, so by Observation (ii):

- either $N[A]$, $N[M]$ are visible quasi stars sharing a triangle. Since $|MB|=3$, $N[B]$ must be also a visible quasi star which share the same triangle with $N[M]$ and $N[A]$. So, $AB$ is connected by an edge of weight $3$ in $MC(G)$, a contradiction.

- or $N[M]$ is $N_4$, $N_5$, or $N'_5$ and $N[A]$, $N[B]$ are visible stars, so we are done.
\end{pf}

\begin{lemma}
\label{lem:MC}
If $G$ is a $4$-leaf basic g-network power, then 

(i) each edge of $G$ is contained in at most $3$ maximal cliques.

(ii) any two maximal cliques have at most $3$ vertices in common.

\end{lemma}

\begin{pf}
Let $N$ be a $4$-leaf basic g-network root of $G$. 

(i) Let $xy$ be an edge of $G$ which is contained in a maximal clique $A$. 

- If $N[A]$ is not a visible or invisible quasi star: suppose that there are other two maximal cliques $B,C$ that contain $xy$. Then 
for any configuration of the family $\mathcal F$ that $N[A],N[B],N[C]$ take, we can check that $N[A \cup B \cup C]$ contains always two non disjoint cycles, contradicting that $N$ is a g-network. So, it this case, there are at most $2$ maximal cliques containing $xy$.

- If $N[A]$ is a visible or invisible quasi star: then there are at most two other maximal cliques different from $A$ that can contain $x,y$ by the same reason. In the case that there are exactly $2$ other maximal cliques $B,C$ that contain $x,y$, then $ABC$ is a triangle of $MC(G)$.

(ii) Let $A,B$ be two maximal cliques having more than $1$ common vertices. Always by the condition of vertex-disjoint cycles in $N$, we deduce that if both $N[A], N[B]$ are either visible stars or invisible (quasi) stars or $N''_5$, or $N_6$, then they have at most $2$ common leaves. If one of them is a visible quasi star or $N_4$ or $N_5$ , or $N'_5$, then they have at most $3$ common leaves. Moreover, when they have exactly $3$ common vertices, either one of them is $N_4$ or $N_5$ or $N'_5$, or both of them are visible quasi stars (Claim (ii) of Observation \ref{ob:2}).
\end{pf}

\begin{corollary}
\label{co:3e}
For  $G=(V, E)$ a $4$-leaf basic g-network power, the number of maximal cliques of $G$ is bounded by $3|E|$. 
\end{corollary}

\begin{theorem}
\label{theo:mc}
For  $G=(V, E)$ a $4$-leaf basic g-network power graph, $MC(G)$ can be computed in $O(|V|. |E|^{2})$.
\end{theorem}

\begin{pf}
It suffices to use any algorithm that generates all maximal cliques of $G$  in $O(|V|. |E|)$ delay using polynomial space as described in \cite{GNS09}. If it has more than $3.|E|$ maximal cliques, we  can stop the algorithm and conclude that it is not a $4$-leaf basic g-network power, otherwise we can calculate the edges of $MC(G)$ within the same complexity. 
\end{pf}

When a clique is a block, let us call it a \textit{block clique}. \index{block clique} Another similar property with $4$-leaf basic tree roots is stated in the following lemma.

\begin{lemma}
\label{lem:blockclique}
If $G$ is a $4$-leaf basic g-network power, then there exists a $4$-leaf basic g-network root $N$ of $G$ such that for any block clique $\mathcal{B}$ of $G$, $N[\mathcal{B}]$ is an invisible star. 
\end{lemma}

\begin{pf}
Let $\mathcal{B}_1, \dots \mathcal{B}_m$ be the blocks of $G$. Let $N_0$  be a $4$-leaf basic g-network root of $G$. For any $\mathcal{B}_i$, let $N_i = N_0[\mathcal{B}_i]$. By the above analysis, $N_0$ is obtained by combining $N_i$ for all $i = 1, \dots , m$. For any block clique $\mathcal{B}_i$ we replace $N_i$ by the invisible star $S_i$ having the vertices of $\mathcal{B}_i$ as its leaf set such that if $N_i$, $N_j$ have $x$ and $p(x)$ as common vertices, then $S_i$ and $N_j$ also have $x$ and $p(x)$ as common vertices. Let the resulting network is $N$. It is obvious that $S_i$ is a $4$-leaf basic g-network root of $\mathcal{B}_i$. We will prove that $N$ is a $4$-leaf root of $G$. It is easy to check that tor any two leaves $x,y$ of $G$, if $x,y$ are not in the same block, then their distance in $N$ are greater than $4$. This corresponds to the fact that $x,y$ are not connected in $G$. If $x,y$ are contained in one block and this block is not a clique, then their distance in $N$ is exactly their distance in $N_0$. If this block is a clique then their distance in $N$ is exactly $4$ because they are two leaves of an invisible star. That corresponds to the fact that $xy$ is connected in $G$ because they are in the same clique. Therefore, $G$ is a $4$-leaf power of $N$.

It is obvious that $N$ is a basic network. We must prove furthermore that the cycles in $N$ are pairwise vertex-disjoint. From the assumption that there is no adjacent invisible vertex in $N$, we deduce that each cycle of $N$ is contained in the root of a block of $G$. Since $N_0$ is a g-network, the cycles in each $N_i$ are pairwise vertex-disjoint. The invisible stars $S_i$ do not contain any cycle, so the cycles in $N$ are always pairwise disjoint, i.e. $N$ is a basic g-network.
\end{pf}

\subsection{Adding extra constraints to the root graphs}
\label{sec:constraint}
\begin{figure}[ht]
\begin{center}
\subfigure[\label{fig:standard_1_l}]{\includegraphics[scale=.7]{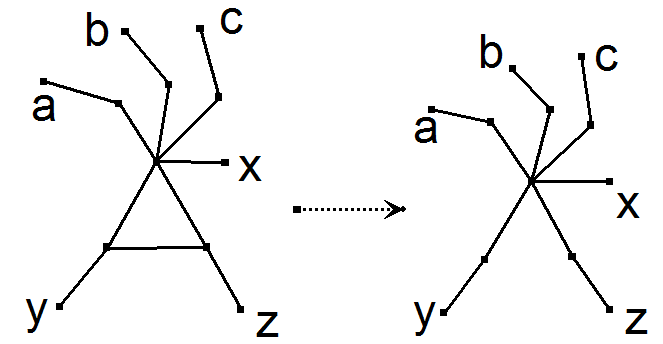}}\;\;\;
\subfigure[\label{fig:standard_2_l}]{\includegraphics[scale=.7]{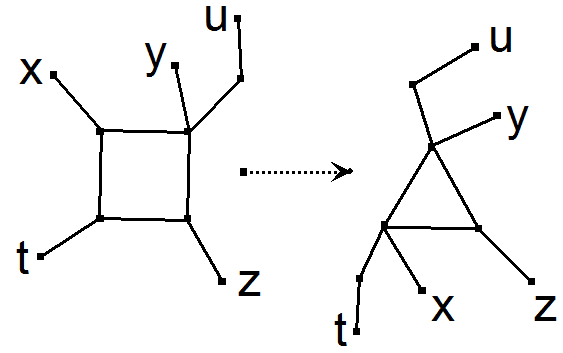}}\;\;\;
\subfigure[\label{fig:standard_3_l}]{\includegraphics[scale=.7]{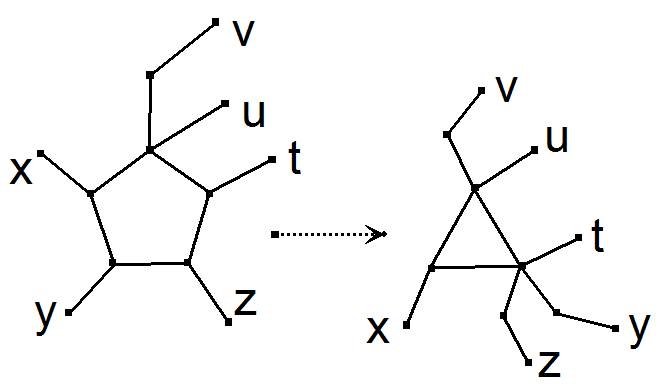}}\;\;\;
\subfigure[\label{fig:standard_4_l}]{\includegraphics[scale=.7]{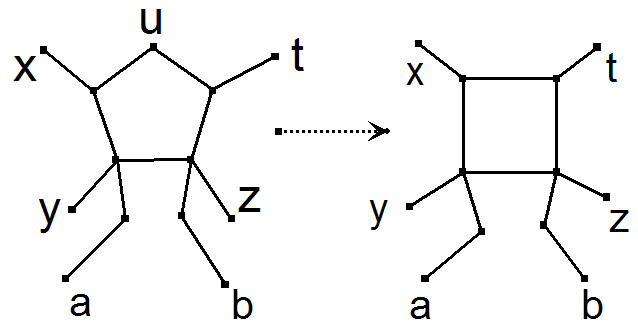}}
\label{fig:standard_l}
\caption{Examples of the simplification}
\end{center}
\end{figure}

Let us suppose that there is a root $N$ in which there exists a $3$ or $4$ or $5$-cycle having only one vertex which is adjacent to other inner vertex not on the cycle. An easy observation shows that in such a case we can simplify the root to another simpler network (in each case the cycle is shortened) without changing the $4$-leaf power and the properties of disjoint cycles of the network as illustrated in Figures \ref{fig:standard_1_l}, \ref{fig:standard_2_l}, \ref{fig:standard_3_l}. If there is a $5$-cycle with one invisible vertex $u$ as in Figure \ref{fig:standard_4_l} such that $p(x),  p(t)$ and $u$ are not adjacent to any other inner vertex not on the cycle, then we can replace it by the $4$-cycle in the right.  We apply such a transformation at most once for each cycle.

\textbf{Therefore  we can suppose that the networks that we construct are connected, do not have any two adjacent invisible vertices and do not contain any such $3,4,5$-cycle. Moreover, according to Lemma \ref{lem:blockclique}, then induced subgraph of the roots on each block clique of $G$ is an invisible star}.

\subsection{Sketch of the analysis of the 4-leaf case}

\begin{figure}[ht]
\begin{center}
\def\svgwidth{14cm}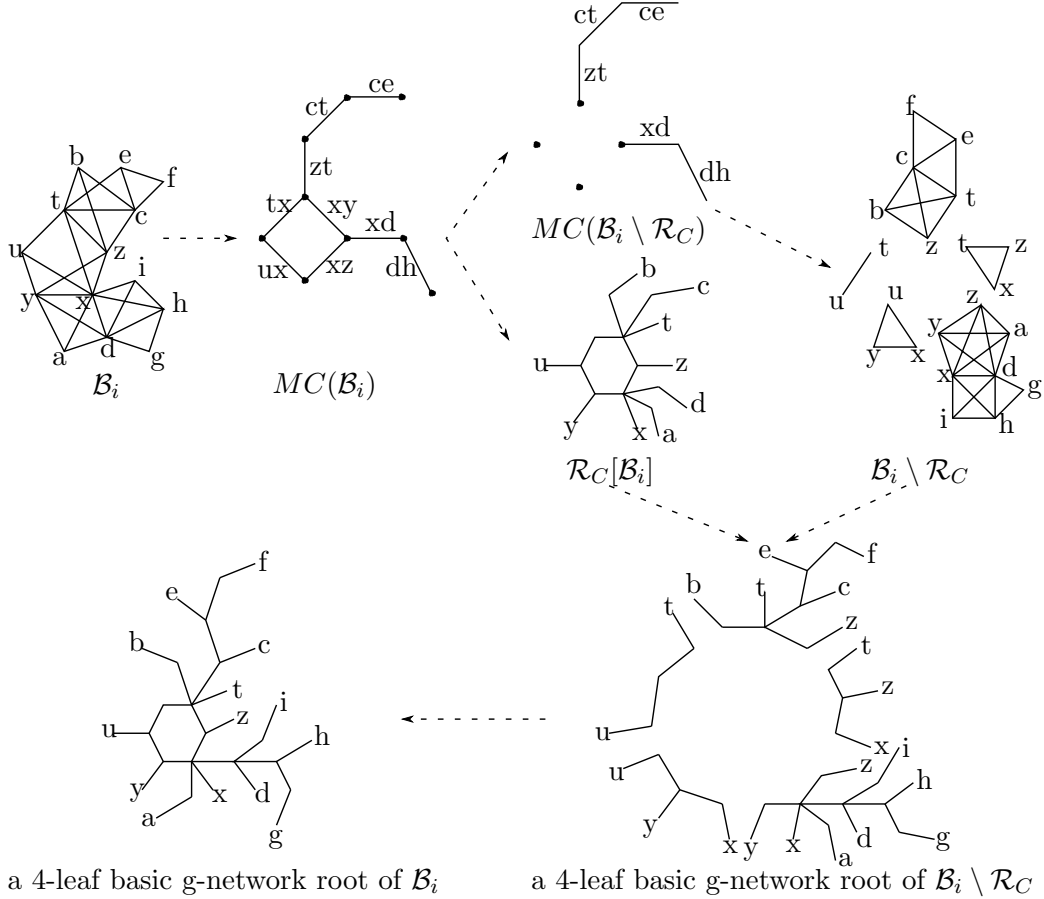
\caption{Construction a $4$-leaf basic g-network root of a block $\mathcal{B}_i$ of $G$}
\label{fig:theorem}
\end{center}
\end{figure}

We consider separately each block of $G$. The steps of the algorithm on one block is resumed in the example of Figure \ref{fig:theorem}.
Firstly, we calculate the maximal cliques graph $MC(\mathcal{B}_i)$ of each block $\mathcal{B}_i$. Next, we find all particular subgraphs of $MC(\mathcal{B})$ such that every root of each one has exactly one cycle. Some conditions must be taken into account on the root set of these subgraphs in order that they are compatible with the others in the same block as well as in the adjacent blocks. After this step, we have the set 
of cycles of a $4$-leaf basic g-network of $G$, namely cycle-root set $\mathcal{R}_c$. 
Therefore, by separating the corresponding subgraphs from $\mathcal{B}_i$ the obtain graph $\mathcal{B}\backslash \mathcal{R}_c$ must be the square of a forest (Theorem \ref{theo:4power}). The later problem is known to be linear. So, it is easy to construct a root of $\mathcal{B}\backslash \mathcal{R}_c$ which is compatible with the cycle-root set $\mathcal{R}_c$. So, by combining them together, we obtain a $4$-leaf basic g-network of $\mathcal{B}_i$.
The final step is combining the root of each block to have the corresponding root of $G$.

Hence, recognition of $4$-leaf basic g-network powers can be done in polynomial time. As a consequence, squares of basic g-networks can also be recognized in polynomial time.

Almost the remaining of this paper is devoted to construct the cycles in the roots of $G$ (Section \ref{sec:cycle}). For this purpose, we introduce some special subgraphs of $MC(G)$ namely $\mathcal{C}_e(G)$, $\mathcal{C}_o(G)$ (when $G$ is biconnected) and $\mathcal{C}_c(G)$ (when $G$ is not biconnected).

\subsection{Detecting the cycles in the roots of $G$.}
\label{sec:cycle}

We call a cycle is \textit{big} \index{big cycle} if it contains at least $7$ vertices, otherwise it is \textit{small} \index{small cycle}. It should be noticed that all cycles that we mention here (cycles of $MC(G)$ or cycles of networks) are \textit{chordless cycles}. We will denote the cycle in $N$ by the lower-case letter $c$ and the cycle in $MC(G)$ by the upper-case letter $C$. 

\begin{minipage}[b]{12.5cm}
Let $N$ be a basic g-network. For any cycle $c$ of $N$, let us define the set of labels of the visible vertices of $c$ as $l(c) = \{x \in \mathcal{L}~|~p(x) \in c\}$ \index{$l(c)$}. Denote by $S(c)$ \index{$S(c)$} the subnetwork of $N$ consisting of $c$ and all (quasi) visible and invisible stars having a vertex of $c$ as their middle vertices. 
\begin{example} 
In the figure in the right, we have a cycle $c$ and $S(c)$. By definition, $l(c) = \{x_1, x_2, x_3, x_4, x_5\}$. 
\end{example}
\end{minipage}\hfill
\begin{minipage}[b]{3cm}
\begin{center}
\includegraphics[scale=.7]{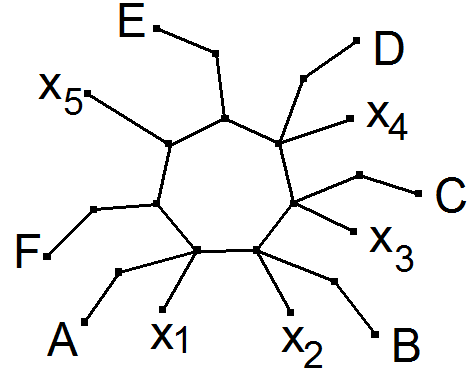}
\end{center}
\end{minipage}

\begin{lemma}
\label{lem:inv_cycle}
If $G$ is biconnected, is not a clique and has a $4$-leaf basic g-network root $N$, then:

(i) Any invisible vertex of $N$ is contained in a cycle. 

For any cycle $c$ of $N$ and for any vertex $u$ on $c$: 

(ii) If $u$ is visible then $u$ is not adjacent to any invisible vertex not on $c$. Furthermore, if $u$ is adjacent to only invisible vertex on $c$, then $u$ is not adjacent to any other inner vertex not on $c$.


(iii) If $u$ is invisible, let $s$ be the (quasi) invisible star having $u$ has its middle vertex, then for any visible vertex $v$ of $s$ which is not on $c$, $v$ is not adjacent to any other inner vertex different from $u$.
\end{lemma}
 
\begin{pf}
(i) Suppose that $N$ has an invisible vertex $u$ which is not contained in any cycle.

\begin{minipage}[b]{12.5cm}
Let $x,y$ be two leaves of the invisible star of $u$, and let $z$ be a leaf not in this invisible star which has distance at most $4$ with $x$.
Since $G$ is biconnected, there is at least a path in $G$ from $y$ to $z$ which does not pass $x$. We deduce that in $N$ there is at least a path from $p(y)$ to $p(z)$ that does not pass $p(x)$. In other words, there is a cycle passing $u$, a contradiction.
\end{minipage}\hfill
\begin{minipage}[b]{3cm}
\begin{center}
\includegraphics[scale=.7]{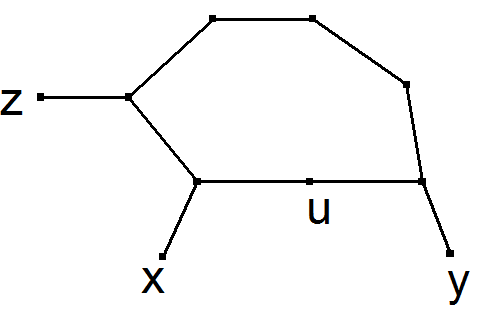}
\end{center}
\end{minipage}

(ii) Let $c$ be a cycle of $N$ and let $u$ be a visible vertex on $c$. 

\begin{minipage}[b]{13.5cm}
\hspace*{.5cm}Suppose that $u$ is adjacent to an invisible vertex $i$ not on $c$. Let $v$ be another vertex adjacent to $i$, so $v$ is visible. Let $x,y$ be the leaves of $u,v$.
Let $t$ be another leaf such that $p(t)$ is on $c$. If in $N$ all path from $t$ to $y$ passes by $u$, then  all paths from $t$ to $y$ in $G$ must pass $x$. This is a contradiction because $G$ is biconnected. So there is another path from $p(t)$ to $p(y)$ which does not pass $u$, this means that there is another cycle different from $c$ intersecting with $c$. This is impossible because $N$ is a g-network.
\end{minipage}\hfill
\begin{minipage}[b]{2cm}
\begin{center}
\includegraphics[scale=.7]{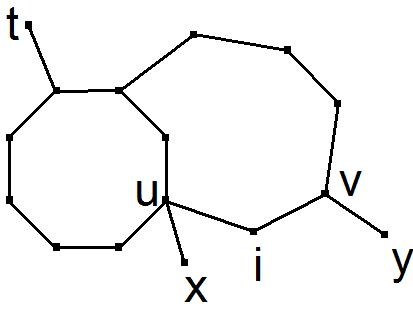}
\end{center}
\vspace*{.5cm}
\end{minipage}

\begin{minipage}[b]{13cm}
\hspace*{.5cm}Suppose that $u$ is adjacent to $2$ invisible vertices on $c$, let $x_1,x_2$ be the leaves of these two invisible vertices. Suppose that $u$ is adjacent to a vertex $v$ not on $c$. As showed above, $v$ must be visible, let $y$ be the leaf of $v$. In order that $y$ is contained in the same block with $x,x_1,x_2$, there must exist a path from $y$ to $x_1$ in $G$ which does not pass $x$. It implied that in $N$ there  exists a path from $p(y)$ to $p(x_1)$ which does not pass $p(x)$. However, that will create a cycle intersecting with $c$, a contradiction.
\end{minipage}\hfill
\begin{minipage}[b]{2.5cm}
\begin{center}
\includegraphics[scale=.7]{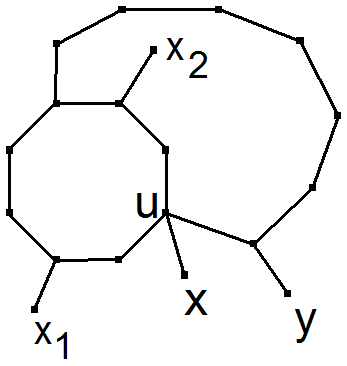}
\end{center}
\vspace*{.2cm}
\end{minipage}

\begin{minipage}[b]{10.5cm}
(iii) Let $c$ be a cycle of $N$ and let $u$ be an invisible vertex on $c$. 
Let $x$ be a leaf of the star $s$ of $u$ such that $p(x)$ is not on $c$. Let $y$ be a leaf of this star such that $p(y)$ is on $c$. Suppose that $p(x)$ is adjacent to an inner vertex different from $u$, then there is a leaf $z$ not in $s$ such that $d_{N_0}(x,z) \le 4$. Then, in $G$ there is a path from $z$ to $y$ which passes $x$.
Because $G$ is biconnected, there must be a path from $z$ to $y$ in $G$ which does not pass $x$. That will produce in $N$ another cycle intersecting with $c$ at $u$ (there are two cases indicated in the figure), a contradiction.
\end{minipage}\hfill
\begin{minipage}[b]{5cm}
\begin{center}
\includegraphics[scale=.7]{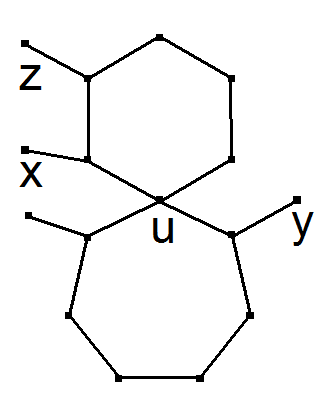}
\includegraphics[scale=.7]{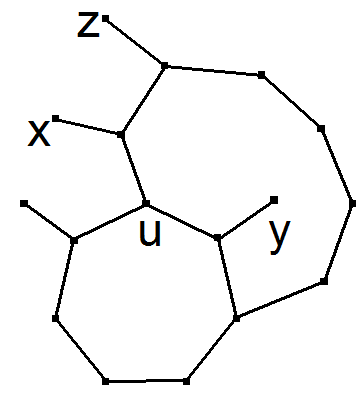}
\vspace*{.5cm}
\end{center}
\end{minipage}
\end{pf} 

\begin{figure}[ht]
\begin{center}
\subfigure[$C_1$\label{fig:C+1}]{\includegraphics[scale=.25]{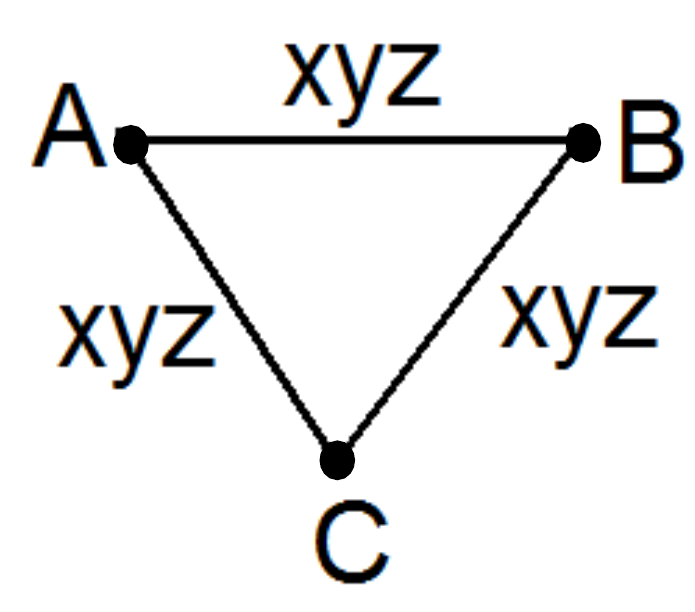}}
\subfigure[$C_2$\label{fig:C+2}]{\includegraphics[scale=.25]{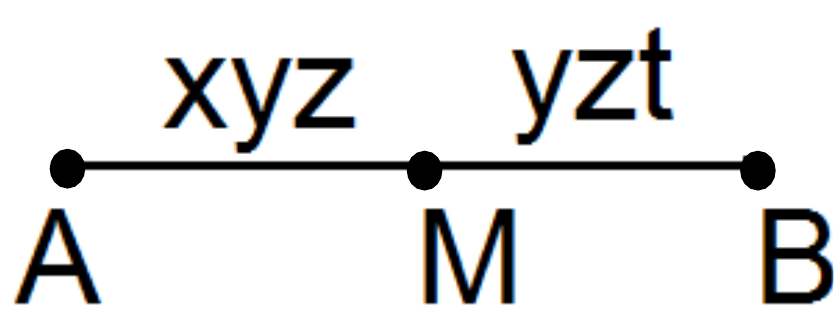}}
\subfigure[$C_3$\label{fig:C+3}]{\includegraphics[scale=.25]{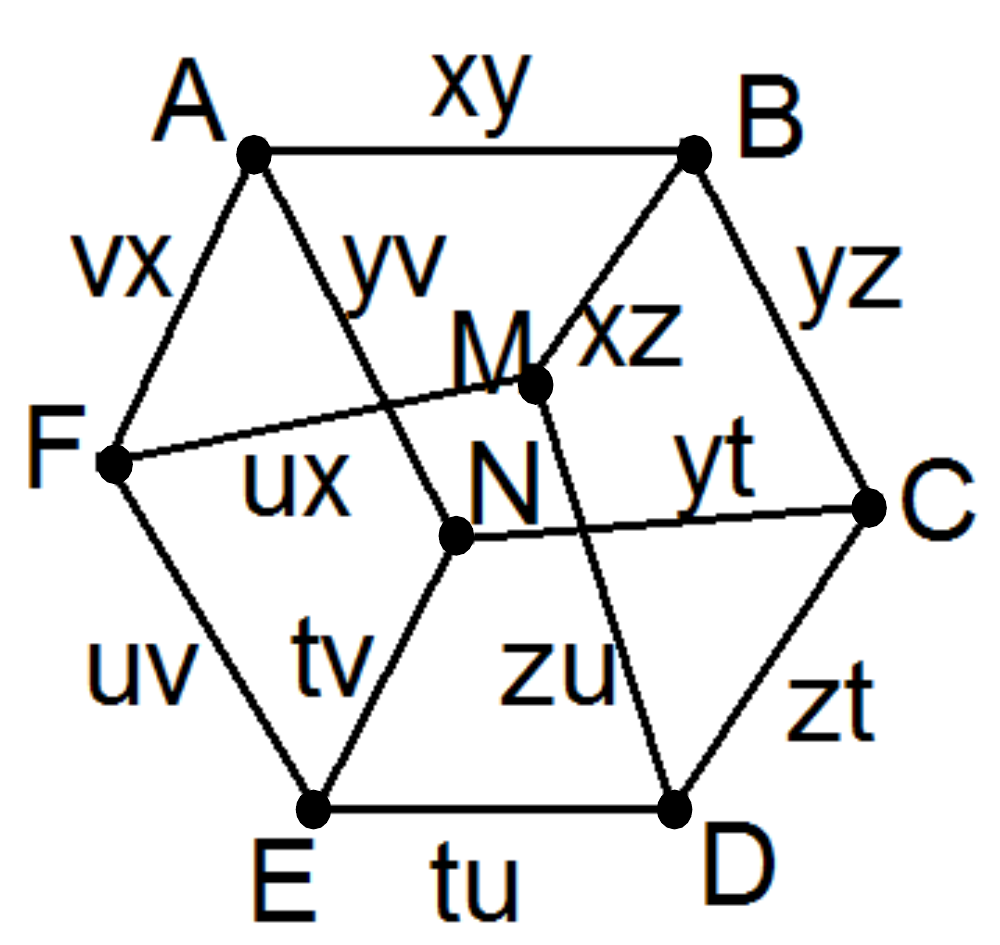}}
\subfigure[$C_4$\label{fig:C+4}]{\includegraphics[scale=.25]{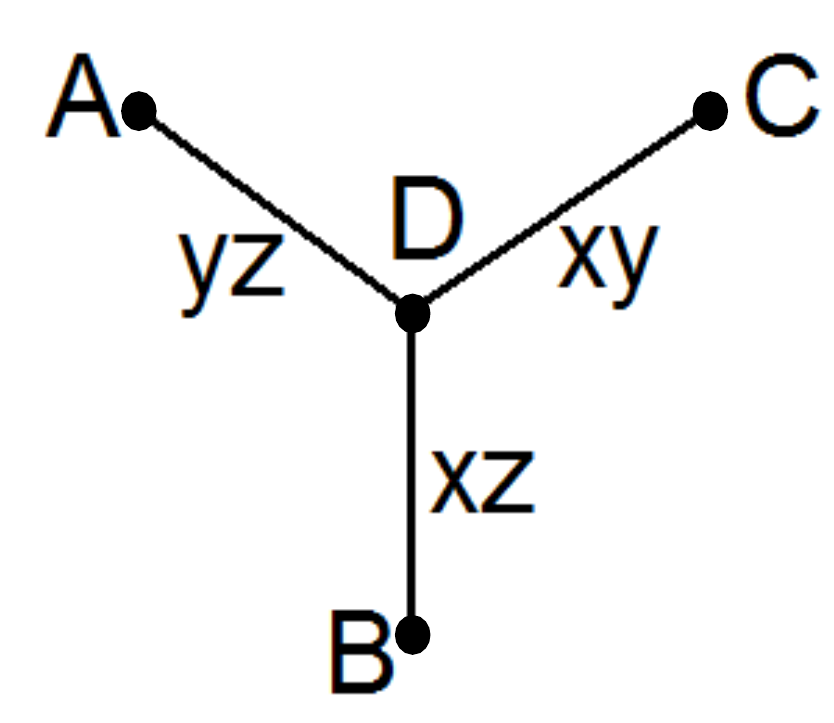}}
\subfigure[$C_5$\label{fig:C+5}]{\includegraphics[scale=.25]{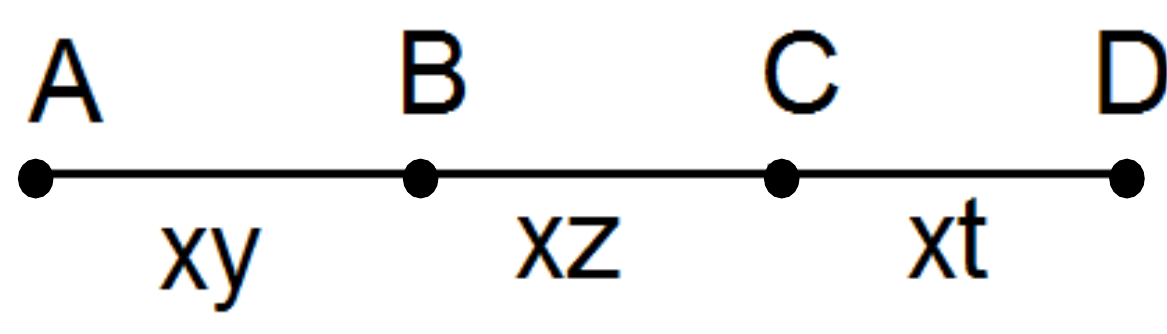}}
\subfigure[$C_6$\label{fig:C+6}]{\includegraphics[scale=.25]{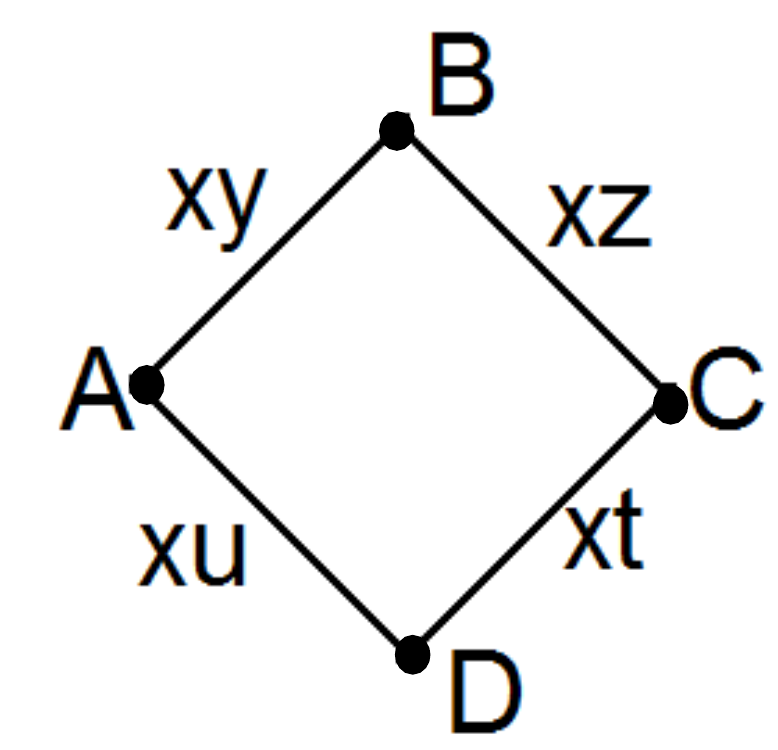}}
\subfigure[$C_7$\label{fig:C+7}]{\includegraphics[scale=.25]{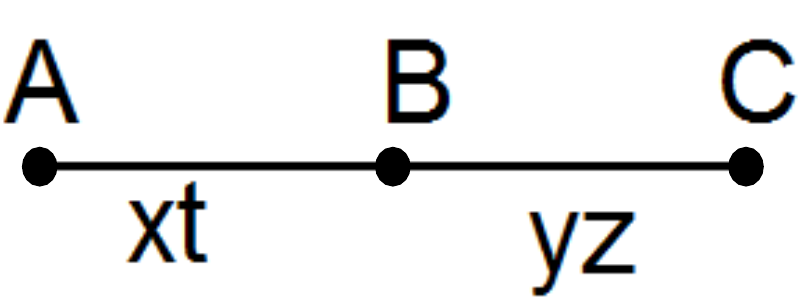}}
\subfigure[$C_8$\label{fig:C+8}]{\includegraphics[scale=.25]{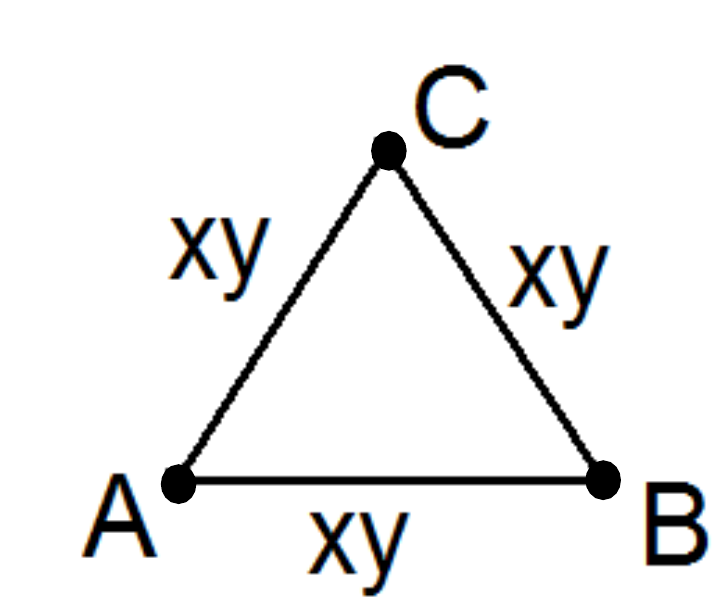}}
\subfigure[$C_9$\label{fig:C+9}]{\includegraphics[scale=.25]{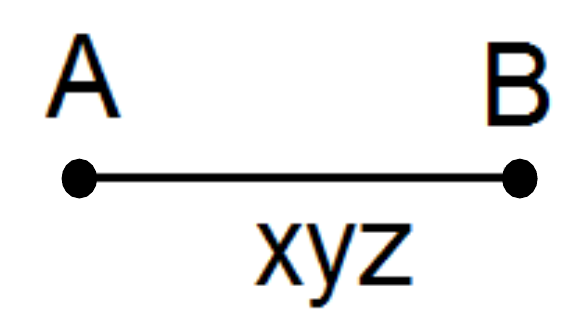}}
\subfigure[$C'$\label{fig:C+'}]{\includegraphics[scale=.25]{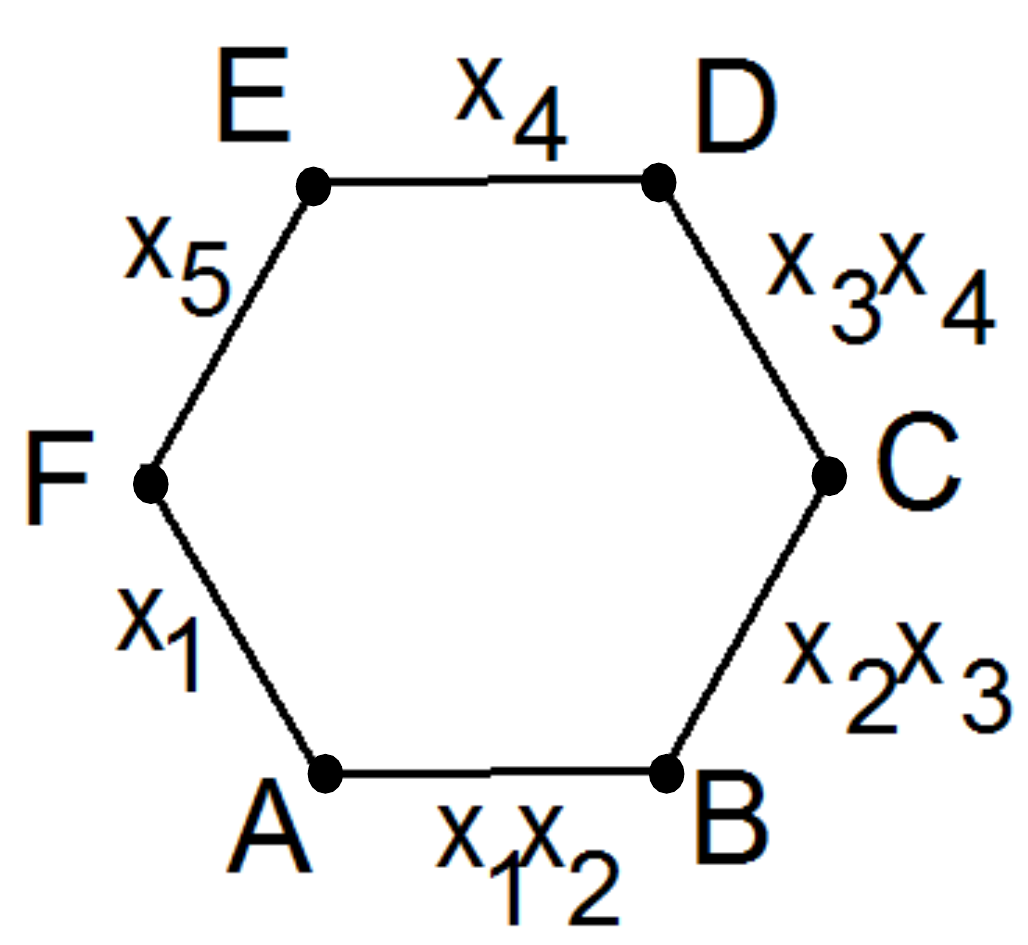}}
\caption{Some subgraphs of $MC(G)$ that imply cycles in the roots of $G$ \index{$C_i$ for $i=1, \dots, 9$}}
\label{fig:subgraphs}
\end{center}
\end{figure}

We introduce a list of subgraphs $C_1, \dots, C_9$ of $MC(G)$ that will be used in the sections below (Figure \ref{fig:subgraphs}) to detect the cycles in the roots of $G$.

$C_2$ is a star of $MC(G)$ consisting of at least two edges such that all of its edges have weight $3$ and non of them  belongs to a subgraph $C_1$. 

$C_6$ is a cycle which is not contained in any subgraph $C_3$. 

$C_9$ is a segment of weight $3$ which is not contained in any subgraph $C_1$ or $C_2$. 

$C'$ \index{$C'$} is a cycle of $MC(G)$ which is different from $C_1,C_6$ and $C_8$. 

The other subgraphs have exactly the forms described in the corresponding figures. 

The following lemma states a relation between the cycles of $MC(G)$ and the cycles in the $4$-leaf basic g-network roots of $G$.

\begin{lemma}
Suppose that $G$ has a $4$-leaf basic g-network root $N$. Then for any cycle $C$ of $MC(G)$, which is not a $4$-cycle in any subgraph $C_3$ of $MC(G)$, there is a cycle $c$ in $N$ such that $l(C)=l(c)$.

And for any subgraph $S$ isomorphic to $C_3$ of $MC(G)$, there is a $6$-cycle without invisible vertex $c$ in $N$ such that $l(S)=l(c)$.
\label{lem:cycle}
\end{lemma}

\begin{pf}
Let $C= [A_1, \dots A_k, A_1]$ be a cycle in $MC(G)$. Let us denote  the label of $A_iA_{i+1}$ by $l_i$. 

Using Observation \ref{ob:1}, $N$ can be constructed by replacing each  $A_i$ by a subnetwork of $\mathcal{F}$ along the cycle in a way such that  if $A_i,A_{i+1}$ has $l_i$ in common then their corresponding subnetworks also have the leaves $l_i$ in common. 
Since all the graphs in  $\mathcal{F}$ are connected, that creates a cycle $c$ in $N$. 

\begin{figure}[ht]
\begin{center}
\subfigure[\label{fig:3_0i}]{\includegraphics[scale=.7]{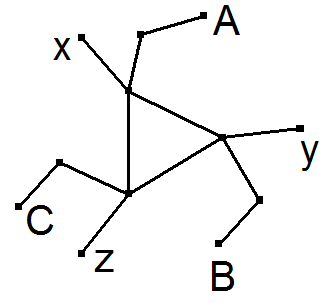}\includegraphics[scale=.25]{Figures/C1.pdf}}\;\;
\subfigure[\label{fig:3_1i}]{\includegraphics[scale=.7]{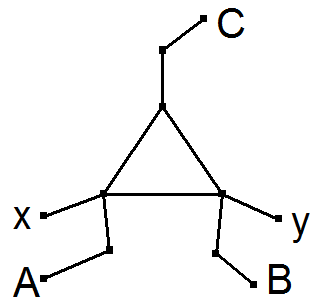}\includegraphics[scale=.25]{Figures/C8.pdf}}\;\;
\subfigure[\label{fig:5_2i}]{\includegraphics[scale=.7]{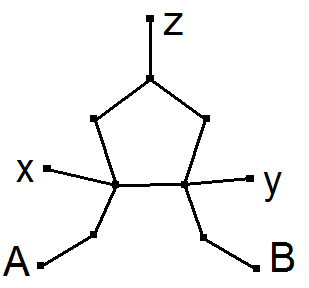}\includegraphics[scale=.25]{Figures/C8.pdf}}\;\;
\subfigure[\label{fig:6_1i}]{\includegraphics[scale=.7]{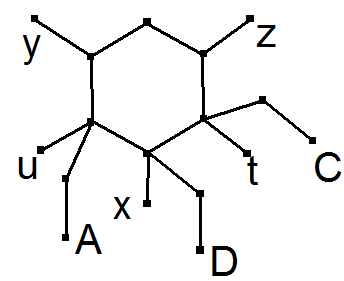}\includegraphics[scale=.25]{Figures/C6.pdf}}
\caption{Some small cycles of $N$ and the corresponding cycles in $MC(G)$}
\label{fig:small_cycle}
\end{center}
\end{figure}

So if there is an $A_i$ such that $N[A_i]$ contains a cycle, then this cycle intersects with $c$. Since the cycles in $N$ must be pairwise disjoint, in this case the cycle of $N[A_i]$ must be exactly $c$. 
As indicated in Observation \ref{ob:1}, the subnetworks of $\mathcal{F}$ contain cycles of at most $6$ vertices. By taking the $4$-leaf power of all possible configurations of $S(c)$, there are only $4$ cases that the corresponding maximal cliques graph is a cycle: A $3$-cycle without any invisible vertex (Figure \ref{fig:3_0i}), a $3$-cycle with exactly one invisible vertex (Figure \ref{fig:3_1i}), a $5$-cycle with exactly two invisible vertices (Figure \ref{fig:5_2i}), a $6$-cycle with one invisible vertex (Figure \ref{fig:6_1i}). It can be verified that in all these cases that $l(c)$ equals $l(C)$. 

\begin{figure}[ht]
\begin{minipage}[b]{80mm}
\begin{center}
\includegraphics[scale=.7]{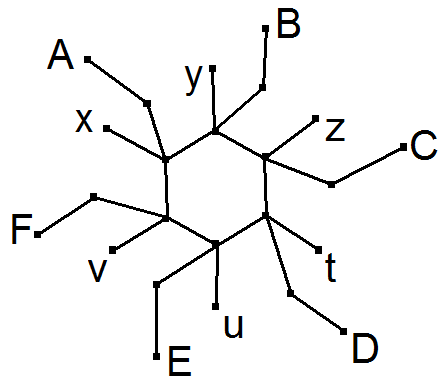}\includegraphics[scale=.25]{Figures/C3.pdf}
\caption{A $6$-cycle without invisible vertex and its maximal cliques graph.}
\label{fig:R3}
\end{center}
\end{minipage}\hfill
\begin{minipage}[b]{80mm}
\begin{center}
\includegraphics[scale=0.25]{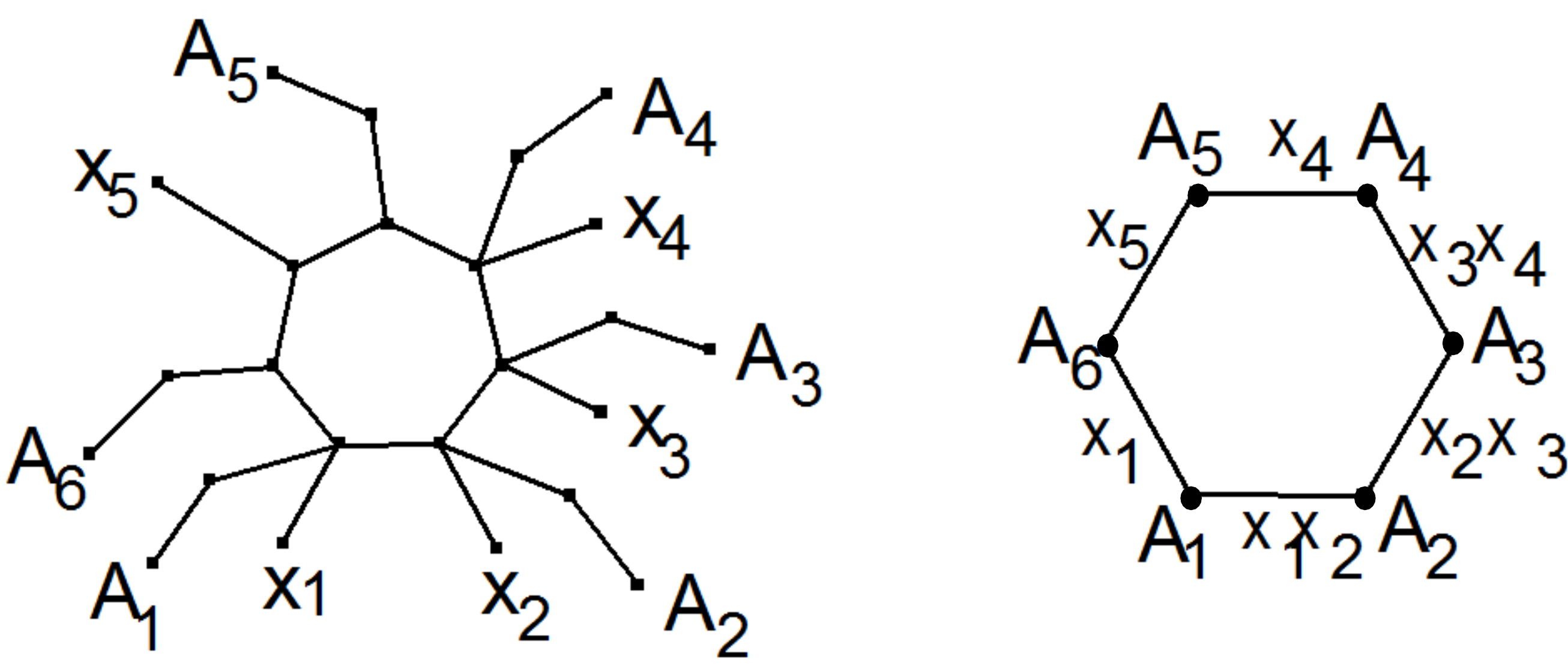}
\caption{$N[A_i]$ is either a visible or invisible star.}
\label{fig:A_i}
\end{center}
\end{minipage}
\end{figure}

Especially, for the case of $6$-cycle without any invisible vertex, the corresponding maximal cliques graph is isomorphic to the graph $C_3$ (Figure \ref{fig:R3}). For the other cases, there is no cycle in the corresponding maximal cliques subgraphs.

On the other hand, if there is not any $A_i$ such that $N[A_i]$ contains a cycle, then $N[A_i]$ is a visible star or an invisible star for all $i$. By replacing each node $A_i$ by a visible or invisible star having $A$ as the leaf set such that if the two maximal cliques have $x,y$ (or $x$) in common then the corresponding stars also have two leaves $x,y$ (or $x$) in common, we obtain a cycle $c$ having $l(c)$ such that  $l(c)$ is equal to $l(C)$ (Figure \ref{fig:A_i}).
\end{pf}

This lemma shows that $MC(G)$ can contain chordless cycles or $G$ can contain chordless cycles. This makes a big difference with the $k$-leaf tree power graphs which were proved to be strongly chordal \index{strongly chordal graph} \cite{BL06}.

\subsubsection{Easy cycles \index{easy cycle} when $G$ is biconnected}

\label{sec:easy}

They consist of all the big cycles, some $6$-cycles with $2$ invisible vertices, and some small cycles without invisible vertices.

Let us denote by $\mathcal{C}_1(G)$ the set of subgraphs isomorphic to $C_1$ of $MC(G)$. So it contains all triangles of $MC(G)$ with $3$ edges of the same label weighted by $3$. 

$\mathcal{C}_2(G)$ the set of all subgraphs $C_2$ of $MC(G)$.

$\mathcal{C}_3(G)$ the set of all subgraphs isomorphic to $C_3$ of $MC(G)$.
 
$\mathcal{C}'(G)$ the set of cycles of $MC(G)$ which are different from $C_1$, $C_6$ and $C_8$.

Let $\mathcal{C}_e(G) = \mathcal{C}'(G) \cup \mathcal{C}_1(G) \cup \mathcal{C}_2(G) \cup \mathcal{C}_3(G)$ ($e$ for easy). \index{$\mathcal{C}_e(G)$}

\begin{lemma} Suppose that $G$ has a $4$-leaf basic g-network root $N$. For any cycle $C=[A_1A_2 \dots A_k,A_1]$ of $\mathcal{C}'(G)$, $N[A_i]$ is either an invisible star or a visible star, and it is an invisible star iff $|A_{i-1}A_i|=|A_iA_{i+1}|=1$.
\label{lem:type1}
\end{lemma}

\begin{pf}
$C$ is not a $4$-cycle in any subgraph of $\mathcal{C}_3(G)$ because $C \in \mathcal{C}'(G)$. By Lemma \ref{lem:cycle}, any root $N$ of $G$ contains a cycle $c$ such that $l(c)=l(C)$. 

The first case of the proof of Lemma \ref{lem:cycle} can not happen because in this case, the corresponding maximal cliques graph is either a triangle or a $4$-cycle having all edges of weight $2$, so can not be in $\mathcal{C}'(G)$.

For the second case, $N[A_i]$ is either a star or an invisible star for any $i$.  
Remark that if an invisible star has two common vertices with another star or invisible star, then a $4$ cycle is created. This cycle intersect with $c$, a contradiction. Hence an invisible star here has at most one common vertex with other stars. In other words, if $N[A_i]$ is an invisible star then $A_{i-1}A_i, A_iA_{i+1}$ have weight $1$. Conversely, suppose that $N[A_i]$ is a visible star, we will prove that either $|A_{i-1}A_i|$ or $|A_iA_{i+1}|$ is greater than $1$. If $N[A_{i-1}]$ (resp. $N[A_{i+1}]$) is also visible star, then  $A_iA_{i-1}$ (resp. $A_iA_{i+1}$) has weight $2$. If both $N[A_{i-1}]$ and $N[A_{i+1}]$ are invisible stars, then by Lemma \ref{lem:inv_cycle} (ii), the middle vertex of  $N[A_i]$ is not adjacent to any other inner vertex not on  $c$. The later implies that the visible star $N[A_i]$ consists of only a middle vertex and a leaf, i.e. it is included in $N[A_{i-1}]$ and $N[A_{i+1}]$, a contradiction. So, $N[A_i]$ is an invisible star if and only if the edges $A_{i-1}A_i, A_iA_{i+1}$ have weight $1$. 
\end{pf}

\begin{figure}[ht]
\begin{center}
\subfigure[A triangle of $\mathcal{C}_1(G)$ and one of its root\label{fig:R_1}]{\includegraphics[scale=.25]{Figures/C1.pdf}\includegraphics[scale=.7]{Figures/R_1,1.png}}\;\;\;\;
\subfigure[A subgraph of $\mathcal{C}_2(G)$ and some of its roots \label{fig:R_2}]{\includegraphics[scale=.25]{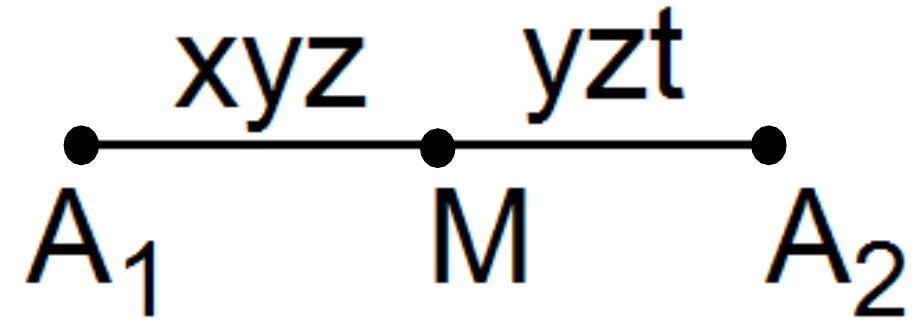}\includegraphics[scale=.7]{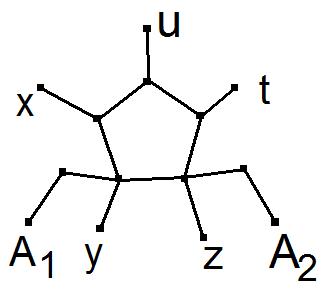}\includegraphics[scale=.7]{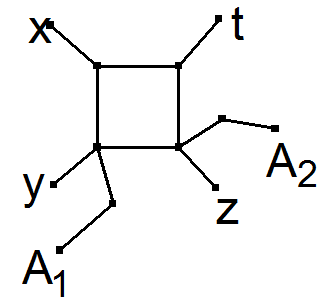}\includegraphics[scale=.7]{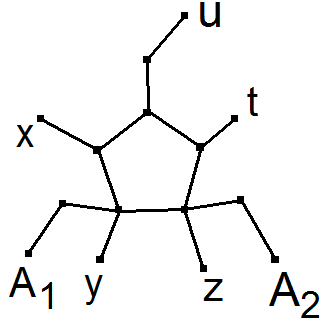}}
\caption{Illustration of Lemma \ref{lem:type1_small}}
\end{center}
\end{figure}

\begin{lemma} Suppose that $G$ has a $4$-leaf basic g-network root $N$, then:

1. For each triangle $A_1A_2A_3$ $\in \mathcal{C}_1(G)$, $N[A_1]$, $N[A_2]$, $N[A_3]$ are visible quasi stars sharing the same triangle.

2. For any star $S  = MA_1\dots A_k$ of $\mathcal{C}_2(G)$ where $M$ is the middle node and $k \ge 2$, each $N[A_i]$ is a visible star, $N[M]$ is a cycle having the label set equals to $l(M)$, and:

- either $|M|=5$, then $N[M]$ is an $N_5$.

- or $|M|=4$, then:

\hspace*{.5cm} If there is not any edge of weight $2$ in $MC(G)$ incident to $M$, then $N[M]$ is an $N_4$. 

\hspace*{.5cm} Otherwise, $N[M]$ is a $N'_5$ and $k=2$. Let $MM'$ be an edge of weight $2$ in $MC(G)$. Then, if $l(MM')$ is disjoint with $l(MA_1) \cap l(MA_2)$ then $N[M']$ is the invisible star having the only invisible vertex on the cycle of $N[M]$ as its middle vertex. Otherwise, $N[M']$ is a visible star.

3. For each subgraph $S$ of $\mathcal{C}_3(G)$, $N[S]$ contains a $6$-cycle $c$ without invisible vertex such that $l(c)=l(S)$. Moreover, for any node $A$ of $S$ which has a neighbour node in $MC(G)$ not in $S$, $N[A]$ is a visible star.
\label{lem:type1_small}
\end{lemma}

\begin{pf}
1. This result is the Claim (iii) of Observation \ref{ob:2}. 

2. For any $i=1, \dots, k$, by Claim (v) of Observation \ref{ob:2} we deduce that either both $N[M]$ and $N[A_i]$ are visible quasi stars, or exactly one of them is an $N_4$ or $N_5$ or $N'_5$. 
If  $N[M]$ is a visible quasi star, then there can not be $2$ edges having weight $3$ incident to $M$, except when $MA_i$ is in a triangle of $\mathcal{C}_1(G)$ which is a contradiction with the definition of $\mathcal{C}_2(G)$. So, $N[M]$ is either an $N_4$ or $N_5$ or $N'_5$. Remark that in these cases the label set of the cycle in $N[M]$ is equal to $l(M)$. So:

- If $|M|=5$ then $N[M]$ is an $N_5$.

- If $|M|=4$ then $N[M]$ is either an $N_4$ or $N'_5$.

\hspace*{.5cm} Remark that if $N[M]$ is an $N_4$, then all edges of $M$ have weight $3$. So, if there is an edge $MM'$ of weight $2$, then $N[M]$ must be a $N'_5$ (the last cycle in Figure \ref{fig:R_2}). In this figure, we can see that only the two stars having $p(y)$ and $p(z)$ as middle vertices can have $3$ common leaves with $N[M]$. Therefore, if $l(MM')$ is disjoint with $l(MA_1) \cap l(MA_2)$, then $l(MM')= \{x,t\}$, and $N[M']$ is the invisible star having the invisible vertex of $N[M]$ as its middle vertex. 

\hspace*{.5cm} If there is not any edge of weight $2$ incident to $M$, then $N[M]$ can be either an $N_4$ or $N'_5$. However, with the constrains that we impose on $N$ (Figure \ref{fig:standard_4_l}), we choose $N[M]$ as an $N_4$.

3. In the proof of Lemma \ref{lem:cycle}, we proved that if  $S$ is a  $C_3$, then  $N[S]$ contains a $6$-cycle $c$ without invisible vertex such that $l(c) = l(S)$. In a  $S(c)$, there are  $8$ networks of the family  $\mathcal F$ consisting of $2$ subnetworks $N_6$ and $6$ visible stars. Observing that if a maximal clique $A$ corresponds to an $N_6$ then it can not have a neighbour node not in $S$, otherwise another cycle intersecting with $c$ will be created. So if a node $A$ of $S$ has a neighbour not in $S$, then it $N[A]$ must be a visible star.
\end{pf}

Lemmas \ref{lem:type1} and \ref{lem:type1_small} are constructive. So it is possible to construct the roots of each subgraph $S \in \mathcal{C}_e(G)$ if we can determine the label of the middle vertex \index{middle vertex} of each visible (quasi) star in the root. This is also easily done with the following remark.

\begin{remark}[Middle vertex]

Let $N$ be a $4$-leaf basic g-network root of $G$. Let $A$ be a node of $MC(G)$ such that $N[A]$ is a visible (quasi) star. Let $B$ be a neighbour of $A$ in $MC(G)$ such that $|AB| \ge 2$ and $N[B]$ is not an $N_6$. Then the label of the middle vertex of $N[A]$ is contained in $l(AB)$.
\label{rem:middle}
\end{remark}

\begin{figure}[ht]
\begin{center}
\subfigure[]{\includegraphics[scale=.7]{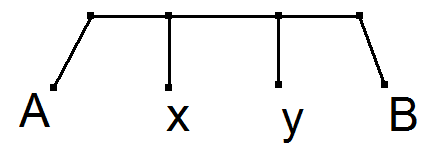}}
\subfigure[]{\includegraphics[scale=.7]{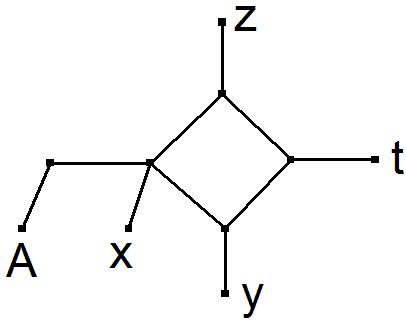}}
\subfigure[]{\includegraphics[scale=.7]{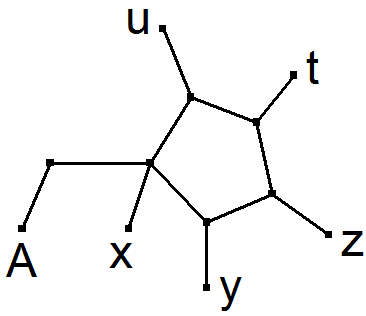}}
\subfigure[]{\includegraphics[scale=.7]{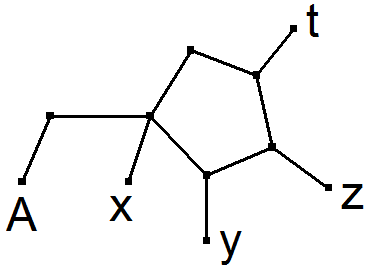}}
\subfigure[]{\includegraphics[scale=.7]{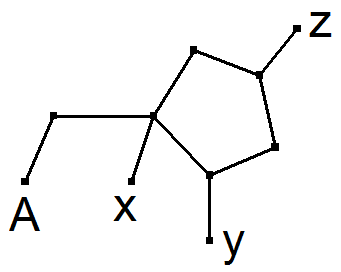}}
\caption{The leaf of the middle vertex of $N[A]$ is a leaf of $N[B]$}
\label{fig:middleA} 
\end{center}
\end{figure}

This remark can be easily proved by checking all the possible configurations of $N[B]$. Indeed, remark that $N[B]$ is not an invisible (quasi) star because otherwise $|AB|=1$. For any other configurations of $N[B]$, the leaf of the middle vertex of $N[A]$ is always a leaf of $N[B]$ (see Figure \ref{fig:middleA}).

So, the choice of the label of the middle vertex of $N[A]$ is unique except in some case when $A$ has at most one such neighbour node. 
For the later case, let $B$ be the only neighbour of $A$ such that $|AB| \ge 2$. Then it depends on the configuration of $N[B]$ that we can deduce easily the label of the middle vertex of $N[A]$. If $B$ does not have neither a further neighbour $C$ such that $|BC| \ge 2$, then the choice of the label of the middle vertex of $N[A]$ does not influence the construction of the remaining part of $N$. 
Remark that for any $S \in \mathcal{C}_e(G)$, except the label of the middle vertices of the visible (quasi) stars, the other part of the roots of $S$ is uniquely determined by Lemmas \ref{lem:type1}, \ref{lem:type1_small}. Hence, with the above remark, we can consider all roots of $S$ as only one root, because they have the same configuration and by replacing one by another, we do not change its $4$-leaf power as well as the properties of disjoint cycles.  With this convention, we have the following result.

\begin{corollary}
\label{co:e}
Each subgraph of $\mathcal{C}_e(G)$ has a unique root.
\end{corollary}

By Lemmas \ref{lem:type1}, \ref{lem:type1_small}, and by the condition of vertex-disjoint cycles in the roots of $G$, we deduce that all subgraphs of $\mathcal{C}'(G)$, $\mathcal{C}_1(G)$, $\mathcal{C}_3(G)$ must be node disjoint and not contain any middle node of any subgraph of $\mathcal{C}_2(G)$.

\subsubsection{Other cycles \index{other cycles} when $G$ is biconnected}
\label{sec:other}
The other cycles in the roots of $G$ when $G$ is biconnected consist of all small cycles which are not detected in the previous section. We will use the subgraphs $C_4, \dots, C_9$ of $MC(G)$ to recognize them. They are not \textit{easy} cycle \index{easy cycle} because each such subgraph can have several roots containing different cycles, and we must consider furthermore conditions to determine which cycles are contained in a $4$-leaf basic g-network root of $G$.

\begin{observation}
For $i = 4, \dots, 9$, every $4$-leaf basic g-network root of $C_i$ contains a small cycle, and any subgraph smaller than $C_i$ does not have this property. 
\end{observation}

\begin{figure}[ht]
\begin{minipage}[b]{85mm}
\begin{center}
\subfigure[$C_4$\label{fig:C4}]{\includegraphics[scale=.25]{Figures/C4.pdf}}
\subfigure[$R_{4,1}(A)$\label{fig:R_4,1}]{\includegraphics[scale=.7]{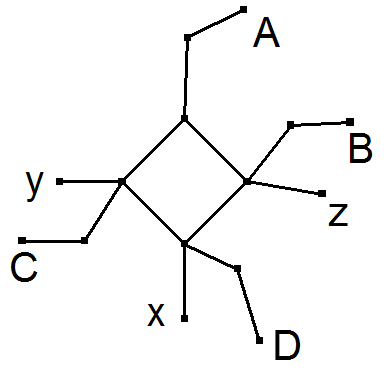}}
\subfigure[$R_{4,2}(A)$\label{fig:R_4,2}]{\includegraphics[scale=.7]{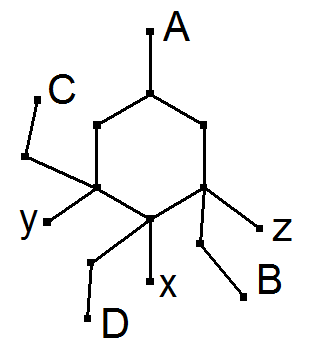}}
\subfigure[$R_{4,3}(xz)$\label{fig:R_4,3}]{\includegraphics[scale=.7]{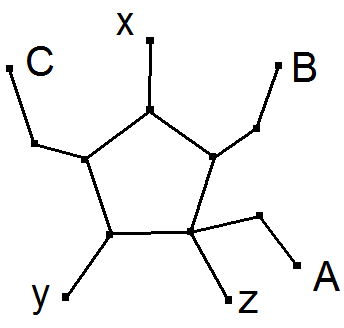}}
\subfigure[$R_{4,4}(x)$\label{fig:R_4,4}]{\includegraphics[scale=.7]{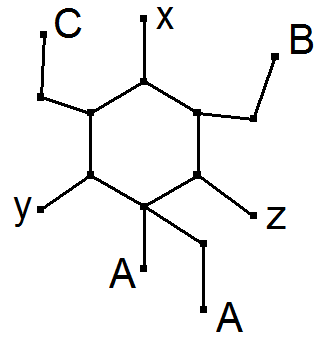}}
\subfigure[$R_{4,5}$\label{fig:R_4,5}]{\includegraphics[scale=.7]{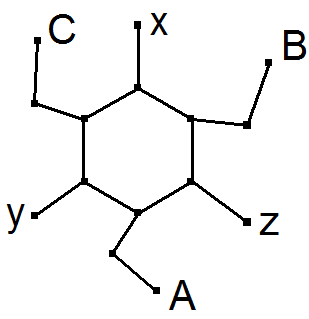}}
\caption{$C_4$ and its $4$-leaf basic g-network roots}
\label{fig:C_4}
\end{center}
\end{minipage}\hfill
\begin{minipage}[b]{75mm}  
\begin{center}
\hspace*{1cm}\subfigure[$C_5$\label{fig:C5}]{\includegraphics[scale=.25]{Figures/C5.pdf}}
\subfigure[$R_{5,1}(A)$\label{fig:R_5,1}]{\includegraphics[scale=.7]{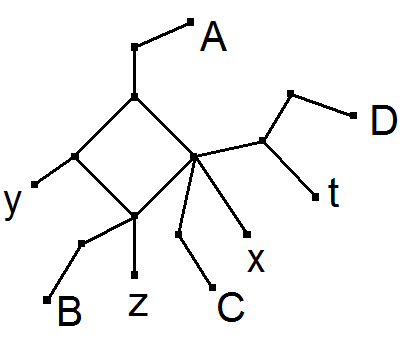}}
\subfigure[$R_{5,2}(A)$\label{fig:R_5,2}]{\includegraphics[scale=.7]{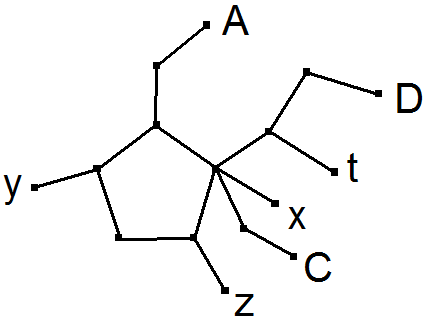}}
\subfigure[$R_{5,3}(A)$\label{fig:R_5,3}]{\includegraphics[scale=.7]{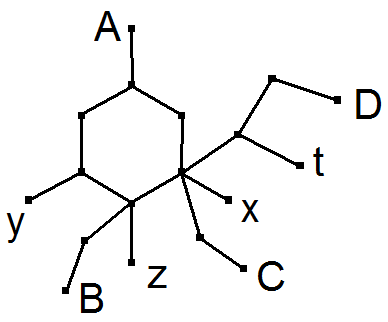}}
\subfigure[$R_{5,4}(A)$\label{fig:R_5,4}]{\includegraphics[scale=.7]{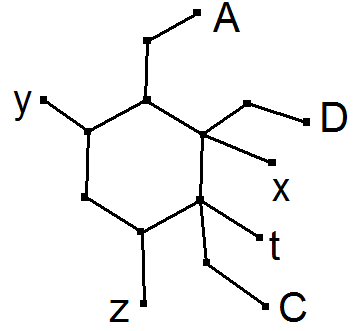}}
\caption{$C_5$ and its $4$-leaf basic g-network roots}
\label{fig:C_5}
\end{center}
\end{minipage}
\end{figure}

\begin{figure}[ht]
\begin{minipage}[b]{80mm}
\begin{center}
\subfigure[$C_8$\label{fig:C8}]{\includegraphics[scale=.25]{Figures/C8.pdf}}
\subfigure[$R_{8,1}(AB)$\label{fig:R_8,1}]{\includegraphics[scale=.7]{Figures/R_8,1.png}}
\subfigure[$R_{8,2}(AB)$\label{fig:R_8,2}]{\includegraphics[scale=.75]{Figures/R_8,2.png}}
\caption{$C_8$ and its $4$-leaf basic g-network roots}
\label{fig:C_8}
\end{center}
\end{minipage}\hfill
\begin{minipage}[b]{80mm}  
\begin{center}
\subfigure[$C_9$\label{fig:C9}]{\includegraphics[scale=.25]{Figures/C9.pdf}}
\subfigure[$R_{9,1}(xy)$\label{fig:R_9,1}]{\includegraphics[scale=.7]{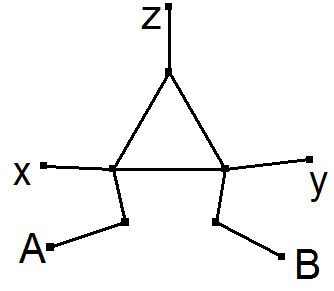}}
\subfigure[$R_{9,2}(Axy)$\label{fig:R_9,2}]{\includegraphics[scale=.8]{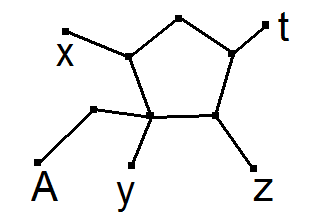}}
\caption{$C_9$ and its $4$-leaf basic g-network roots}
\label{fig:C_9}
\end{center}
\end{minipage}
\end{figure}

Indeed, using Observation \ref{ob:1} and by considering all the possible configurations, we find for each $C_i$ a finite list of its possible roots. They are the networks in Figures \ref{fig:C_4}, ..., \ref{fig:C_9}. \textit{Each root has exactly one cycle.} Note that $C_9$ can actually have some more roots but by the constraints imposed on the networks, its root have only $2$ forms $R_{9,1}, R_{9,2}$. However, not all networks listed in Figures \ref{fig:C_4}, ..., \ref{fig:C_9} are always roots of the corresponding $C_i$. We must check furthermore the size of each maximal clique. For example, with $C_4$, if the maximal clique $D$ contains more than $3$ vertices, then $R_{4,3}$ can not be its root because in this network $D$ corresponds to the maximal clique having $3$ vertices $x,y,z$.

\begin{figure}[ht]
\begin{minipage}[b]{50mm}
\subfigure[$C_6$\label{fig:C6}]{\includegraphics[scale=.25]{Figures/C6.pdf}}
\subfigure[$R_{6,1}(yztu)$\label{fig:R_6,1}]{\includegraphics[scale=.75]{Figures/R_6,1.png}}
\caption{$C_6$ and its $4$-leaf basic g-network roots}
\label{fig:C_6}
\end{minipage}\hfill
\begin{minipage}[b]{90mm}  
\subfigure[$C_7$\label{fig:C7}]{\includegraphics[scale=.25]{Figures/C7.pdf}}
\subfigure[$R_{7,1}(Ay)$\label{fig:R_7,1}]{\includegraphics[scale=.7]{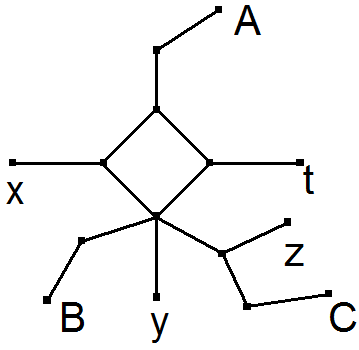}}
\subfigure[$R_{7,2}(xz)$\label{fig:R_7,2}]{\includegraphics[scale=.7]{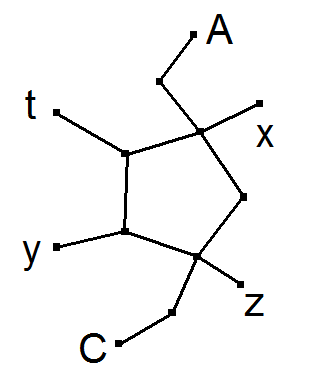}}
\subfigure[$R_{7,3}(Ay)$\label{fig:R_7,3}]{\includegraphics[scale=.7]{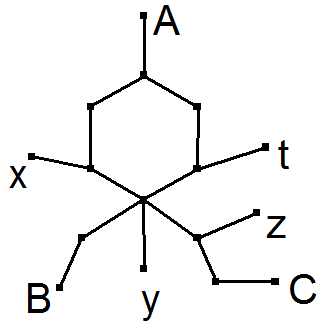}}
\caption{$C_7$ and its $4$-leaf basic g-network roots}
\label{fig:C_7}
\end{minipage}
\end{figure}

\begin{example}[Roots of $C_i$]

$C_5$ has $4$ forms or roots: $R_{5,1}$, $R_{5,2}$, $R_{5,3}$, $R_{5,4}$. Each form can have several derivations: for the form $R_{5,1}$, we have $R_{5,1}(A)$ as in Figure \ref{fig:R_5,1} and also $R_{5,1}(D)$ by permuting $A$ and $D$, $B$ and $C$. In general, the list of roots of each $C_i$ is generated up to permutation with respect to symmetry.  
\end{example}

These subgraphs in $MC(G)$ are not necessarily node-disjoint, and they are not necessarily node-disjoint with the subgraphs of $\mathcal{C}_e(G)$. Moreover, two different subgraphs can contain same cycle in their roots. We give here $2$ examples.

\begin{minipage}[b]{7.5cm}
\textit{Example} $1$: the subgraph in the following figure consists of two subgraphs $C_4$ and its root contains only one cycle. 

\begin{center}
\includegraphics[scale=.25]{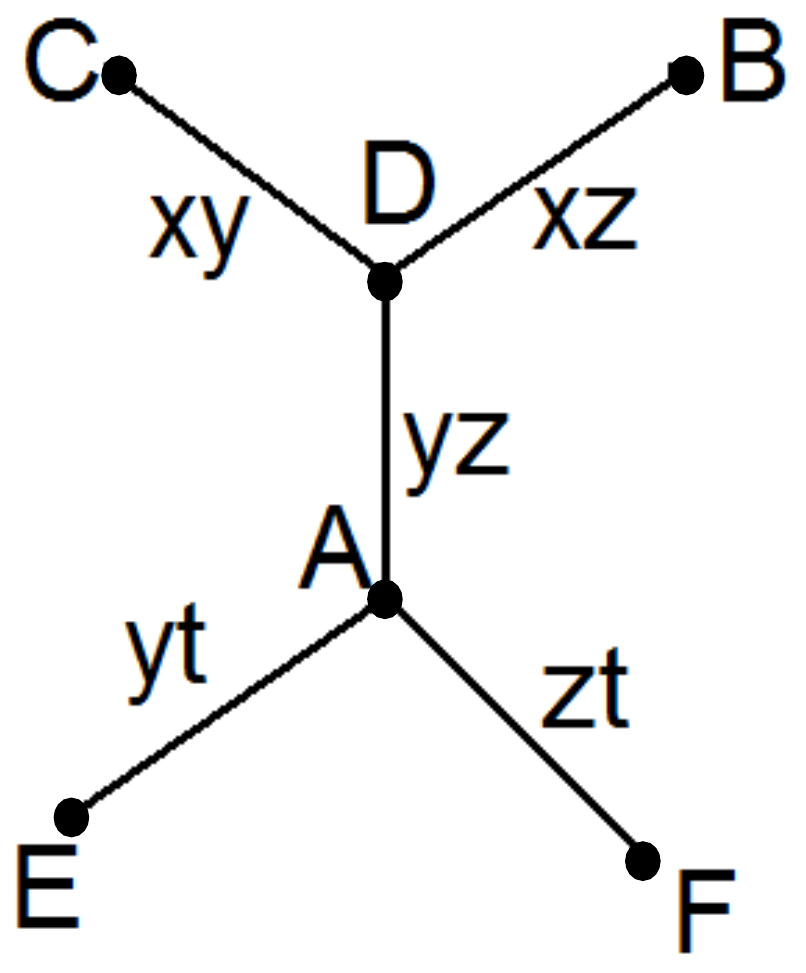}\;\;\;\;
\includegraphics[scale=.8]{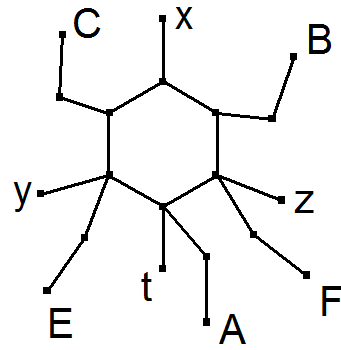}
\end{center}
\end{minipage}\hfill
\begin{minipage}[b]{7.5cm}
\textit{Example} $2$: the subgraph in the following figure consists of a subgraph $C_4$ and a subgraph $C_8$. Its root contains only one cycle.

\begin{center}
\includegraphics[scale=.25]{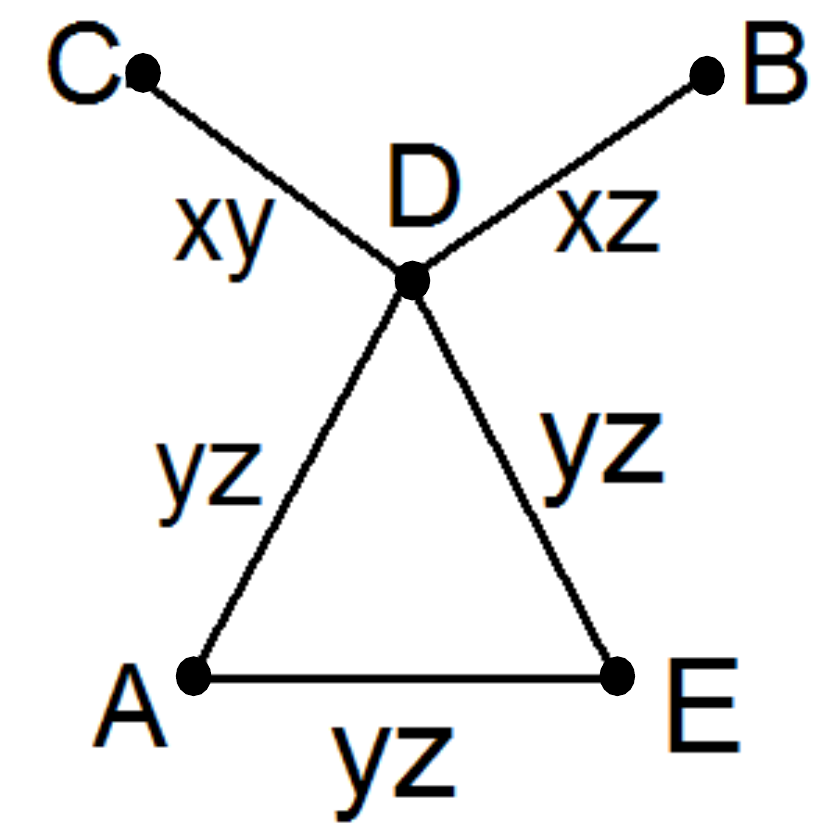}\;\;\;\;\includegraphics[scale=.8]{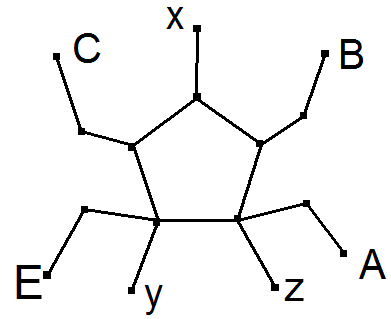}
\vspace*{.3cm}
\end{center}
\end{minipage} 

\begin{figure}[ht]
\begin{center}
\subfigure[$\textbf{C}_\textbf{4}+C_4$]{\includegraphics[scale=.25]{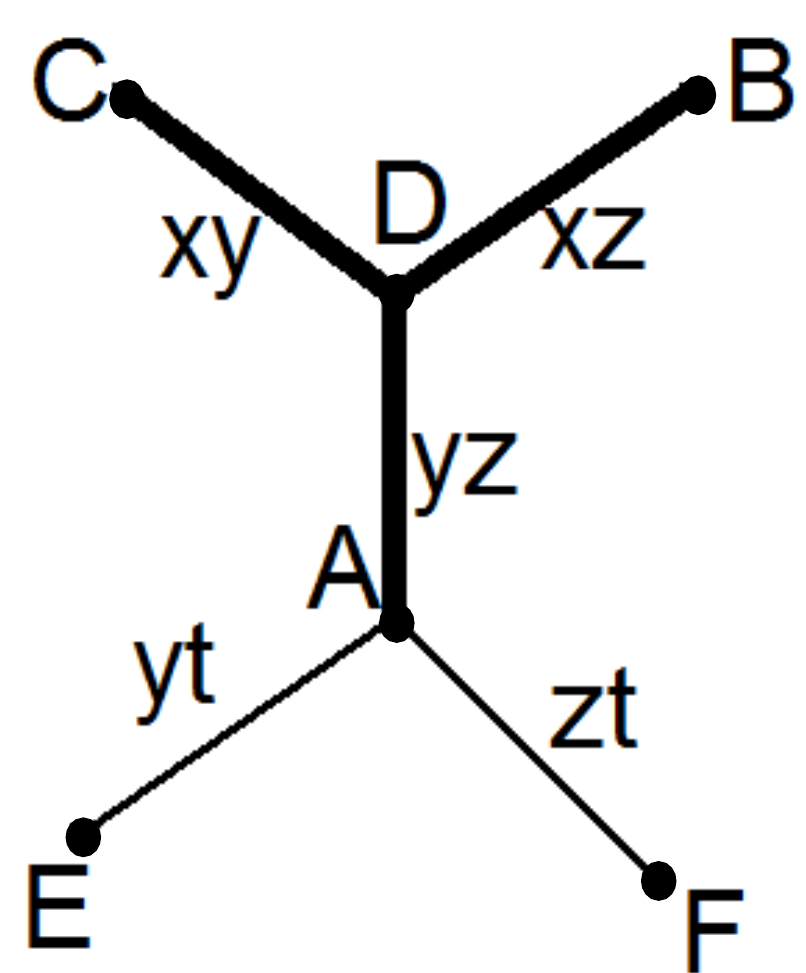}}
\subfigure[$\textbf{C}_\textbf{4}+C_5$\label{fig:C45}]{\includegraphics[scale=.25]{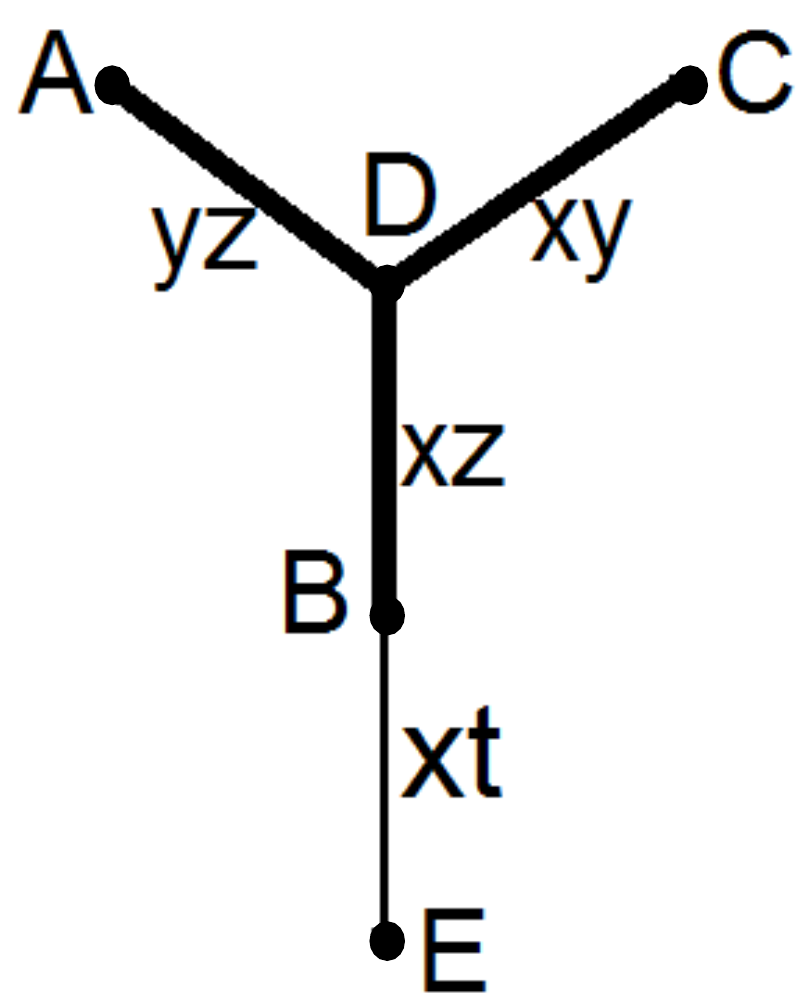}}
\subfigure[$\textbf{C}_\textbf{4}+C_6$\label{fig:C46}]{\includegraphics[scale=.25]{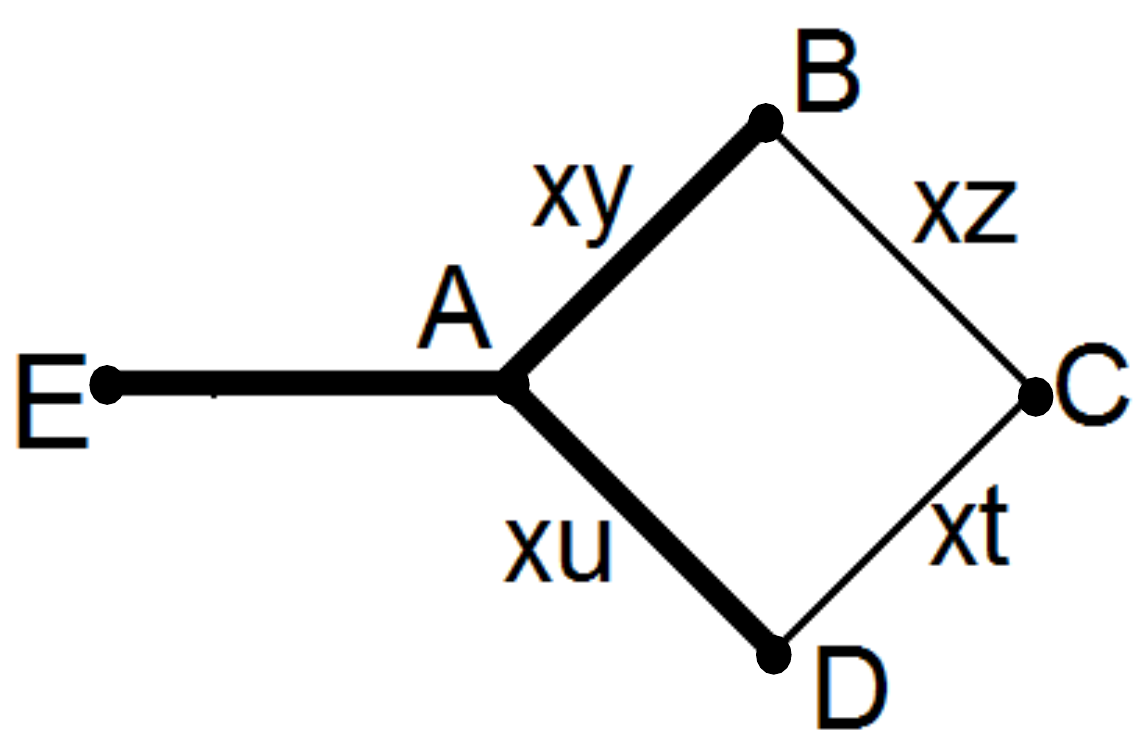}}
\subfigure[$\textbf{C}_\textbf{4}+C_7$\label{fig:C47}]{\includegraphics[scale=.25]{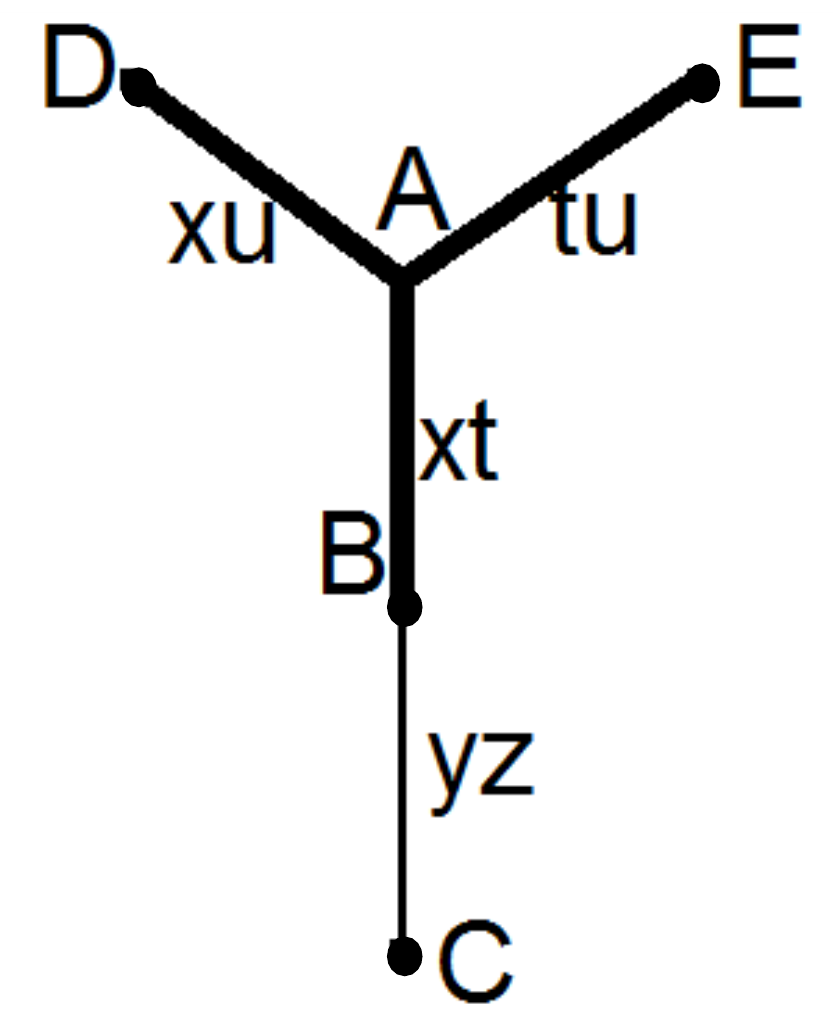}}
\subfigure[$C_4+\textbf{C}_\textbf{8}$\label{fig:C48}]{\includegraphics[scale=.25]{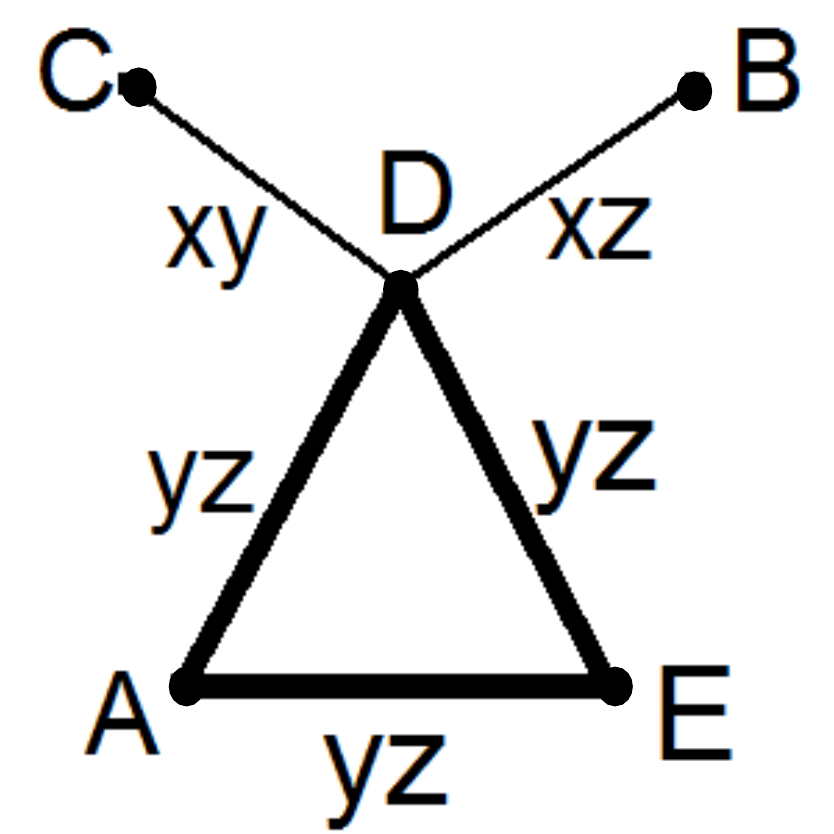}}
\subfigure[$C_5+\textbf{C}_\textbf{7}$]{\includegraphics[scale=.25]{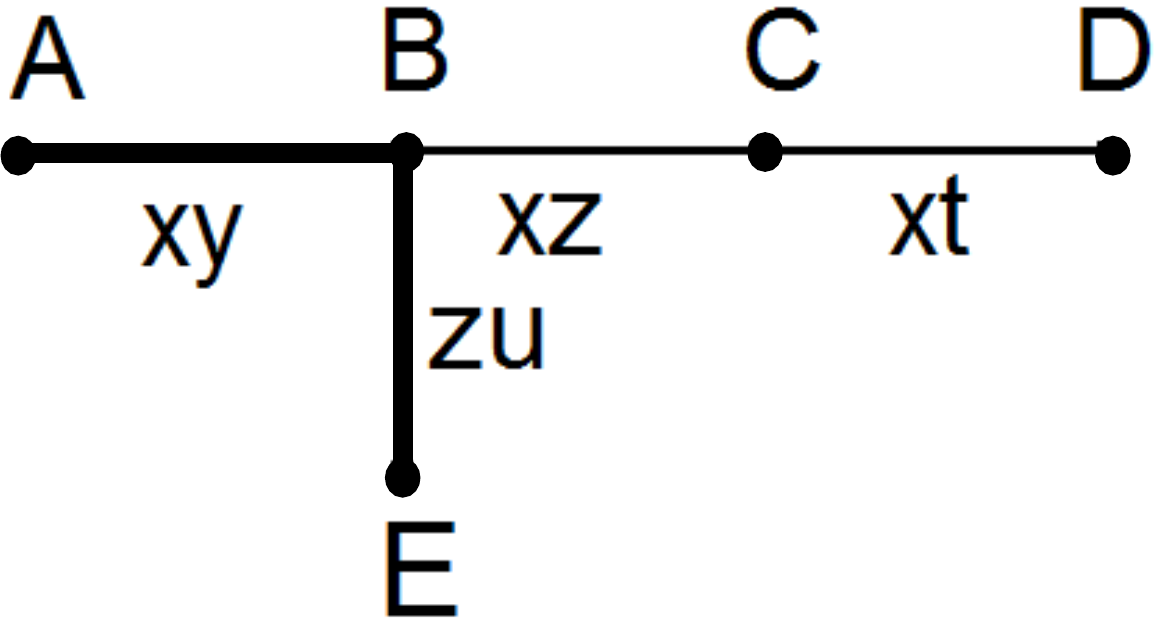}}
\subfigure[$C_5+\textbf{C}_\textbf{8}$\label{fig:C58}]{\includegraphics[scale=.25]{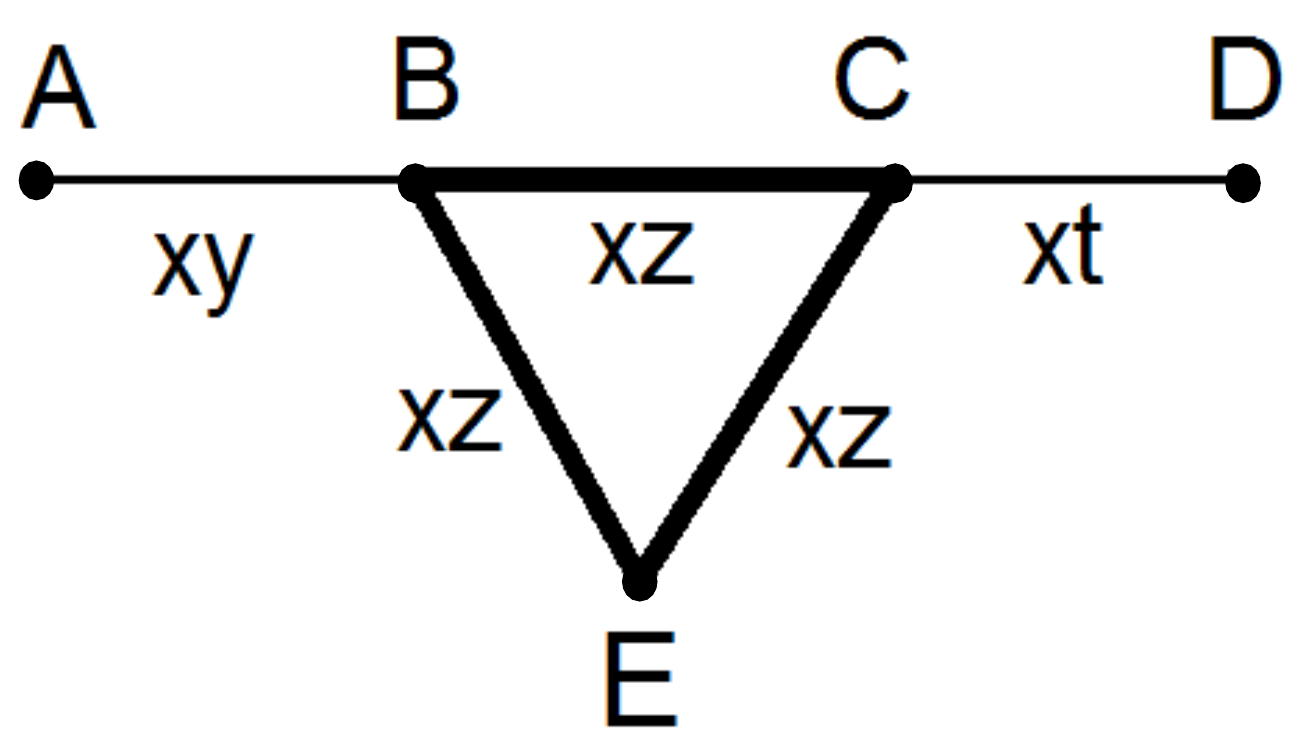}}
\subfigure[$C_7+\textbf{C}_\textbf{9}$\label{fig:C79}]{\includegraphics[scale=.25]{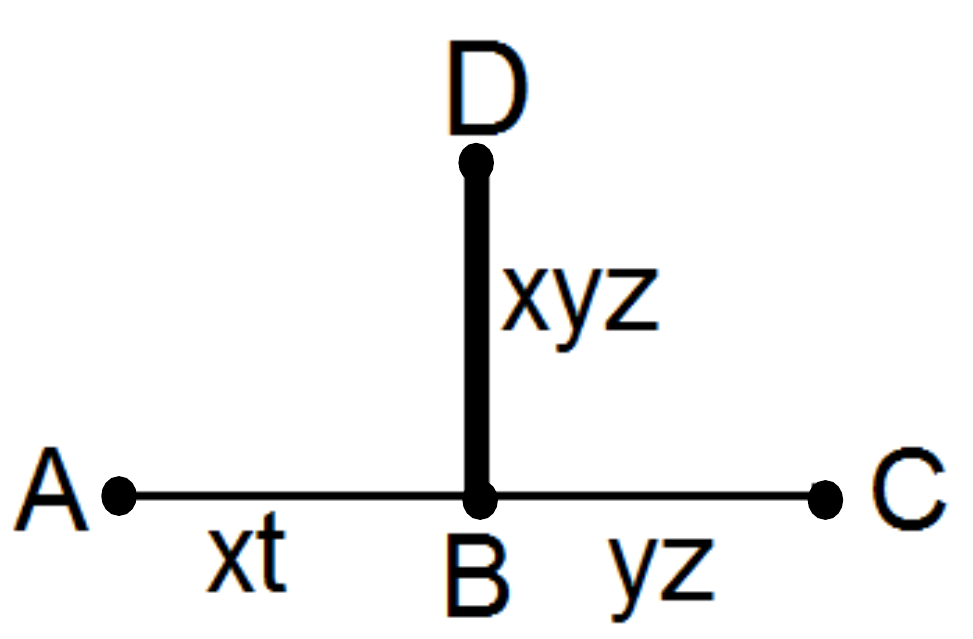}}
\subfigure[$C_7+\textbf{C}_\textbf{2}$\label{fig:C72}]{\includegraphics[scale=.25]{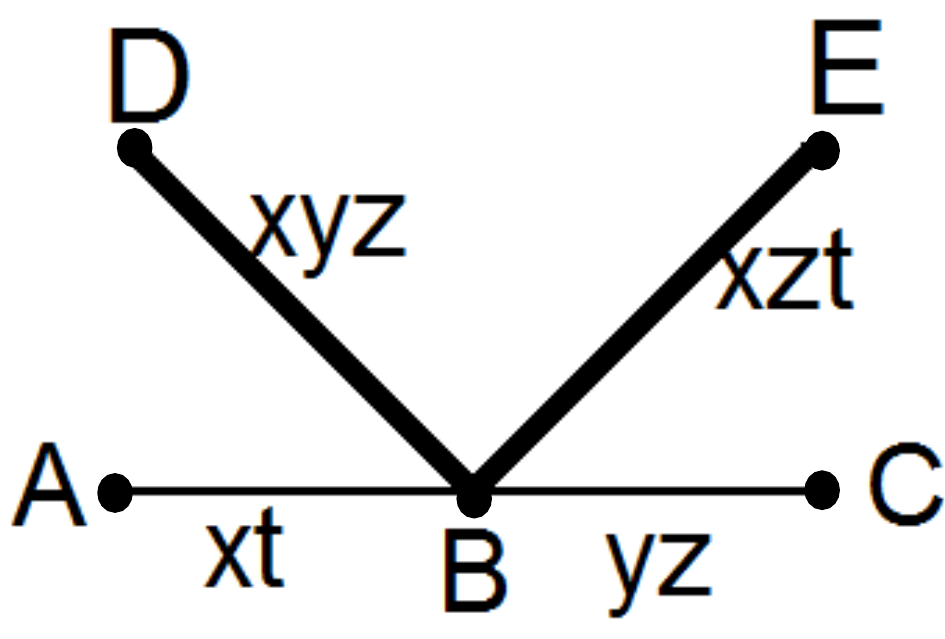}}
\caption{The list of subgraphs in which each one consists of two $C_i,C_j$ but infers only one cycle in its roots. The bold one is the one chosen to put into the corresponding list.}
\label{fig:2subgraphs}
\end{center}
\end{figure}

By considering all possible cases, we obtain a list of subgraphs in Figure \ref{fig:2subgraphs} such that each one consists of $2$ intersecting $C_i,C_j$ but its root contains exactly one cycle. Denote by $\mathcal{C}_i(G)$ the set of all subgraphs in $MC(G)$ isomorphic to $C_i$ and not contained in any subgraphs of $\mathcal{C}_e(G)$. Moreover, to avoid the redundancy of the cycles that they infer, for each subgraph isomorphic to one in Figure \ref{fig:2subgraphs}, we count only one of the two $C_i,C_j$ into the corresponding list. In the figure, the chosen one is in bold.

Let $\mathcal{C}_o(G) = \cup~\mathcal{C}_i(G)$ for $i=4, \dots ,9$ ($o$ for other), then we have the following remark. 
\textit{Suppose that $G$ has a $4$-leaf basic g-network root $N$. Let $S_1,S_2$ be two distinct subgraphs of $\mathcal{C}_e(G) \cup \mathcal{C}_o(G)$, then the cycles in $N[S_1]$ and $N[S_2]$ are distinct.}

\begin{lemma}
\label{lem:replace}
Suppose that $G$ is biconnected and has a $4$-leaf basic g-network root. Then there is a $4$-leaf basic g-network root $N$ of $G$ such that  any cycle of $N$ is contained in a root of a subgraph of $\mathcal{C}_e(G) \cup \mathcal{C}_o(G)$. 
\end{lemma}


\begin{pf}Let $c$ be a cycle of $N$. We consider each possible configuration of $c$.

If $c$ is a big cycle, or a $6$-cycle with $2$ invisible vertices of distance $3$, then the maximal cliques graph of the $4$-leaf power of $S(c)$ is a cycle of $\mathcal{C}'(G)$. Hence, $c$ is contained in a root of a subgraph of $\mathcal{C}'(G)$.

If $c$ is a triangle without invisible vertex, then by the constraint imposed on the roots of $G$, $c$ has at least two vertices which are adjacent to other inner vertices not on $c$. The corresponding maximal cliques graph is then either $C_1$ or $C_9$. So $c$ is contained in a root of $\mathcal{C}_1(G)$ or $\mathcal{C}_9(G)$.

Similarly, we show the other cases.

If $c$ is a $6$-cycle with one invisible vertex, then $c$ is contained in a root of a subgraph of $\mathcal{C}_6(G)$ of $\mathcal{C}'(G)$. 

If $c$ is a $4,5$-cycle without invisible vertex, then  $c$ is  contained in a root of a  subgraph of $\mathcal{C}_2(G)$. 

If $c$ is a $6$-cycle without invisible vertex, then $c$ is contained in a root of a  subgraph of $\mathcal{C}_3(G)$.

\begin{figure}[ht]
\begin{center}
\subfigure[\label{fig:3_1}]{\includegraphics[scale=.7]{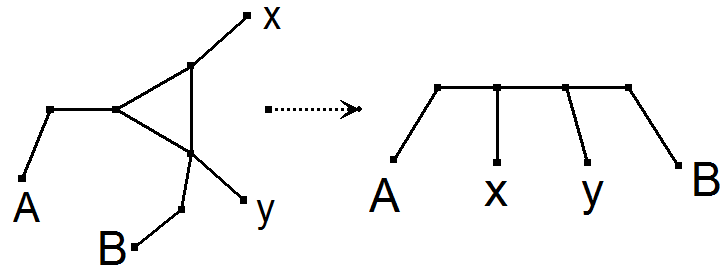}}\;\;\;\;\;
\subfigure[\label{fig:4_1}]{\includegraphics[scale=.7]{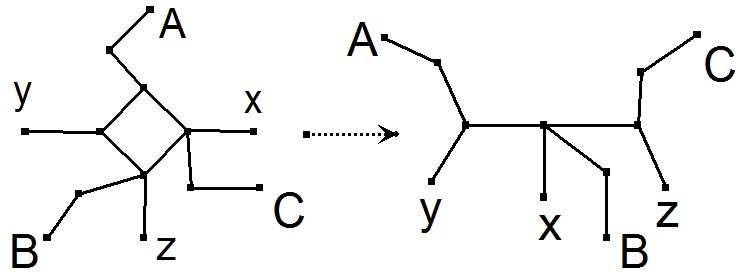}}\;\;\;\;\;
\subfigure[\label{fig:4_2}]{\includegraphics[scale=.7]{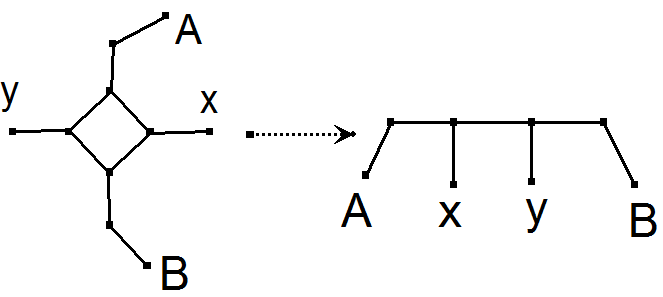}}
\subfigure[\label{fig:5_1}]{\includegraphics[scale=.7]{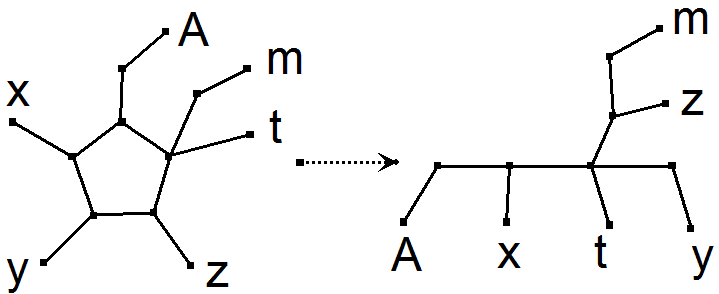}}\;\;\;\;\;
\subfigure[\label{fig:5_2}]{\includegraphics[scale=.7]{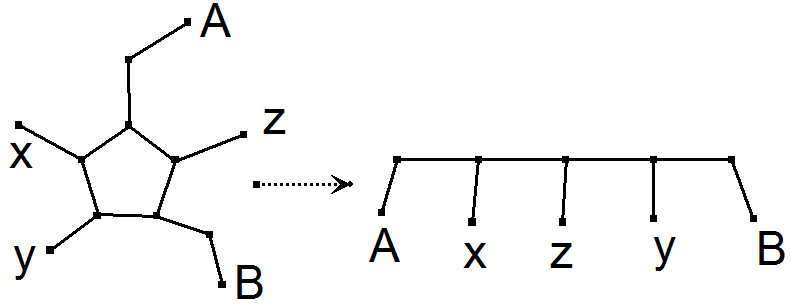}}\;\;\;\;\;
\subfigure[\label{fig:6_1}]{\includegraphics[scale=.7]{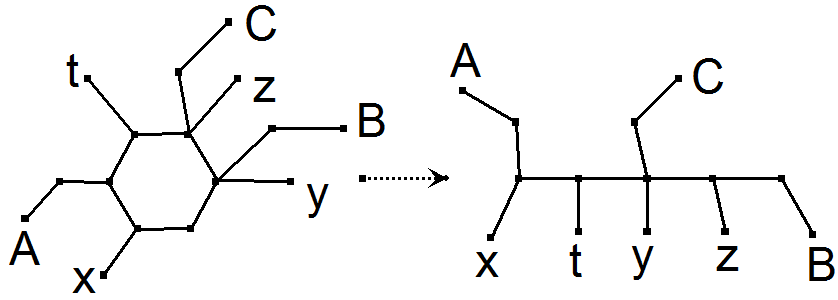}}
\caption{Replacing some subnetworks cycles by subnetworks without cycle}
\label{fig:replace}
\end{center}
\end{figure}

If $c$ is a triangle with one invisible vertex: If it is contained in a network as $R_{8,1}$, then $c$ is contained in a root of a subgraph of $\mathcal{C}_8(G)$. Otherwise, we can replace it by the network in Figure \ref{fig:3_1} which does not contain any cycle and the resulting network is still a root of $G$. 

If $c$ is a $4$-cycle with one invisible vertex: If $C$ is contained in a network  as  $R_{4,1}$ or  $R_{5,1}$ or $R_{7,1}$, then $c$ is contained in a root of a subgraph of $\mathcal{C}_4(G)$ or $\mathcal{C}_5(G)$ or $\mathcal{C}_7(G)$. Otherwise, we can replace it by the network in Figure \ref{fig:4_1} right. 

If $c$ is a $4$-cycle with two invisible vertices (Figure \ref{fig:4_2} left), then $MC(G)$ consists of only nodes $AB$ connected by an edge of weight $2$. So, $N$ in the Figure \ref{fig:4_2} right is a root of $G$ and it does not contain any cycle.

If $c$ is a $5$-cycle with one invisible vertex: If $c$ is contained in a network corresponding to a root of $C_2$ or as $R_{5,2}$ or $R_{7,2}$, then $c$ is contained in a root of a subgraph of $\mathcal{C}_2(G)$ or $\mathcal{C}_7(G)$. Otherwise, we can replace it by the network in Figure \ref{fig:5_1} right.

If $c$ is a $5$-cycle with two invisible vertices: If $c$ is contained in a network as $R_{4,3}$ or $R_{5,2}$ or $R_{8,2}$, then $c$ is contained in a root of a subgraph of $\mathcal{C}_4(G)$ or $\mathcal{C}_5(G)$ or $\mathcal{C}_8(G)$. Otherwise, we can replace it by the network in Figure \ref{fig:5_2} right.

If $c$ is a $6$-cycle with two invisible vertices of distance $2$: If $c$ is contained in a network as $R_{4,2}$ or $R_{4,4}$ or $R_{5,3}$ or $R_{5,4}$ or $R_{7,3}$, then $c$ is contained in a root of a subgraph of $\mathcal{C}_4(G)$ or $\mathcal{C}_5(G)$ or $\mathcal{C}_7(G)$. Otherwise, we can replace it by the network in Figure \ref{fig:6_1} right.

Finally, if $c$ is a $6$-cycle with $3$ invisible vertices, then $c$ is contained in a root of a subgraph of $\mathcal{C}_4(G)$.
\end{pf}

So, if $G$ is biconnected, the sets $\mathcal{C}_e(G)$ and $\mathcal{C}_o(G)$ capture all the cycles that appear in a root of $G$, and different cycle is contained in a root of a different subgraph. There may exist some roots of $G$ which contain other cycles but we can always replace them by the subnetworks without cycles as in Figure \ref{fig:replace}. 

\subsection*{Further conditions for the roots of $C_i$, $i=4, 5, \dots, 9$}

With a given $S \in \mathcal{C}_o(G)$, not every root of $S$ can always appear in some basic g-network root of $G$, if there is any. For example, 
let $S \in \mathcal{C}_4(G)$ as in Figure \ref{fig:C4}. If in $MC(G)$, $A$ has a neighbour node not in $S$, then by Lemma \ref{lem:inv_cycle} (iii), $N[A]$ can not be an invisible (quasi) star. So, the root $R_{4,1}(A)$ of $S$ can not be presented in any $4$-leaf basic g-network  root of $G$. 
A further example is the following.

\begin{example}
\label{ex:r_S}
Let $MC(G)$ be the graph in Figure \ref{fig:ex_r_S} where $A$ is a maximal clique of $G$ consisting of $\{m,k,u,s\}$. 
$S$ consists of $M,A$ is a subgraph of $\mathcal{C}_9(G)$. Let us consider the two roots $R_1,R_2$ of $S$. $R_1$ is an induced subgraph of a $4$-leaf basic g-network root of $G$. However, if we choose $R_2$ as a root of $S$ to reconstruct a root of $G$, we obtain a network having intersecting cycles. So, $R_2$ will be not chosen to construct $4$-leaf basic g-network roots of $G$.  

\begin{figure}
\begin{center}
\includegraphics[scale=.24]{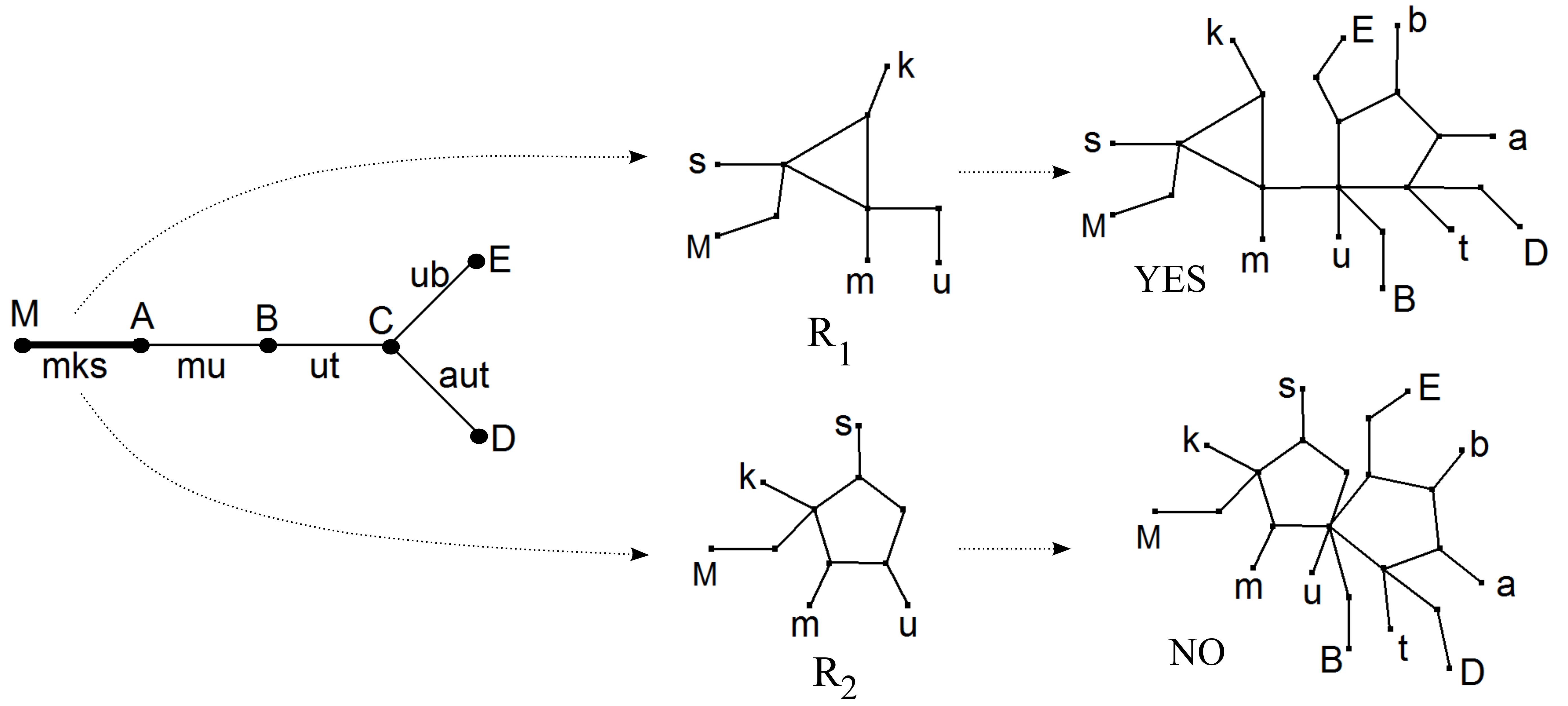}
\caption{An example of $r(S)$}
\label{fig:ex_r_S}
\end{center}
\end{figure}
\end{example}

Before presenting the way to choose the appropriate roots for each subgraph of $\mathcal{C}_o(G)$, we show some properties that these roots must have.

For any $S \in \mathcal{C}_o(G)$, we denote by $\textbf{ext(S)}$ \index{$ext(S)$} the induced subgraph of $MC(G)$ consisting of $S$ and all nodes of $MC(G)$ having distance at most $2$ to at least a node of $S$. For example in Figure \ref{fig:ex_r_S}, if $S$ is the edge $MA$, then $ext(S)$ is the path $[M,A,B,C]$.

\begin{lemma}
\label{lem:prop_root}
Let $A$ be a node of a subgraph $S\in \mathcal{C}_o(G)$. Let us consider the nodes around $S$ in $ext(S)$. For any $4$-leaf basic g-network root $N$ of $G$:

(i) If $A$ has a neighbour node not in $S$, then $N[A]$ is not an invisible (quasi) star.

(ii) Suppose that there are two nodes $B,C \not\in S$ such that $B$ is a neighbour of $A$, and $C$ is a neighbour of $B$ but not of $A$:

\hspace*{.6cm}- If $S \not\in \mathcal{C}_9(G)$, then $N[A]$ is a visible star. 

\hspace*{.6cm}- If $S \in \mathcal{C}_9(G)$, then $N[A]$ is either a visible (quasi) star or $N'_5$. Moreover, if $N[A]$ is an $N'_5$, then $N[B]$ is a visible star.
\end{lemma}

\begin{pf}
(i) This claim is deduced from Lemma \ref{lem:inv_cycle} (iii).

(ii) By Claim (i), $N[A]$ can not be an invisible (quasi) star because otherwise it can not have the neighbour $B$. 

- Suppose that $S \not\in \mathcal{C}_9(G)$: By observing the lists of roots of each $C_i$ where $i=4, \dots, 8$, we see that $N[A]$ is either a visible star, or $N'_5$ or $N''_5$ or $N_6$. 

The only configuration of $N[S]$ such that $N[A]$ is an $N'_5$ is $R_{7,2}$ (Figure \ref{fig:R_7,2}), i.e. $S \in \mathcal{C}_7(G)$. Then $N[B]$ is necessarily a visible star having the middle vertex on the cycle of $N[A]$, and $B$ must have $3$ common vertices with $A$. So, $(A,B)$ corresponds to a subgraph $S_1$ in either $\mathcal{C}_9(G)$ or $\mathcal{C}_2(G)$. It means that $S \cup S_1$ is isomorphic to the subgraph in either Figure \ref{fig:C79} or \ref{fig:C72}. However, by definition, the cycle of $N[A]$ is implied from $S_1$, not from $S$, and $S$ is not chosen to be in $\mathcal{C}_7(G)$, a contradiction.

If $N[A]$ is $N''_5$, then the possible configurations of $N[S]$ are $R_{4,3}, R_{5,2}, R_{8,2}$. Moreover, $N[S]$ can not be $R_{8,2}$ because in this network, all $2$ possible neighbours of the maximal cliques corresponding to $N''_5$ are in $S$. It means that $A$ can not have such a neighbour $B$. So, $N[S]$ has form of either  $R_{4,3}$ or $R_{5,2}$, i.e. $S$ is either in $\mathcal{C}_4(G)$ or in $\mathcal{C}_5(G)$. $N[B]$ can not be an invisible (quasi) star because otherwise it can not have such a neighbour $C$ (by Lemma \ref{lem:inv_cycle} (iii)). $N[B]$ can not contain a cycle  because otherwise its cycle intersects with the cycle of $N[A]$. So, $N[B]$ must be a visible star having the middle vertex on $N[A]$. 
However, that will create  a triangle $ABX$ in $MC(G)$, where $X$ is the neighbour of $A$ in $S$ such that $N[X]$ is a visible star having the middle vertex on $N[A]$. 
Let $S_1=(ABX)$, so $S_1$ is a subgraph of $\mathcal{C}_8(G)$. It means that  $S \cup S_1$ is isomorphic to the subgraph in Figure \ref{fig:C48} or \ref{fig:C58}. However, by definition, in this case $S$ is not chosen to be in $\mathcal{C}_4(G)$ or $\mathcal{C}_5(G)$, and the corresponding cycle in $N[A]$ is implied from $S_1$, not from $S$.

Similarly, if $N[A]$ is $N_6$, by analysing the lists of roots as above, we deduce a contradiction.

Hence, $N[A]$ is a visible star. 

- Suppose that $S \in \mathcal{C}_9(G)$: So $N[S]$ has form of either $R_{9,1}$ or $R_{9,2}$ where $N[A]$ is either a visible (quasi) star or isomorphic to $N'_5$. In the latter case, if $N[B]$ contains a $4,5,6$-cycle, it will intersect with the cycle of of $N[A]$. If $N[B]$ is an invisible (quasi) star then its middle vertex must be on the cycle of $N[A]$, and $B$ does not have the neighbour $C$, a contradiction. If $N[B]$ is a visible quasi star then its triangle intersect with the cycle of $N[A]$. Hence, $N[B]$ must be a visible star.
\end{pf}

By observing the lists of roots of $\mathcal{C}_o(G)$, we deduce the following.

\begin{observation}

(i) For $i=4,6,8,9$, $\forall S \in \mathcal{C}_i(G)$, every root of $S$ contains a cycle $c$ such that $l(S) \subseteq l(c)$.

(ii) $\forall S \in \mathcal{C}_5(G)$, denote by $lc(S)$ \index{$lc(S)$} the only common vertex of the $4$ maximal cliques of $S$, then every root of $S$ contains a cycle $c$ such that $lc(S) \in l(c)$. 
\label{ob:cycle_label}
\end{observation}

\begin{example}
(i) The subgraph $C_4$ in Figure \ref{fig:C4} has several roots, but the cycles in all of its roots always have the label set contains $\{x,y,z\}$, i.e. equals to $l(S)$.

(ii) The subgraph $C_5$ in Figure \ref{fig:C5} has several roots, but the cycles in all of its roots always have the label set containing $x$, which is the common vertex of the $4$ maximal cliques of this subgraph.
\label{ex:cycle_label}
\end{example}

The following definition shows how to determine the appropriate roots of each subgraph of $\mathcal{C}_o(G)$.

\begin{definition}

Let $S \in \mathcal{C}_o(G)$, we define $\textbf{r(S)}$ \index{$r(S)$} as the set of roots $r$ of $S$ such that:

1. The cycle $c$ in $r$ is vertex-disjoint with the cycle in the root of every subgraph of $\mathcal{C}_e(G)$. Moreover, $l(c)$ is disjoint with $l(S')$ for any $S' \in \mathcal{C}_i(G), i=4,6,8,9$ and it is disjoint with $lc(S')$ for any $S' \in \mathcal{C}_5(G)$.

2. $r$ is the induced subgraph of a $4$-leaf basic g-network $r'$ where $r'$ is a root of $ext(S)$ which satisfies the properties in Lemma \ref{lem:prop_root}.
\label{def:r_S}
\end{definition} 

The following properties of $r(S)$ can be deduced from the above definition.

\begin{lemma}
\label{lem:AB}
Let $S \in \mathcal{C}_o(G)$, and let $r_0$, $r_1$ be two distinct networks of $r(S)$. For $i=0,1$, let $r'_i$ be a root of $ext(S)$ which satisfies the properties in Lemma \ref{lem:prop_root} such that $r_i$ is the induced subgraph of $r'_i$ on $S$ . Let $A$ be a node of $S$ which has a neighbour $B$ and $B$ has a neighbour $C$ not adjacent with $A$ such that $B,C \not\in S$, then:

1. $r'_0[A \cap B]$ and $r'_1[A \cap B]$ have the same configuration.

2. Denote by $r'_{0,1}$ the network obtained by overlaying the common part of $r'_0[S]$ and $r'_1[B \cup C]$ on $A \cap B$. Then they cycles in $r'_{0,1}$  are vertex-disjoint.
\end{lemma}

\begin{pf}
Remark that $d_{MC(G)}(A,B)=1$, $d_{MC(G)}(A,C)=2$, so $B,C \in ext(S)$.

1. Let $M$ be a neighbour of $A$ in $S$, and let $r'$ be a certain $4$-leaf basic g-network root of $ext(S)$ which satisfies the properties in Lemma \ref{lem:prop_root}. We will prove that $r'[A \cap B]$ has only one configuration by considering all possible cases:

\hspace*{.5cm}(i) If $S \not\in \mathcal{C}_9(G)$, let $l(AM)=\{m,k\}$. By Lemma \ref{lem:prop_root} (ii), $r'[A]$ is a visible star. 

\begin{minipage}[b]{12cm}
\hspace*{1cm} \textit{(i)-a If $|AB|=1$, let $l(AB)=\{m\}$}, then $r'[A]$ has $p(m)$ as its middle vertex, and $r'[B]$ must be an invisible (quasi) star (according to Observation \ref{ob:2} (iv)). The configuration of $r'[A \cap B]$ is as the figure on the right.
\end{minipage}\hfill
\begin{minipage}[b]{2.5cm}
\includegraphics[scale=.25]{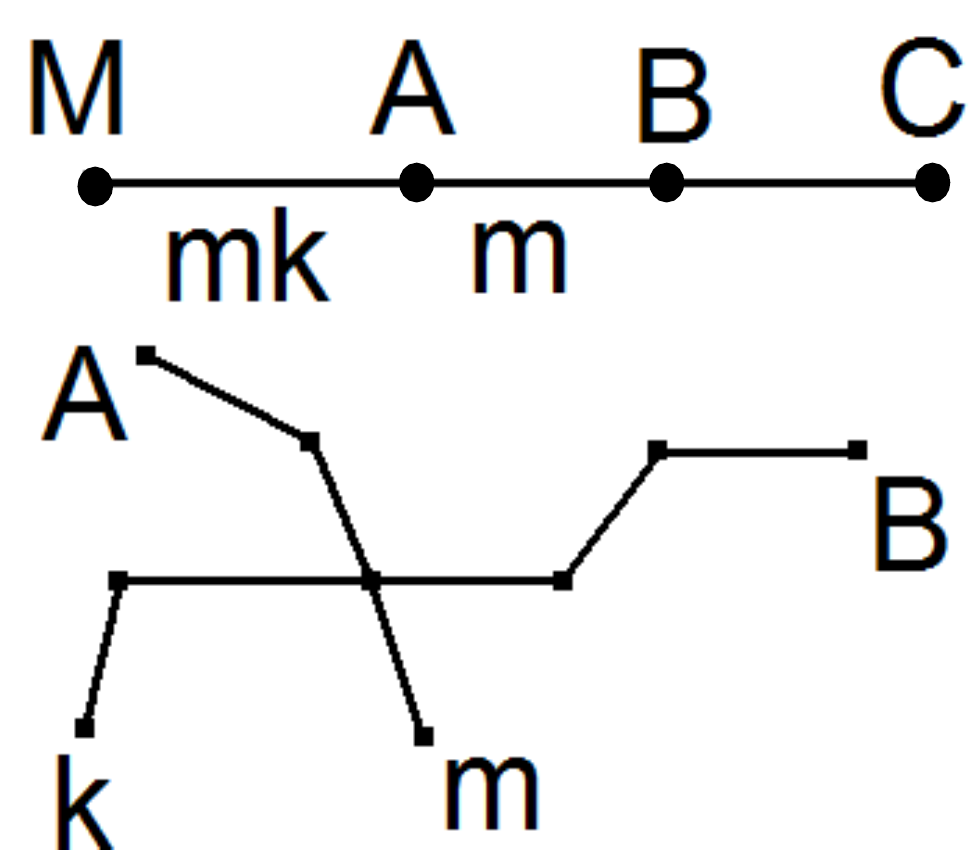}
\end{minipage}

\begin{minipage}[b]{12cm}
\hspace*{1cm} \textit{(i)-b If $|AB|=2$ and $l(AM) \cap l(AB) = \{m\}$, let $l(AB)=\{m,q\}$}, then due to Remark \ref{rem:middle}, $p(m)$ must be the middle vertex of $r'[A]$. We have the configuration as the figure on the right.
\end{minipage}\hfill
\begin{minipage}[b]{2.5cm}
\includegraphics[scale=.25]{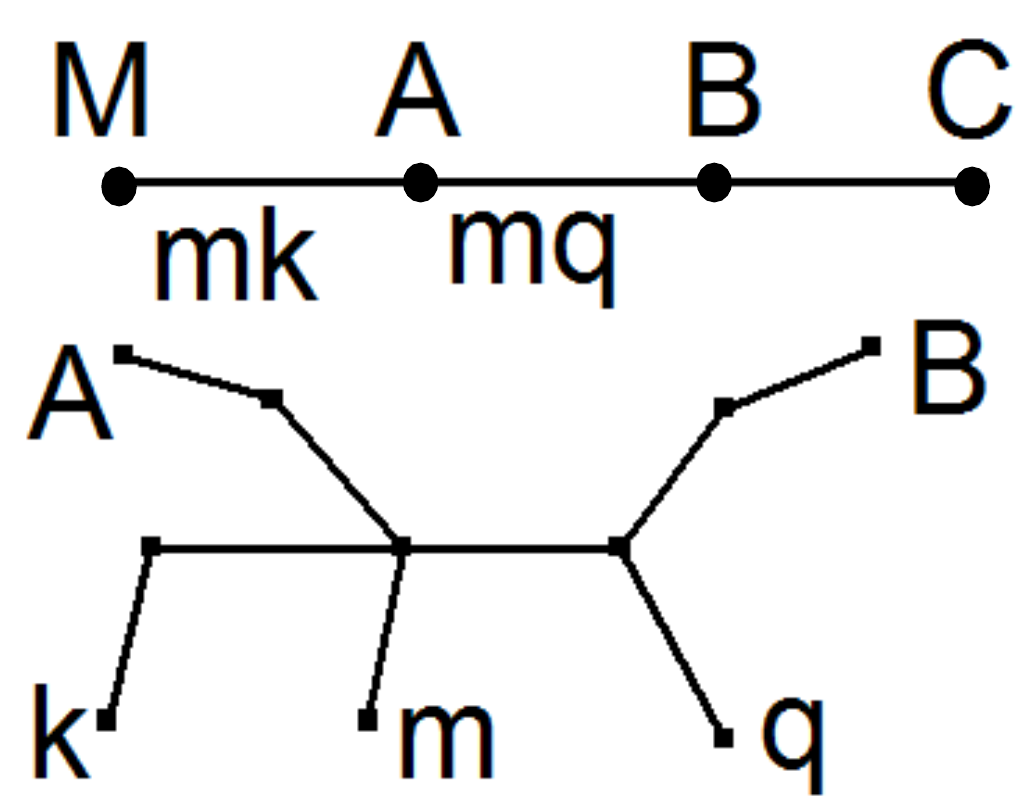}
\end{minipage}

\begin{minipage}[b]{10.5cm}
\hspace*{1cm} \textit{(i)-c If $|AB|=2$ and $l(AM) \cap l(AB) = \emptyset$}, then  $S'=(M,A,B)$ is a subgraph $C_7$.

If $S \cup S'$ has the configuration $C_4+C_7$ as in Figure \ref{fig:C47} where $S$ is a $C_4$, then $r'[M]$ is a $N_6$, $r'[A]$ is a visible star. We have the configuration as the figure in the right.
\end{minipage}\hfill
\begin{minipage}[b]{4cm}
\includegraphics[scale=.45]{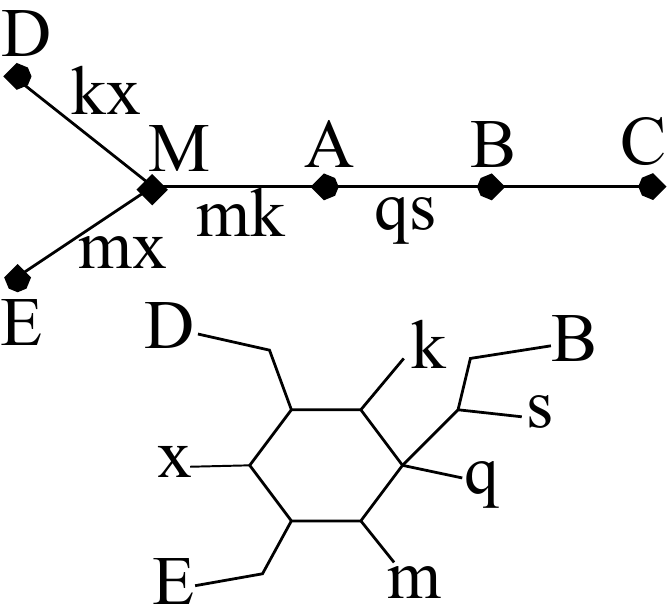}
\end{minipage}

Otherwise, the cycles in the root of $S$ and $S'$ are distinct. 
$S'$ can have $3$ forms of roots: $R_{7,1}$, $R_{7,2}$, $R_{7,3}$. 

\hspace*{1cm}- If $r'[M,A,B]$ has form $R_{7,1}$, and $r'[M]$ is an invisible star, then the cycle in the root of $S$ and $S'$ is the same, a contradiction. So in this case $r'[B]$ is an invisible star, i.e. $B$ can not have a neighbour $C$, a contradiction. 

\hspace*{1cm}- If $r'[M,A,B]$ has form $R_{7,2}$, then all the four vertices $p(m)$, $p(k)$, $p(q)$, $p(s)$ are contained in the cycle of this root. However, by observing the lists of roots of each $C_i$, we see that either $p(m)$ or $p(k)$ must be contained in the cycle of every root of $S$. So, these two cycles intersect, contradicting $r'$ is a g-network.

\begin{minipage}[b]{8.5cm}
\hspace*{1cm}- Hence, $r'[M,A,B]$ has form $R_{7,3}$ where $r'[B]$ is an $N_6$ as in the figure on the right and the middle vertex of $r'[A]$ is on the cycle of $r'[B]$. 
\end{minipage}\hfill
\begin{minipage}[b]{6cm}
\includegraphics[scale=.25]{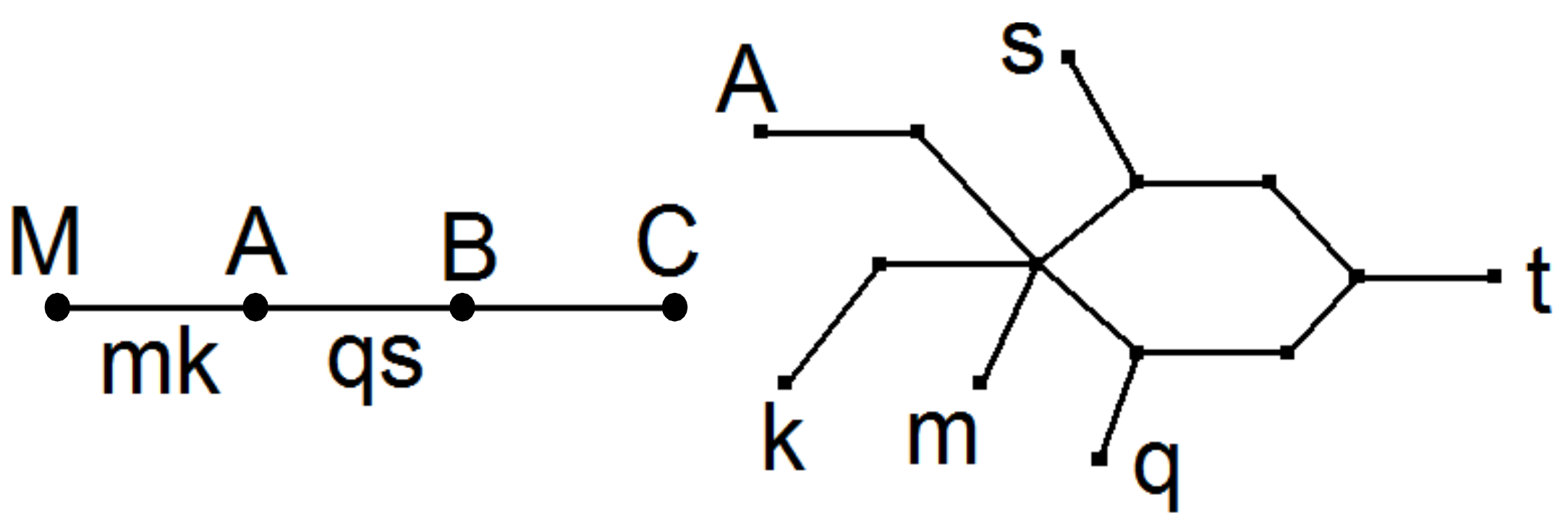}
\end{minipage}

\hspace*{1cm} \textit{(i)-d If $|AB|=3$, let $l(AB)=\{m,q,s\}$}, then according to Observation \ref{ob:2} (ii), $r'[B]$ is either a visible quasi star or $N_4$ or $N_5$ or $N'_5$. By Remark \ref{rem:middle}, $p(m)$ is the middle vertex of $r'[A]$. We have the configuration in the following figure.

\begin{figure}[ht]
\begin{center}
\includegraphics[scale=.25]{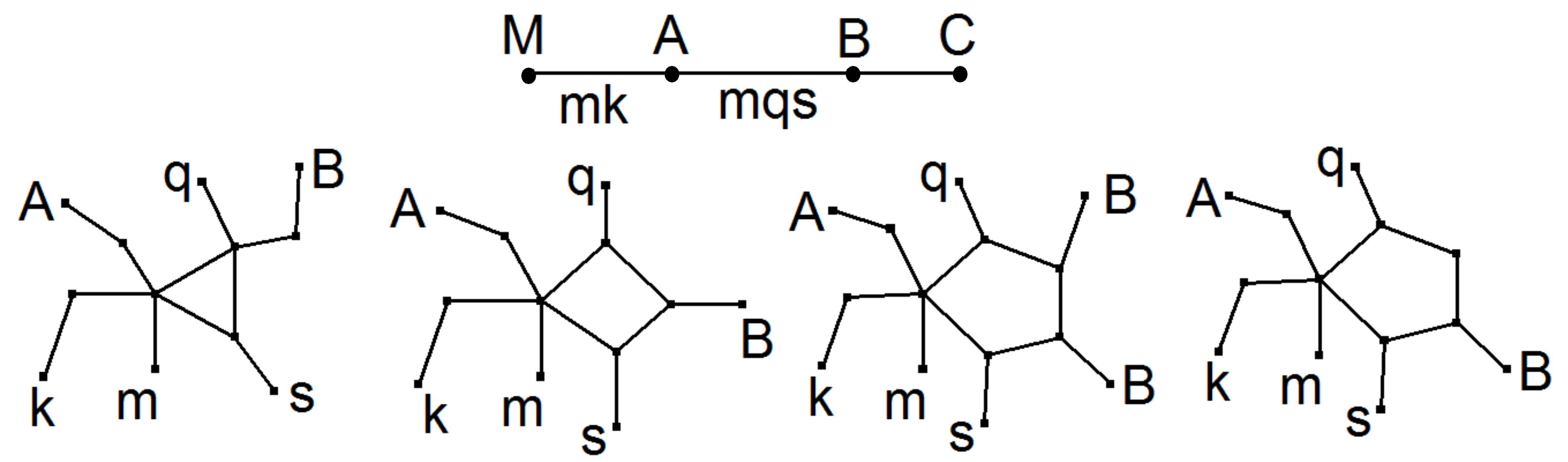}
\end{center}
\end{figure}

\hspace*{.5cm}(ii) If $S \in \mathcal{C}_9(G)$, let $l(AM) = \{m,k,s\}$, then $|AB|=2$ and $l(AB) \cap l(AM) \neq \emptyset$. The roots of $S$ have two forms: $R_{9,1}$ and $R_{9,2}$. There are $2$ following cases.

\begin{minipage}[b]{10.5cm}
\hspace*{1cm} \textit{(ii)-a $|l(AM) \cap l(AB)| = 1$, let $l(AB)= \{m,u\}$.} $r'[B]$ must be a  visible (quasi) star  having $p(u)$ as its middle vertex.  
If $r'[S]$ has form $R_{9,1}$ where $r'[A]$ and $r'[M]$ are two visible quasi star sharing the same triangle, then by Remark \ref{rem:middle}, the middle vertex of $r'[A]$ is $p(m)$. If $r'[S]$ has form $R_{9,2}$ then $r'[A]$ is an $N'_5$ and  $r'[B]$ is a visible star. Moreover, if $r'[B,C]$ contains a cycle passing $p(u)$, then $N[S]$ has form $R_{9,1}$ because otherwise the cycle in $r'[B,C]$ will intersect with the cycle in $r'[S]$. 
\end{minipage}\hfill
\begin{minipage}[b]{4.5cm}
\includegraphics[scale=.25]{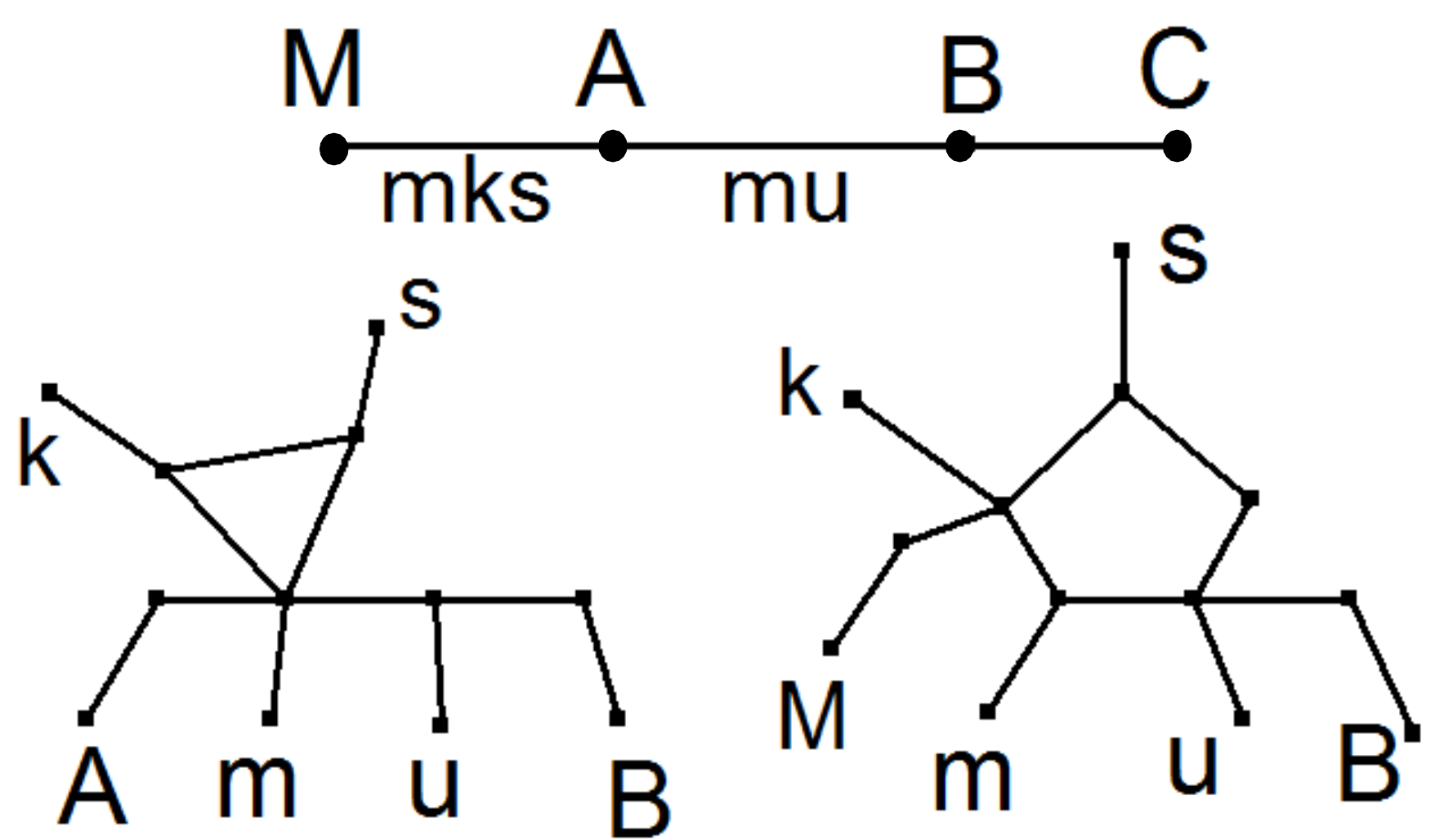}
\vspace*{.3cm}
\end{minipage}

\begin{minipage}[b]{11.5cm}
\hspace*{1cm} \textit{(ii)-b $|l(AM) \cap l(AB)| = 2$, let $l(AB) = \{m,k\}$.} Suppose that $r'[S]$ has form $R_{9,1}$, then for any configuration of $r'[B]$, $|l(MA) \cap l(AB)| \le 1$, a contradiction. So, $r'[S]$ has form $R_{9,2}$ where $r'[A],r'[B]$ are two visible stars having either $p(m)$ or $p(k)$ as their middle vertices. Moreover, if $l(AB) \cap l(BC) = \{k\}$, then $r'[B]$ has $p(k)$ as its middle vertex, and $r'[A]$ has $p(m)$ as its middle vertex. 
\end{minipage}\hfill
\begin{minipage}[b]{3.5cm}
\includegraphics[scale=.25]{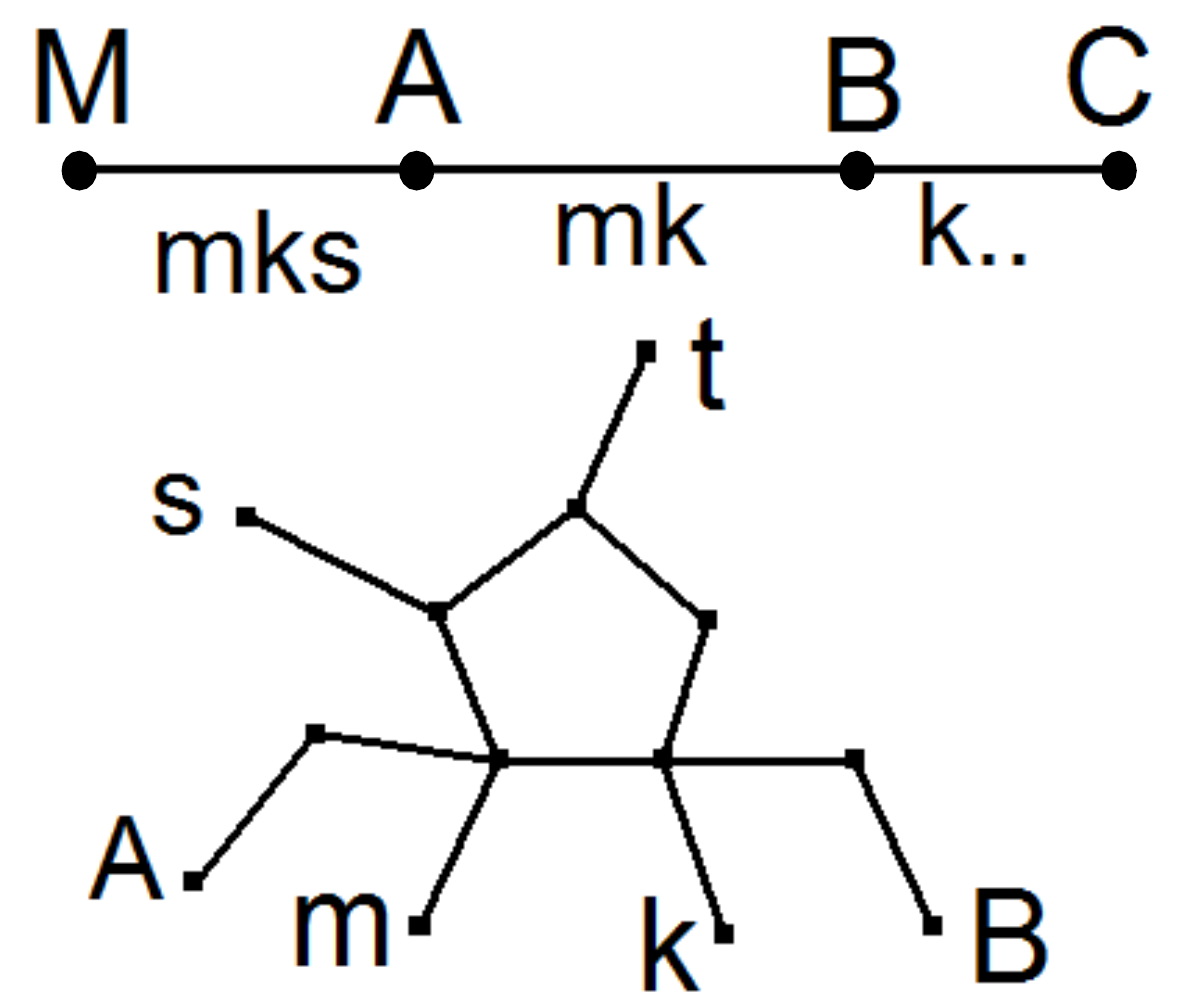}
\vspace*{.2cm}
\end{minipage}

We see that in each case the possible configurations of $r'$ on $A \cap B$ are unique even in the case that it has several possible configurations on $M,A,B,C$ as in the cases (i)-d, (ii)-a. In other words, $r'_0[A \cap B]$ and $r'_1[A \cap B]$ have the same configuration.

2. Due to Claim $1$, we can combine $r'_0[S]$ with $r'_1[B,C]$ by overlaying their common part on $A \cap B$ to obtain the network  $r'_{0,1}$. 
So we have $r'_{0,1}[S]=r'_0[S]$ and $r'_{0,1}[B,C]=r'_1[B,C]$. Suppose that $r'_{0,1}$ has two intersecting cycles $c_0,c_1$. Then $c_0$ is contained in $r'_{0,1}[S]$ and $c_1$ is contained in $r'_{0,1}[B,C]$. This claim will also be proved by considering all possible cases that are analysed above.

\hspace*{.5cm}(i) If $S \not\in \mathcal{C}_9(G)$: In this case $r'_{0,1}[A]$ is a visible star. So $c_0$ is contained in $r'_{0,1}[M]$, and it passes the middle vertex of $r'_{0,1}[A]$. In order that $c_0$ and $c_1$ intersect, the cycle $c_1$ must be contained in $r'_{0,1}[B]$ and $c_0,c_1$ intersect at the middle vertex of $r'_{0,1}[A]$.

\hspace*{1cm}(i)-a As proved above, $r'_{0,1}[B]$ must be an invisible (quasi) star. Then in order that $r'_{0,1}[B]$ contains a cycle, it must be an invisible quasi star and its triangle contains $p(m)$. However, the other leaf of this triangle is also contained in $A$. It means that $A,B$ have more than one common vertices, a contradiction.

\hspace*{1cm}(i)-b We consider furthermore the node $C$ to determine the possible forms of $r'[B]$ where $r'$ is a certain $4$-leaf basic g-network root of $G$ which satisfies the properties in Lemma \ref{lem:prop_root}. In this case, $r'[B]$ is either a visible (quasi) star, $N_5'$, $N''_5$ or $N_6$. It can be checked that each possible configuration of $BC$ brings about a unique configuration of $r'[B]$. So $r'_0[B]$ and $r'_1[B]$ are the same. Since $r'_0$ is a g-network, $r'_0[M]$ and $r'_0[B]$ do not contain intersecting cycles. Hence, the cycles in $r'_{0,1}[M]=(r'_0[M])$ and $r'_{0,1}[B]=(r'_1[B]=r'_0[B])$ do not intersect.

\hspace*{1cm} (i)-c For a certain $r'$, it can be seen that $r'[B]$ has only one configuration which is $N_6$ and this cycle passes the middle vertex of $r'[A]$. Hence, the cycles in $r'_0[B]$ and $r'_1[B]$ are the same. Similarly with the above case, we deduce that the cycles in $r'_{0,1}[M]$ and $r'_{0,1}[B]$ do not intersect.

\hspace*{1cm} (i)-d It can be seen that in all possible configurations, $r'_0[B]$ and $r'_1[B]$ contain a cycle passing $p(m)$. If $r'_{0,1}[M]$ and $r'_{0,1}[B]$ have $p(m)$ as a common vertex, then $r'_0[M]=(r'_{0,1}[M])$ and $r'_0[B]$ also have $p(m)$ as a common vertex. The later is a contradiction because $r'_0$ is a g-network.

\hspace*{.5cm}(ii) If $S \in \mathcal{C}_9(G)$:

\hspace*{1cm} (ii)-a Suppose that $r'_0$ corresponds to the first configuration and $r'_1$ corresponds to the second configuration of the case (ii)-a in Claim $1$. Then $r'_{0,1}[A]$ is a visible quasi star having $p(m)$ as its middle vertex and $r'_{0,1}[B]$ is a visible star having $p(u)$ as its middle vertex. 
It means that the cycle $c_0$ is the triangle of $r'_{0,1}[A]$, and $c_1$ must be contained in $r'_{0,1}[C]$. However, these two cycles do not intersect.

\hspace*{2cm} Suppose that $r'_0$ is the second configuration and $r'_1$ is the first configuration of the case (ii)-a in Claim $1$. Then $c_0$ is contained in cycle $N'_5$ of $r'_{0,1}[A]$. We know that $r'_{0,1}[B,C]$ is a visible (quasi) star. In order that $r'_{0,1}[B,C]$ contains a cycle intersecting with $c_0$,  either $r'_{0,1}[B]$ is a visible quasi star or $r'_{0,1}[C]$ contains a cycle passing the middle vertex of $r'_{0,1}[B]$. In the first case, we deduce that $r'_1[A]$ and $r'_1[B]$ have $3$ common leaves, a contradiction because $|AB|=2$. In the second case, we deduce that $C$ is contained in a subgraph $S_1$ of $\mathcal{C}_i(G)$ where $i \in \{4,5,6,8,9\}$. So, the label of the middle vertex of $r'_1[B]$ intersects  with either $l(S_1)$ if $i \in \{4,6,8,9\}$ or $lc(S_1)$ if $i=5$. The later shows that $r'_1$ does not satisfy the property $1$ of Definition \ref{def:r_S}, i.e $r'_1 \not\in r(S)$, a contradiction.

\hspace*{1cm} (ii)-b $r'[A,B]$ has a unique configuration as indicated in the case (ii)-b of Claim $1$. In this configuration $r'[A]$ contains the cycle $c_0$ and $r'[B]$ is a visible star. So, $c_0$ is contained in $r'_{0,1}[A]$ and $c_1$ is contained in $r'_{0,1}[C]$. Moreover, $r'_0[A]=r'_1[A]$. If $c_0$ intersects with $c_0$, then the cycle in $r'_1[C]=(r'_{0,1}[C])$ also intersects with the cycle in $r'_1[A]=(r'_0[A]=r'_{0,1}[A])$. The later is a contradiction because $r'_1$ is a g-network.
\end{pf}

\begin{theorem}
\label{theo:G_bc}
Let $G$ be a biconnected graph which has a $4$-leaf basic g-network root $N_0$, and $S \in \mathcal{C}_o(G)$. 

1. Let $r_0=N_0[S]$, then $r_0 \in r(S)$ 

2. For any other network $r_1 \in r(S)$ different from $r_0$, we can replace $r_0$ in $N_0$ by $r_1$ and the obtained network $N_1$ is also a $4$-leaf basic g-network root of $G$.
\end{theorem}

\begin{pf}
(i) It is obvious that $r_0$ is a root of $S$. We have to show that $r_0$ satisfies the $2$ conditions in Definition \ref{def:r_S}. Let $c$ be the cycle in $r_0$. 
The first condition: we know that the cycle in the root of each subgraph in $\mathcal{C}_e(G)$ is contained in $N_0$, then $c$ can not intersect with any of them. For any $S_1 \in \mathcal{C}_o(G)$ different from $S$, $N_0[S_1]$ contains a cycle different from $c$. So, $l(c)$ can not intersect with the leaf set $l(S_1)$ if $S_1 \in \mathcal{C}_i(G)$ where $i=4,6,8,9$ or with $lc(S_1)$ if $S_1 \in \mathcal{C}_5(G)$.
The second condition is obviously satisfied because $r'_0=N_0[ext(S)]$ is a root of $ext(S)$ and it obviously has the properties in Lemma \ref{lem:prop_root} because $N_0$ is a $4$-leaf basic g-network root of $G$. Moreover, $r_0$ is an induced subgraph of $r'_0$ on $S$.
Therefore, $r_0 \in r(S)$.

(ii) Let $r_1$ be another root of $r(S)$ different from $r_0$ (Figure \ref{fig:theorem}), so there is basic g-network $r'_1$ which is a root of $ext(S)$ and satisfies the properties in Lemma \ref{lem:prop_root} such that $r_1=r'_1[S]$. We will construct a basic g-network $N_1$ as follows. Let $H_1, \dots, H_m$ be the connected components of $MC(G)$ obtained after deleting $S$. Then for each $H_i$, there are two possible cases.

Case (a): If all nodes of $H_i$ are adjacent to at least a node of $S$, then denote $N^i=r'_1[H_i]$.

Case (b): Otherwise, let $N^i=N_0[H_i]$. 

\begin{figure}[ht]
\begin{center}
\includegraphics[scale=.25]{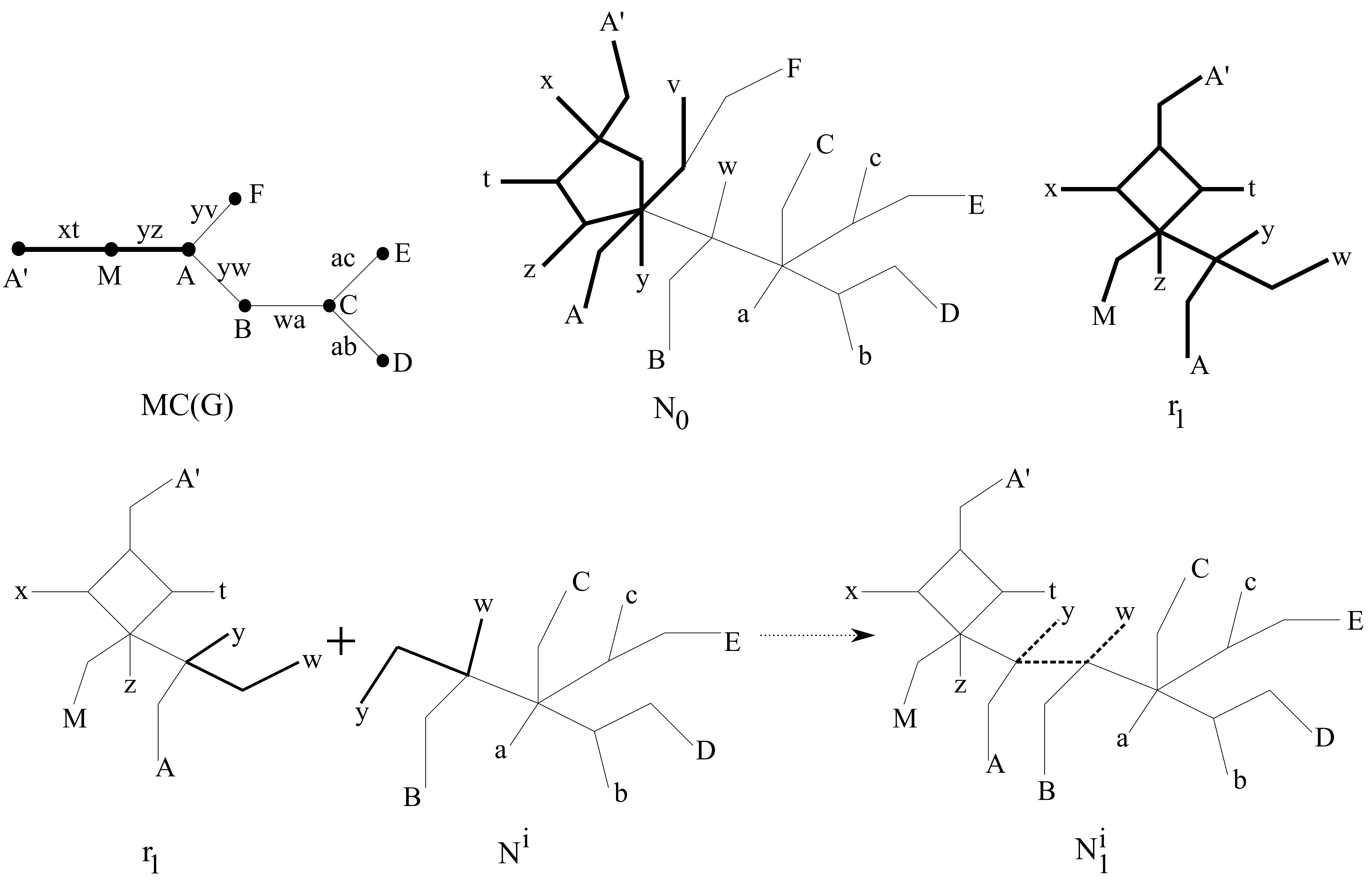}
\caption{$S=(A',M,A)$. The two subnetworks $r_0, r_1$ of $r(S)$ are in bold.}
\label{fig:theorem1}
\end{center}
\end{figure}

In case (a), both $N^i$ and $r_1$ are induced subgraphs of $r'_1$, then we can combine $N^i$ with $r_1$ by overlaying their common part.
In case (b), let $A$ be a node of $S$ with has a neighbour $B$ and $B$ has a neighbour $C$ such that $B,C \in H_i$. So the common vertices of $H_i$ and $S$ are contained in $A \cap B$. According to Claim $1$ of Lemma \ref{lem:AB}, $r'_0[A \cap B]$ and $r'_1[A \cap B]$ are the same. Moreover, $r'_0[A \cap B]=N_0[A \cap B]=N^i[A \cap B]$, and $r'_1[A \cap B]=r_1[A \cap B]$. Then  $N^i[A \cap B]$ and $r_1[A \cap B]$ are the same. So, we can also combine $N^i$ with $r_1$ by sticking their common part on $A \cap B$ (Figure \ref{fig:theorem}). Denote the obtained network by $N^i_1$.

It is easy to see that $N^i_1$ is a $4$-leaf root of $S \cup H_i$. Indeed, for case (a), both $r_1$ and  $N^i$ is an induced subgraphs of $r'_1$, i.e. $N^i_1 = r'_1[S \cup H_i]$. We know that $r'_1$ is a root of $ext(S)$, so $N^i_1$ is a root of $S \cup H_i$. For case (b), we have $(N^i_1[S])^{4l}=r_1^{4l}=S$, and $(N^i_1[H_i])^{4l} = (N^i)^{4l} = H_i$. Moreover, for any $x \in S \setminus H_i, y \in H_i \setminus S$, $d_{N_0}(x,y)$ and $d_{N^i_1}(x,y)$ are both greater than $4$. Hence,  $(N^i_1)^{4l} = (N_0[S \cup H_i])^{4l}=S \cup H_i$. Then $N^i_1$ is a $4$-leaf root of $S \cup H_i$.

Let $N_1$ be the network obtained by combining all $N^i$ to $r_1$ as described above for every $H_i$. Hence, $N_1$ is a $4$-leaf basic network root of $G$. 

We must show furthermore that $N_1$ is a g-network, i.e. its cycles are pairwise vertex disjoint. Let $c$ be the cycle of $r_1$. All cycles of $N_1$ different from $c$ are pairwise vertex-disjoint because by construction, they are also the cycles of $N_0$ and $N_0$ is a g-network. 
So, it remains to show that not any cycle in $N_1$ different from $c$ intersects with $c$.
By Definition \ref{def:r_S}, $c$ does not intersect with the cycle in the root of any subgraph of $\mathcal{C}_e(G)$. 
Suppose that there is a cycle $c'$ in $N_1$ that intersects with $c$. Because $c'$ does not contained in the root of any subgraph of $\mathcal{C}_e(G)$, it is a small cycle and different from a $6$-cycle without invisible vertex.
For all possible configurations of $c'$, it can be checked that $c'$ is contained in $N_1[X]$ where $X$ is a node in $ext(S)$. Moreover, $X$ can not be in $S$ because otherwise $r_1$ contains two cycles $c,c'$. So, $X \in ext(S) \setminus S$. Suppose that $X$ is a node of a certain $H_i$ in case (a). Then by construction, $N^i_1$ is an induced subgraph of $r'_1$. Since $r'_1$ does not contain intersecting cycles, therefore $N^i_1$ can not have intersecting cycles, contradicting the two intersecting cycles $c,c'$ be both in $N^i_1$.
Otherwise $X$ is a node of a certain $H_i$ in case (b). So, there exist two nodes $B,C$ in $H^i$ such that $B, C$ are adjacent and $B$ has a neighbour $A$ in $S$ where either $X=B$ or $X=C$. However, by Lemma \ref{lem:AB}, $r'_1[S]$ and  $r'_0[B,C]$ do not have intersecting cycles. It means that $N^i_1[S]=(r'_1[S])$ and $N^i_1[B,C]=(r'_0[B,C])$ do not have intersecting cycles. In other words,  $c, c'$ do not intersect.
Hence, $N_1$ is a basic g-network.

Therefore, $N_1$ is a $4$-leaf basic g-network root of $G$ and $N_1$ is obtained by replace $r_0$ in $N_0$ by $r_1$.
\end{pf}

Using Definition \ref{def:r_S} and Lemma \ref{lem:prop_root}, one can calculate for each subgraph $S$ of $\mathcal{C}_o(G)$ the root set $r(S)$ basing on the lists of roots in Figures \ref{fig:C_4}, ..., \ref{fig:C_9} and the analysis of configurations in the proof of Lemma \ref{lem:AB}. 
However, Theorem \ref{theo:G_bc} does not hold when $G$ is not biconnected.

\subsubsection{Other cycles \index{other cycles} when $G$ is not biconnected}
\label{sec:Q}
When $G$ contains several blocks, we consider separately each block and construct $4$-leaf basic g-networks roots of each one. However, there are some conditions must be taken into account on the roots of two adjacent blocks.

\begin{lemma}
\label{lem:bc}
Let $\mathcal{B}_1, \dots, \mathcal{B}_m$ be the blocks of $G$. Then $G$ is a $4$-leaf basic g-network power if and only if each $\mathcal{B}_i$ is a $4$-leaf basic g-network power and for any two blocks $\mathcal{B}_i, \mathcal{B}_j$ having a common vertex $x$, there exist some roots $N_i, N_j$ of $\mathcal{B}_i, \mathcal{B}_j$ such that: 

- either all the neighbours of $p(x)$ in $N_i$ are invisible or all the neighbours of $p(x)$ in $N_j$ are invisible.

- if $p(x)$ is contained in a cycle of $N_i$ (resp. $N_j$) then it is not contained in any cycle of $N_j$ (resp. $N_i$).
\end{lemma}

\begin{pf}
$\Rightarrow$ Suppose that $G$ has a $4$-leaf basic g-network power $N$. Let $N_i = N[\mathcal{B}_i]$ for any $i = 1, \dots k$, so each $N_i$ is a $4$-leaf basic g-network root of $\mathcal{B}_i$. Suppose that there is a neighbour $C_i$ of $p(x)$ in $N_i$ and a neighbour $C_j$ of $p(x)$ in $N_j$ which are both visible. Let $y_i, y_j$ be the leaves attached to $C_i, C_j$. So the distance of $y_i,y_j$ is $4$, or $y_iy_j$ are connected in $G$, which is a contradiction with the fact that $\mathcal{B}_i, \mathcal{B}_j$ are  $2$ distinct blocks. 

The second condition is implied from the fact that $N$ is a g-network and so its cycles are pairwise disjoint.

$\Leftarrow$ Suppose that each $\mathcal{B}_i$ has a $4$-leaf basic g-network rot $N_i$ and the two conditions are satisfied. We construct $N$ by combining all $N_i$ by the same way in \cite{BLS08}: For any cut-vertex $x$ corresponding to a leaf which is a cut-vertex of $G$, we glue $p(x)$ together and remove all copies of $x$ except one. So $N$ is a $4$-leaf root of $G$ and $N$ is a g-network. Indeed, the distance between any $2$ leaves in one block does not change, the distance between any $2$ leaves of $2$ distinct blocks which do not equal the cut-vertex of these blocks are always greater than $4$ (from the first condition), and the cycles in $N$ are pairwise disjoint (from the second condition and from the fact that each $N_i$ is a basic g-network).
\end{pf}


By Lemma \ref{lem:bc}, the conditions at the cut-vertices of $G$ impose some further cycles in the roots of $G$ apart from those in the roots of the subgraphs in $\mathcal{C}_e(G)$ and $\mathcal{C}_o(G)$. The subgraphs  $C_{10}, C_{11}, C_{12}$ in the following allow us to recognize them.

\begin{minipage}[b]{14cm}
Let $\mathcal{C}_{10}(G)$ be the set of subgraphs of $MC(G)$ which is not contained in any subgraphs of $\mathcal{C}_e(G) \cup \mathcal{C}_o(G)$ and isomorphic to $C_{10}$ in the figure on the right where $A$ contains a cut-vertex of $G$ different from $y,z$.
\end{minipage}\hfill
\begin{minipage}[b]{1.5cm}
\includegraphics[scale=.8]{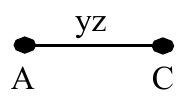}
\vspace*{.15cm}
\end{minipage}

\begin{minipage}[b]{14cm}
Let $\mathcal{C}_{11}(G)$ be the set of subgraphs of $MC(G)$ which is not contained in any subgraphs of $\mathcal{C}_e(G) \cup \mathcal{C}_o(G)$ and isomorphic to $C_{11}$ in figure on the right where $x$ is a cut-vertex of $G$, and  $A, C$ do not have any other neighbours in the same block.
\end{minipage}\hfill
\begin{minipage}[b]{1.5cm}
\includegraphics[scale=.8]{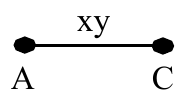}
\vspace*{.15cm}
\end{minipage}

\begin{minipage}[b]{14cm}
Let $\mathcal{C}_{12}(G)$ be the set of subgraphs of $MC(G)$ which is not contained in any subgraphs of $\mathcal{C}_e(G) \cup \mathcal{C}_o(G)$ and isomorphic to $C_{12}$ in the figure on the right where $x$ is a cut-vertex of $G$, and $A,B,C$ do not have any other neighbours in the same block.
\end{minipage}\hfill
\begin{minipage}[b]{1.5cm}
\includegraphics[scale=.8]{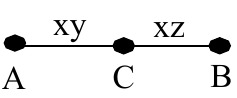}
\vspace*{.15cm}
\end{minipage}

And  $\mathcal{C}_c(G) = \mathcal{C}_{10}(G) \cup \mathcal{C}_{11}(G) \cup \mathcal{C}_{12}(G)$ ($c$ for cut-vertex).

The following lemma shows that $\mathcal{C}_c(G)$ allows us to recognize the remaining cycles.

\begin{lemma}Let $G$ be a $4$-leaf basic g-network power and let $\mathcal{B}$ be a block of $G$ which is not a clique and contains a cut-vertex $x$.
Suppose that $\mathcal{B}$ has a basic g-network root $N_\mathcal{B}$ in which $p(x)$ is adjacent to only invisible vertex. Then $N_\mathcal{B}$ has a cycle $c$  which is contained in a root of a subgraph of $\mathcal{C}_e(\mathcal{B}) \cup \mathcal{C}_o(\mathcal{B}) \cup \mathcal{C}_c(\mathcal{B})$ such that $S(c)$ \index{$S(c)$} contains $x$.
\label{lem:cut_vertex}
\end{lemma}

\begin{pf}
By Lemma \ref{lem:inv_cycle} (i), each of invisible vertices adjacent to $p(x)$ is contained in a cycle of $N_\mathcal{B}$. Because $N_\mathcal{B}$ does not have intersecting cycles, these invisible vertices are contained in an unique cycle. 

\begin{minipage}[b]{13cm}
So,  there are only $2$ possible cases as in the figures on the right, where $u,v$ are invisible vertices contained in the cycle $c$. In both two cases, $x$ is contained in $S(c)$. In the first case, $x$ is contained in exactly one maximal clique of $\mathcal{B}$, corresponding to the invisible star of $u$. In the second case, $x$ is contained in at most $3$ maximal cliques of $\mathcal{B}$. Two of them correspond to the two invisible stars of $u, v$, and it may exist the third one corresponding to the cycle $c$ passing $p(x), u, v$. In the latter case, $c$ must be either $N''_5$ or $N_6$ because only these networks of $\mathcal F$ can have $2$ invisible vertices on their cycle.
\end{minipage}\hfill
\begin{minipage}[b]{2.5cm}
\includegraphics[scale=.7]{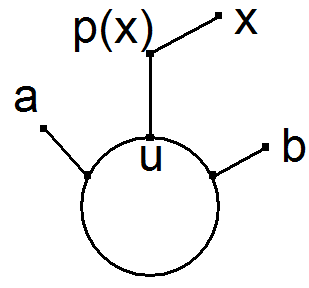}
\includegraphics[scale=.7]{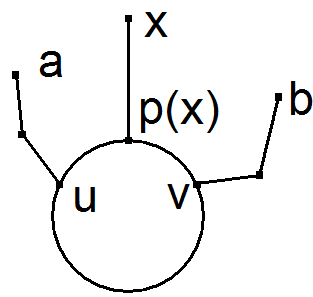}
\end{minipage}

Suppose that $x$ is not contained in any subgraph  of $\mathcal{C}_e(\mathcal{B}) \cup \mathcal{C}_o(\mathcal{B})$, then the cycle $c$ is not contained in any root of $\mathcal{C}_e(\mathcal{B})$ or $\mathcal{C}_o(\mathcal{B})$. So $c$ is one of the cycle in the networks that are replaced by the networks without cycle in the proof of Lemma \ref{lem:replace} (Figures \ref{fig:3_1}, \ref{fig:4_1}, \ref{fig:4_2}, \ref{fig:5_1}, \ref{fig:5_2}, \ref{fig:6_1}). By calculating the corresponding maximal cliques graphs, there are $3$ cases:

- $x$ is contained in exactly one maximal clique $A$ of $\mathcal{B}$, then $x$ is contained in a subgraph of $\mathcal{C}_{10}(\mathcal{B})$.

- $x$ is contained in exactly two maximal cliques $A,C$ of $\mathcal{B}$, then $x$ is contained in a subgraph of $\mathcal{C}_{11}(\mathcal{B})$. Indeed, if $A, C$ have neighbours in  $\mathcal{B}$, then by checking the root list we see that the corresponding subgraph does not have any root in which $p(x)$ is adjacent to only invisible vertex, a contradiction. 

- $x$ is contained in exactly three maximal cliques $A,B,C$ of $\mathcal{B}$, then  $x$ is contained in a subgraph of $C_{12}(\mathcal{B})$. Indeed, if $A,B,C$ have neighbours in $\mathcal{B}$, then either there is not any root in which $p(x)$ is adjacent to only invisible vertex, or the corresponding subgraph is in $\mathcal{C}_e(G) \cup \mathcal{C}_o(G)$, contradictions. 
\end{pf}

\begin{figure}[ht]
\begin{center}
\subfigure[$C_{10}$\label{fig:C_10_}]{\includegraphics[scale=.8]{Figures/C10.pdf}}
\subfigure[$R_{10,1}(y)$\label{fig:R_10,1}]{\includegraphics[scale=.75]{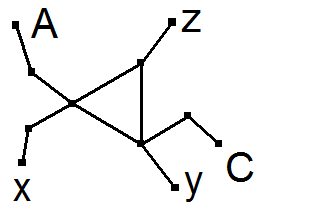}}
\subfigure[$R_{10,2}$\label{fig:R_10,2}]{\includegraphics[scale=.75]{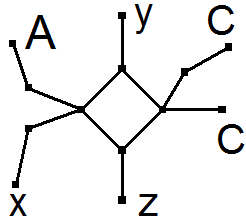}}
\subfigure[$R_{10,3}(y)$\label{fig:R_10,3}]{\includegraphics[scale=.75]{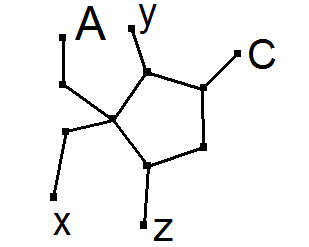}}
\subfigure[$R_{10,4}$\label{fig:R_10,4}]{\includegraphics[scale=.75]{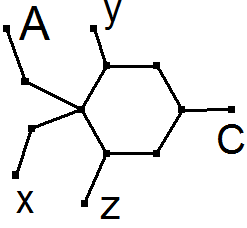}}
\subfigure[$R_{10,5}(y)$\label{fig:R_10,5}]{\includegraphics[scale=.7]{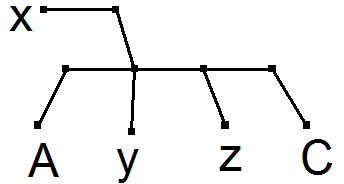}}
\caption{$C_{10}$ and  its $4$-leaf basic g-network roots, $x$ is a cut-vertex contained in $A$.}
\label{fig:C_10}
\end{center}
\end{figure}

\begin{figure}[ht]
\begin{minipage}[b]{70mm}
\begin{center}
\subfigure[$C_{11}$\label{fig:C_11_}]{\includegraphics[scale=.8]{Figures/C11.pdf}}
\subfigure[$R_{11,1}$\label{fig:R_11,1}]{\includegraphics[scale=.7]{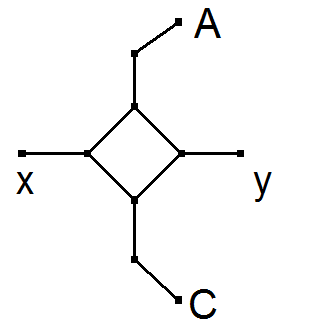}}
\subfigure[$R_{11,2}(x)$\label{fig:R_11,2}]{\includegraphics[scale=.7]{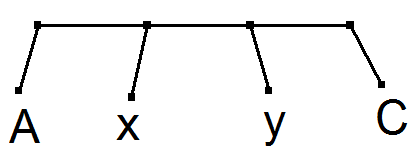}}
\caption{$C_{11}$ and  its $4$-leaf basic g-network roots, $x$ is a cut-vertex.}
\label{fig:C_11}
\end{center}
\end{minipage}\hfill
\begin{minipage}[b]{70mm}
\begin{center}
\subfigure[$C_{12}$\label{fig:C12_}]{\includegraphics[scale=.8]{Figures/C12.pdf}}
\subfigure[$R_{12,1}$\label{fig:R_12,1}]{\includegraphics[scale=.7]{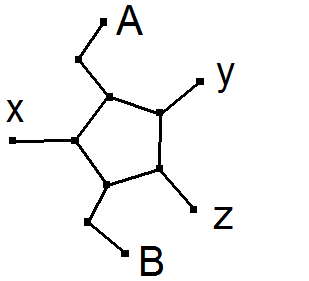}}
\subfigure[$R_{12,2}$\label{fig:R_12,2}]{\includegraphics[scale=.7]{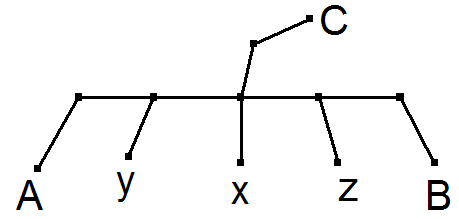}}
\caption{$C_{12}$ and  its $4$-leaf basic g-network roots, $x$ is a cut-vertex.}
\label{fig:C_12}
\end{center}
\end{minipage}
\end{figure}

For each subgraph $C_{10}, C_{11}, C_{12}$, we find also a finite list of its corresponding roots (Figures \ref{fig:C_10}, \ref{fig:C_11}, \ref{fig:C_12}). Beside the roots without cycle (Figures \ref{fig:R_10,5}, \ref{fig:R_11,2}, \ref{fig:R_12,2}), we have also the roots containing cycles for the case that all neighbour vertices of $p(x)$ are invisible (figures \ref{fig:R_10,1}, \ref{fig:R_10,2}, \ref{fig:R_10,3}, \ref{fig:R_10,4}, \ref{fig:R_11,1}, \ref{fig:R_12,1}). These cycles do not necessarily appear in the roots of one block (because we can replace it by a cycless network as proved in Lemma \ref{lem:replace}), but when $p(x)$ needs to be adjacent to only invisible vertices, they necessarily appear. 

For any subgraph $S \in \mathcal{C}_c(G)$, we also restrain the choices of roots of $S$ from the full provided lists as in Definition \ref{def:r_S}. Hence, the obtained root set $r(S)$ has the same properties as the root set of the subgraphs of $\mathcal{C}_o(G)$ in Theorem \ref{theo:G_bc}. 
In fact, only the subgraphs of $\mathcal{C}_{10}(G)$ are concerned here because by definition the subgraphs of $\mathcal{C}_{11}(G)$ and $\mathcal{C}_{12}(G)$ do not have other neighbours in the same blocks.

Denote $\mathcal{S}(G) = \mathcal{C}_e(G) \cup \mathcal{C}_o(G) \cup \mathcal{C}_c(G)$ \index{$\mathcal{S}(G)$}, we have the following lemma.

\begin{corollary}
\label{coro:Q}
Suppose that $G$ has at least one $4$-leaf basic network root. Then there is a $4$-leaf basic network root $N$ of $G$ such that any cycle of $N$ is contained in a root of a subgraph of $\mathcal{S}(G)$.
\end{corollary}

\begin{pf}
It can be seen that each cycle listed in Figures \ref{fig:3_1}, \ref{fig:4_1}, \ref{fig:4_2}, \ref{fig:5_1}, \ref{fig:5_2}, \ref{fig:6_1} of Lemma \ref{lem:replace} is contained in a root of a subgraph  of $\mathcal{C}_c(G)$ in Figures \ref{fig:C_10}, \ref{fig:C_11}, \ref{fig:C_12}. Combining with the proof of Lemma \ref{lem:replace}, we are done.
\end{pf}

So, we are going to construct the roots of $G$ in which each cycle is contained in a root of a subgraph of $\mathcal{S}(G)$ \index{$\mathcal{S}(G)$}.

For each block $\mathcal{B}$ which contains a cut-vertex $x$, we introduce two variables $c_\mathcal{B}(x)$ and $i_\mathcal{B}(x)$ with the convention that: $c_\mathcal{B}(x) = true$ in a root of $\mathcal{B}$ if $p(x)$ is not contained in any cycle of this root, $false$ otherwise; $i_\mathcal{B}(x)=true$ in a root of $\mathcal{B}$ if $p(x)$ is adjacent to only invisible vertices in this root and $false$ otherwise. The $2$ conditions in Lemma \ref{lem:bc} can be rewritten as: in any root of $G$ we have $c_\mathcal{B}(x) \vee c_{\mathcal{B}'}(x)=true$ and $i_\mathcal{B}(x) \vee i_{\mathcal{B}'}(x)=true$ for any two blocks $\mathcal{B}, \mathcal{B}'$ sharing a cut-vertex $x$. We say that these values are \textit{compatible} in this case. 

If $\mathcal{B}$ is a block clique, then by the constraints imposed on the roots of $G$, we have $c_\mathcal{B}(x)=true$ and $i_\mathcal{B}(x)=true$ because the root of $\mathcal{B}$ is an invisible star. So, there is no constraint imposed by $\mathcal{B}$ on the root of $\mathcal{B}'$. Hence, we need to consider only the case that neither $\mathcal{B}$ nor $\mathcal{B}'$ is a clique. 

If $x$ is not in any subgraph of $\mathcal{S}(\mathcal{B})$, then by Lemma \ref{lem:cut_vertex}, Corollary \ref{coro:Q}, $p(x)$ is not contained in any cycle and  it is adjacent to at least one visible vertex. So in this case, we can impose $c_\mathcal{B}(x) = true$ and $i_\mathcal{B}(x) =false$. 

If $x$ is contained in at least a subgraph of $\mathcal{S}(\mathcal{B})$ which has several roots and each one gives a different values for these variables, then we must choose the appropriate one. So, the conditions in Lemma \ref{lem:bc} restrain the choices of roots of each subgraph of $\mathcal{S}(G)$ which contains a cut-vertex. For example, the root set of a subgraph $S_1$ may restrain the choice of roots of a subgraph $S_2$ which is contained in an adjacent block of the block of $S_1$. This restriction may influent the choice of roots of other subgraphs and so on. However, we propose in the following theorem an efficient way to apply these conditions. 

\begin{theorem}
\label{theo:condition} Given $\mathcal{S}(G)$ and the root set of each subgraph in $\mathcal{S}(G)$, one can find in linear time  a root for each subgraph of $\mathcal{S}(G)$, if there is any, such that the conditions in Lemma \ref{lem:bc} are satisfied at any cut-vertex of $G$.
\end{theorem}

\begin{pf}
Let $BC(G)$ be the \textit{block tree} of $G$. We choose a certain node $R$ of $BC(G)$ as its root. At the beginning, we impose $root(S)=r(S)$, i.e. the root set of $S$ for any subgraph $S$ of $\mathcal{S}(G)$. We apply the condition in Lemma \ref{lem:bc} as follows:

- Begin with the blocks corresponding to the leaves of $BC(G)$: for any subgraph $S \in \mathcal{S}(G)$ of these blocks, we keep $root(S)=r(S)$.

For each block $\mathcal{B}$ of $G$, denote by $BC_\mathcal{B}(G)$ the subtree of  $BC(G)$ rooted at $\mathcal{B}$.

Induction hypothesis: Let $\mathcal{B}$ is the currently considered block, then for any root in $root(S)$ of a subgraph $S$ of $\mathcal{B}$, there exists at least a root in $root(S')$ for each subgraph $S'$ in any block of $BC_\mathcal{B}(G)$ such that the conditions in Lemma \ref{lem:bc} are satisfied at any cut-vertex between two adjacent blocks in $BC_\mathcal{B}(G)$. 

- Consider a block $\mathcal{B}$ such that all of its descendants have been already considered. 
Let $\mathcal{B}'$  be a block in $BC_\mathcal{B}(G)$ which has a common cut-vertex $x$ with $\mathcal{B}$. For any subgraph $S$ of $\mathcal{S}(\mathcal{B})$ which contains $x$, we keep only the roots $r$ in  $root(S)$ such that for any subgraph $S'$ of $\mathcal{S}(\mathcal{B}')$ which contains $x$, there is at least a root $r'$ in $root(S')$ which gives the compatible values to $r$.

- We continue this process up to the root $R$.  Suppose that $root(S) \neq \emptyset$ for every subgraph $S$ of the block $R$. Then by the induction hypothesis, there is at least a root in $root(S)$ for each subgraph $S$ of any block of $BC(G)$ such that the conditions in Lemma \ref{lem:bc} are satisfied at any cut-vertex of $G$. Otherwise, if there exists a subgraph $S$ of the block $R$ such that $root(S) = \emptyset$, then $G$ is not a $4$-leaf basic g-network power. 

Finally, we can easily choose for each subgraph $S$ of $\mathcal{S}(G)$ a root in $root(S)$ as follows:

- Choose for each subgraph $S$ of $R$ a certain root in $root(S)$.

- Next, for each subgraph $S'$ of $\mathcal{S}(\mathcal{B})$ where $\mathcal{B}$ is a child block of $R$, choose a certain root in $root(S)$ which is compatible with the chosen roots on $R$ if they share a common cut-vertex. There exists always at least such a root due to the above proof.

- We continue to do that down to the leaf block of $BC(G)$. 

Hence, the obtained collection of roots verify the property stated in the theorem.
\end{pf}

We call each collection of roots calculated in Theorem \ref{theo:condition} a \textbf{subroot set} \index{subroot set} of $G$. So, if $\mathcal{R}$ is a subroot set of $G$, then each subgraph of $\mathcal{S}(G)$ has exactly one root in $\mathcal{R}$. Moreover, the cycles in the networks of $\mathcal{R}$ are pairwise vertex disjoint, and the conditions in Lemma \ref{lem:bc} are satisfied at any cut-vertex $x$ of $G$. Therefore, according to Corollary \ref{co:e} and Theorem \ref{theo:G_bc}, if $G$ is a $4$-leaf basic g-network power then there is always a basic g-network $N$ containing the subnetworks of $\mathcal{R}$ as induced subgraphs. 

Let $\mathcal{R}_c=\{S(c)|~c$ is a cycle in a network of $\mathcal{R}\}$, namely a \textbf{cycle-root set} \index{cycle-root set} of $G$. 

After constructing the collection of cycles, we consider the remaining parts of the roots in the next section.

\subsection{Putting the things together and constructing the roots}

Let $N$ be a basic g-network without invisible vertices. Denote by $X(N)$ \index{$X(N)$} the graph obtained from $N$ by labelling each inner vertex following the label of its leaf and removing all of the leaves. So $X$ is a bijection from the set of basic g-networks without invisible vertices to the set of g-networks. It is easy to see that the $4$-leaf power of $N$ is exactly the square of $X(N)$.

\begin{proposition}
Let $H$ be the square of a tree and its vertices are distinctly labelled. 

If $H$ is a clique then the square tree roots of $H$ are all possible stars having vertices set equal to the vertices set of $H$.

If $H$ contains $2$ maximal cliques then $H$ has two square tree roots, each one consisting of two adjacent stars. These roots have the same forms but are labelled differently.

If $H$ contains at least $3$ maximal cliques then $H$ has a unique square tree root.
\label{propo:squaretree}
\end{proposition}

\begin{figure}[ht]
\begin{center}
\subfigure[A clique and its square tree roots.\label{fig:squaretree1}]{\includegraphics[scale=.7]{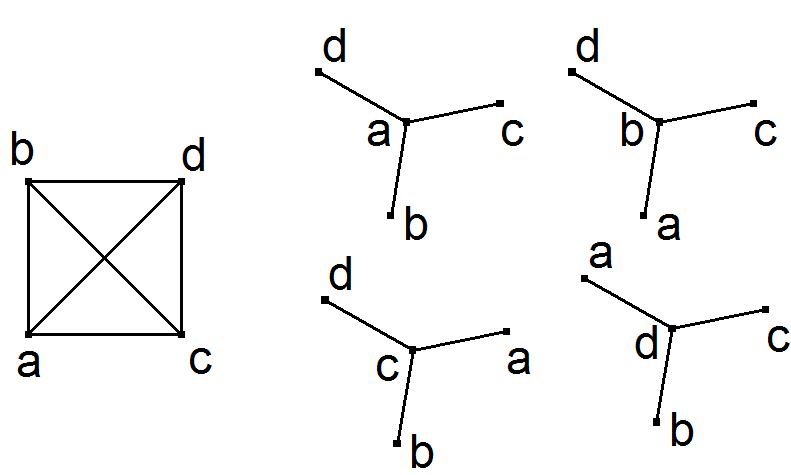}}\;\;\;
\subfigure[A graph consisting of $2$ maximal cliques and its two square tree roots.\label{fig:squaretree2}]{\includegraphics[scale=.7]{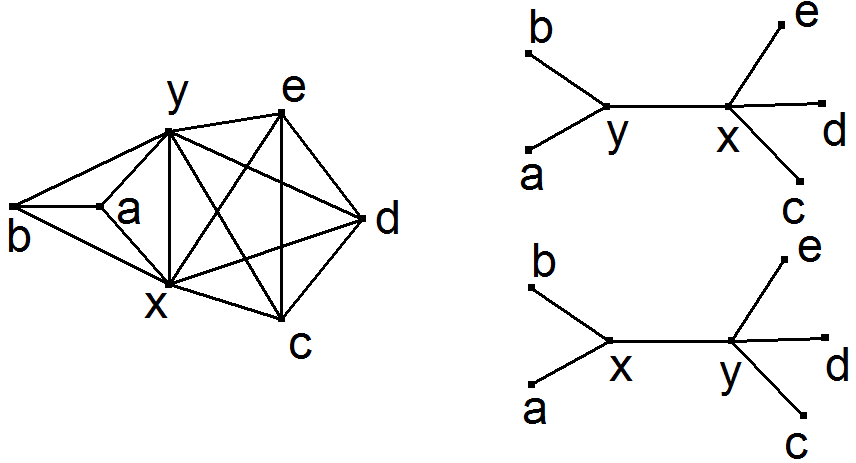}}\;\;\;
\subfigure[A graph and its unique square tree roots.\label{fig:squaretree3}]{\includegraphics[scale=.7]{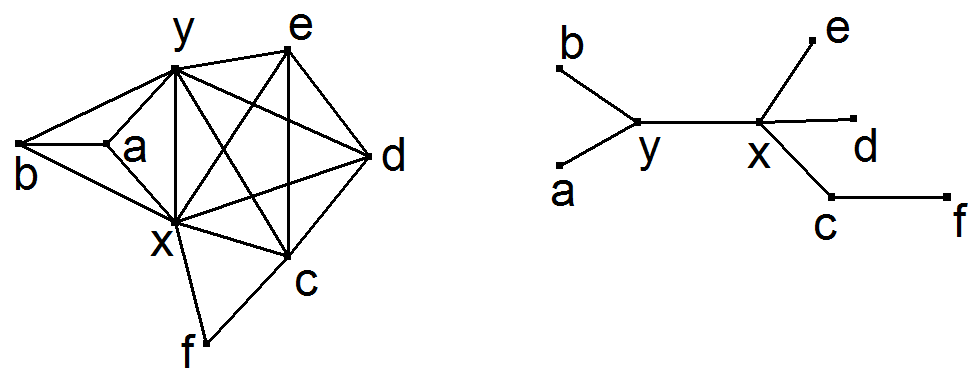}}
\caption{Tree square roots.}
\label{fig:squaretree}
\end{center}
\end{figure}

\begin{pf}
If $H$ is a clique, it is easy to see that its tree roots must be stars. These stars are different by the labels of their middle vertices (Figure \ref{fig:squaretree1}).

\cite{BL06} represents a characteristic of the square of a tree and the way to construct its roots: each maximal clique corresponds to a star, and the middle vertex of a star is uniquely determined (Figure \ref{fig:squaretree3}) except when $H$ contains only two maximal cliques. For example $H$ has $2$ maximal cliques $A,B$, so they have exactly $2$ common vertices $x,y$. So, the star corresponding to $A$ can take either $x$ or $y$ as its middle vertex, and the star corresponding to $B$ take the other vertex as its middle vertex. Hence, in this case $H$ has two roots (Figure \ref{fig:squaretree2}).
\end{pf}

Given a cycle-root set $\mathcal{R}_c$ \index{cycle-root set} \index{$\mathcal{R}_c$} of $G$, and a block $\mathcal{B}$ of $G$, denote by $\mathcal{R}_c[\mathcal{B}]$  the cycles of $\mathcal{R}_c$ restricted on $\mathcal{B}$. 

\begin{definition}Let $\mathcal{B}$ be a block of $G$ and let $\mathcal{R}_c$ be a cycle-root set of $G$. We define $\mathcal{B} \setminus \mathcal{R}_c$ \index{$\mathcal{B} \setminus \mathcal{R}_c$} as the graph  such that $MC(\mathcal{B} \setminus \mathcal{R}_c)$ is constructed as follows: 

For each cycle $c$ in $\mathcal{R}_c[\mathcal{B}]$, let $A_1, \dots , A_k$ be the nodes of $MC(\mathcal{B})$ corresponding to the visible, invisible (quasi) stars of $S(c)$. Let $M$ be the node corresponding to the subnetwork of $\mathcal F$ contains $c$ if there is any. 
Then, delete $M$ and all edges between $A_i$ from $MC(\mathcal{B})$.
\label{def:B-C}
\end{definition}

\begin{example}
For instance, see the two examples in Figures \ref{fig:B_C}, \ref{fig:B_C1}. In Figure \ref{fig:B_C}, $MC(\mathcal{B})$ contains a cycle which is a subgraph of $\mathcal{C}'(\mathcal{B})$. By Lemma \ref{lem:type1}, there is a cycle in any root of $\mathcal{B}$ in which each node $A,B,C,D,E,F$ corresponds to a visible or invisible star of this cycle. So, in $MC(\mathcal{B} \setminus \mathcal{R}_c)$, we delete the edges between them. In Figure \ref{fig:B_C1}, $MC(\mathcal{B})$ contains  subgraph $AMB$ in $\mathcal{C}_2(\mathcal{B})$. By Lemma \ref{lem:type1_small}, there is a cycle in any root of $\mathcal{B}$ in which each node $A,B$ corresponds to a visible star and $M$ corresponds to the cycle. So, in $MC(\mathcal{B} \setminus \mathcal{R}_c)$, we delete $M$ and the edges $MA,MB$. 

\begin{figure}[ht]
\begin{center}
\def\svgwidth{10cm}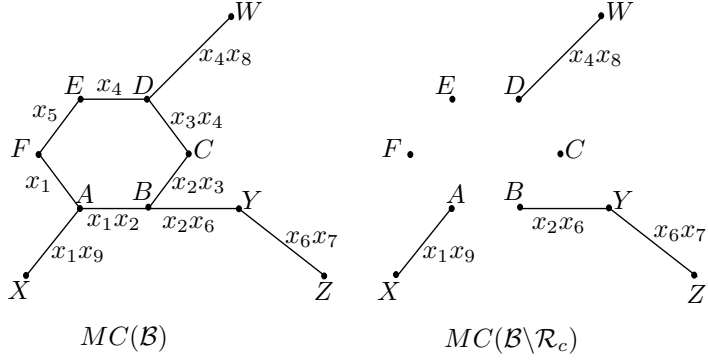
\caption{Example $1$}
\label{fig:B_C}
\end{center}
\end{figure}

\begin{figure}[ht]
\begin{center}
\def\svgwidth{8cm}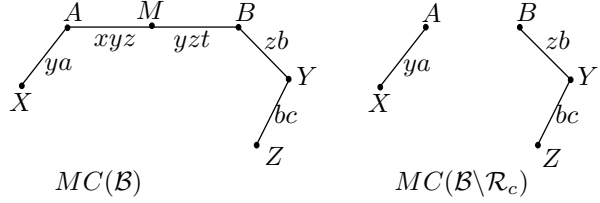
\caption{Example $2$}
\label{fig:B_C1}
\end{center}
\end{figure}

\end{example}

\begin{minipage}[b]{8.8cm}
\begin{definition}[Separating the stars]
Let $c$ be a cycle in a network $N$. The operation of \textbf{separating the stars} \index{separating the stars} of $S(c)$ is  illustrated in the figure, i.e. the stars which have the middle vertices on $c$ are separated.
Denote by $\textbf{T(N)}$ \index{$T(N)$} the graph obtained from $N$ by this operation on all cycles of $N$. 
\label{def:T_N}
\end{definition}
So, for any network $N$, $T(N)$ is a forest. 
By Lemma \ref{lem:inv_cycle}-(i), if $N$ has an invisible vertex, then it must be contained in a cycle of $\mathcal{R}_c[\mathcal{B}_i]$.
\end{minipage}\hfill
\begin{minipage}[b]{6.7cm}
\includegraphics[scale=.2]{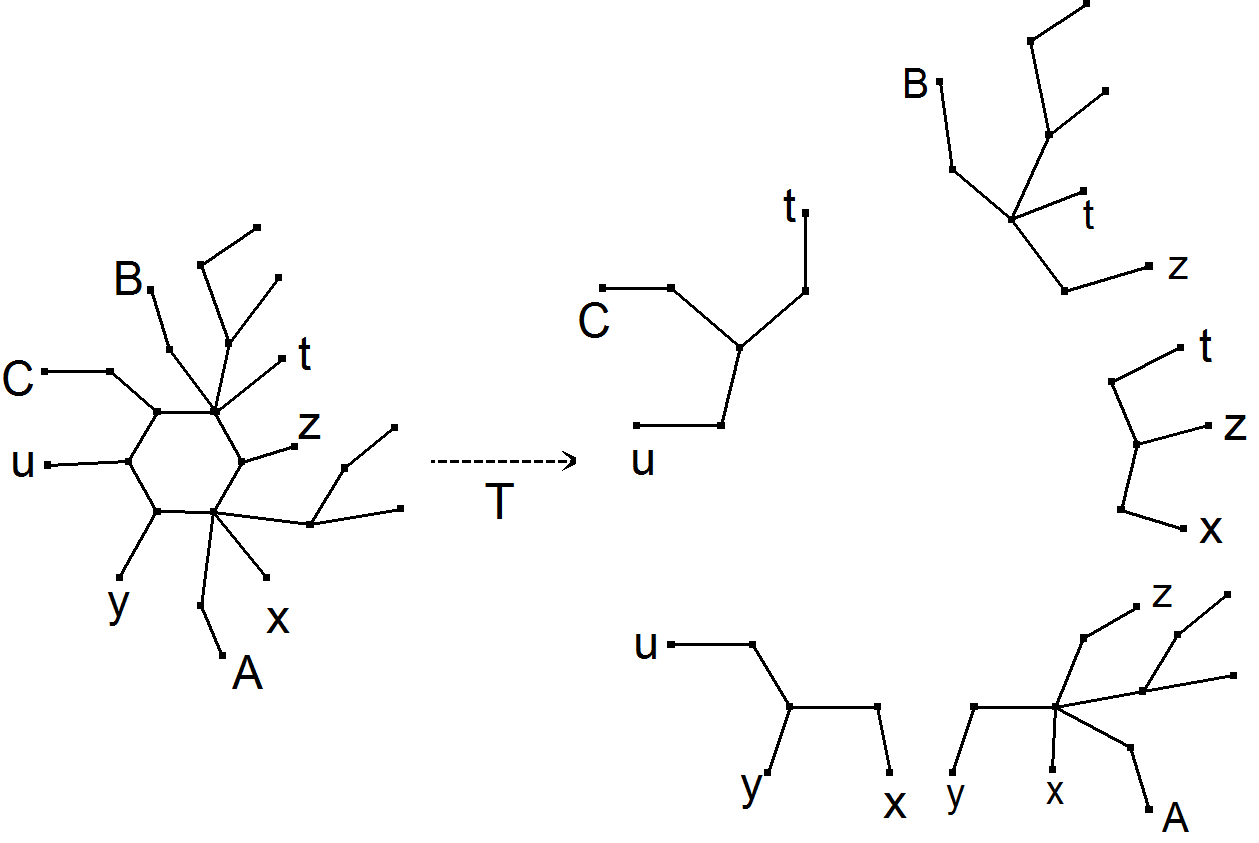}
\end{minipage}

So, each tree of $T(N)$ is either an invisible star or  a tree without invisible vertex. 

\begin{theorem}
\label{theo:4power}
Let $\mathcal{B}_1, \dots, \mathcal{B}_k$ be the blocks of $G$, then $G$ is a $4$-leaf basic g-network power iff $G$ has a cycle-root set  \index{cycle-root set} $\mathcal{R}_c$ such that for every $1\leq i \leq k$, $\mathcal{B}_i \setminus \mathcal{R}_c$ is the square of a forest.
\end{theorem}

\begin{pf}
$\Rightarrow$ Suppose that $G$ has a $4$-leaf basic g-network root $N$. Then, the collection of cycles of $N$ forms a cycle-root set $\mathcal{R}_c$ of $G$. 
For each block $\mathcal{B}_i$ of $G$, let $N_i = N[\mathcal{B}_i]$. 

If $\mathcal{B}_i$ is a clique then $N[\mathcal{B}_i]$ is an invisible star, i.e. it does not contain any cycle. So, $\mathcal{B} \setminus \mathcal{R}_c = \mathcal{B}$, which is obviously the square of a tree.

Otherwise, for any $i =  1, \dots, k$, we calculate $T(N_i)$. By Definitions \ref{def:B-C} and \ref{def:T_N}, each connected component $H$ of $\mathcal{B}_i \setminus \mathcal{R}_c$ is a $4$-leaf power of a tree $T$ in the forest $T(N_i)$. 
If $T$ is an invisible star, then $H$ is a clique, so it is obviously the square of a tree. 
Otherwise, $T$ is without invisible vertex, then $H$ has $X(T)$ as a square tree root. In other words, $\mathcal{B}_i \setminus \mathcal{R}_c$ is the square of a forest.

$\Leftarrow$ Suppose that $G$ has a cycle-root set $\mathcal{R}_c$ such that $\mathcal{B}_i \setminus \mathcal{R}_c$ is the square of a forest for any $i$. 

If $\mathcal{B}_i$ is a clique, then we construct $N_i$ as an invisible star having the leaf set equal to the vertex set of $\mathcal{B}_i$. 

Otherwise, we will show a way to construct a network $N_i$ which is a $4$-leaf basic g-network root of $\mathcal{B}_i$. Let $H$ be a connected component of $\mathcal{B}_i \setminus \mathcal{R}_c$, so $H$ is the square of a tree. There are two cases:

- $H$ is a clique which corresponds to an invisible (quasi) star having the middle vertex on a cycle of $\mathcal{R}_c[\mathcal{B}_i]$. In this case we construct a $4$-leaf root $r$ of $H$ which is an invisible star. It is the case of the clique $(u,t)$ in Figure \ref{fig:theorem}.

- otherwise, we will construct a $4$-leaf root $r$ of $H$ which is a basic tree without invisible vertex. It is equivalent with constructing a square tree root $t=X^{-1}(r)$ of $H$. The root $r$ of $H$ must be \textit{compatible} with the cycles in $\mathcal{R}_c[\mathcal{B}_i]$, it means that it must satisfy the following conditions: For any maximal clique $A$ of $H$ such that there is a cycle $c$  of $\mathcal{R}_c[\mathcal{B}_i]$ and $S(c)[A]$ is a visible (quasi) star, then $r[A]$ must be the visible star having the same label of the middle vertex of $S(c)[A]$. In other words, $t[A]$ is a star having the same label of the  middle vertex of $S(c)[A]$. 

For example, in Figure \ref{fig:theorem}, let consider the connected component $H$ which contains the maximal clique $A=(b,c,t,z)$ of $\mathcal{B}_i \setminus \mathcal{R}_c$. There is a cycle $c$ in $\mathcal{R}_c[\mathcal{B}_i]$ where $S(c)[A]$ is a visible star having $t$ as the label of the middle vertex. So, we construct the root of $H$ as a visible star also having $t$ as the label of the middle vertex.

There is always such a root of $H$ because:

\hspace*{.5cm}. If $H$ is a clique, i.e. $H=A$, then $S(c)[A]$ is certainly a root of $H$. So we need only to choose $r=S(c)[A]$.

\hspace*{.5cm}. If $H$ contains $2$ maximal cliques $A,B$, then the label of the middle vertex of $S(c)[A]$ is one of the common labels of $A,B$ (see Remark \ref{rem:middle}). It is also the way that we choose the middle vertex for the root of $H$ in this case (see Proposition \ref{propo:squaretree}).

\hspace*{.5cm}. If $H$ contains at least $3$ maximal cliques, then by Proposition \ref{propo:squaretree}, it has a unique square root $t$. 
The label of the middle vertex of $t[A]$, or $r[A]$, and $S(c)[A]$ must have the same label. This claim is followed from the fact that when we construct the roots for each subgraph $S$ in $\mathcal{S}(\mathcal{B})$, we consider also the nodes around $S$ (see the way of determining the middle vertex in Remark \ref{rem:middle} and of considering the nodes around each subgraph up to distance $2$ in Definition \ref{def:r_S}).

Denote by $F_i$ the forest consisting of the roots of $\mathcal{B}_i \setminus \mathcal{R}_c$ such that the root of each connected component of $\mathcal{B}_i \setminus \mathcal{R}_c$ is constructed as above. 
We calculate $N_i$ by overlaying the common parts of the stars in the trees of $F_i$. In fact, $N_i = T^{-1}(F_i)$, because this is the inverse operation of separating the cycles in Definition \ref{def:T_N}. It is easy to see that $N_i$ is a $4$-leaf root of $\mathcal{B}_i$, and each cycle in $\mathcal{R}_c[\mathcal{B}_i]$ is a cycle of $N_i$. 

\textit{We must prove furthermore that $N_i$ is a basic g-network}. 
For each block $\mathcal{B}_i$, each cycle of $\mathcal{R}_c[\mathcal{B}_i]$ is a cycle of $N_i$, so these cycles are vertex disjoint. Suppose that $N_i$ has a cycle $c$ not in $\mathcal{R}_c[\mathcal{B}_i]$. $N_i$ is a $4$-leaf root of $\mathcal{B}_i$, so by Corollary \ref{coro:Q}, $c$ must be contained in a root of a subgraph $S$ of $\mathcal{S}(\mathcal{B}_i)$. However, by definition of cycle-root set, $S$ must has a root $r$ whose cycle is contained in $\mathcal{R}_c[\mathcal{B}_i]$. This cycle must be exactly $c$ because otherwise $r$ has two distinct cycles, a contradiction. Hence, $N_i$ is a basic g-network. 

Finally, we combine $N_i$ for any $i=1, \dots, k$ by gluing all $p(x)$ for any cut-vertex $x$ of $G$ and deleting all copies of $x$ except one.
So, the resulting network $N$ is a $4$-leaf basic g-network root of $G$ due to Lemma \ref{lem:bc} and the property of the cycle-root set $\mathcal{R}_c$, i.e. no cycles in the roots of two different blocks of $G$ can intersect.
\end{pf}

We can resume the method to construct a $4$-leaf basic g-network root of $G$ in the following algorithm. See Figure \ref{fig:theorem} for an illustration.

\restylealgo{boxed}\linesnumbered
\begin{algorithm}[ht]
\caption{Construction of $4$-leaf basic g-network root}
\label{algorithm2}
\KwData{A connected undirected graph $G$, and a fixed $k$}
\KwResult{A $4$-leaf basic g-network root of $G$, if there exists one}
\For {(each block $\mathcal{B}$ of $G$)}{
	Calculate $MC(\mathcal{B})$\;
	Calculate $\mathcal{S}(\mathcal{B}) = \mathcal{C}_e(\mathcal{B}) \cup \mathcal{C}_o(\mathcal{B}) \cup \mathcal{C}_c(\mathcal{B})$\;	
	Construct the root set $r(S)$ of each subgraph  $S \in \mathcal{S}(\mathcal{B})$\;
	\If {($\exists S \in \mathcal{S}(\mathcal{B})$ such that $r(S)=\emptyset$)}{
		\Return \textit{null}\;
	}
}
\If {($G$ does not have any cycle-root set)}{
		\Return \textit{null}\;
}
Calculate a cycle-root set $\mathcal{R}_c$ of $G$.

\For {(each block $\mathcal{B}$ of $G$)}{
Calculate $MC(\mathcal{B} \setminus \mathcal{R}_c)$ and $\mathcal{B} \setminus \mathcal{R}_c$\;
\If{($\mathcal{B} \setminus \mathcal{R}_c$ is not the square of a forest)}{
\Return \textit{null}\;
}
Construct the square root of each connected component of $\mathcal{B} \setminus \mathcal{R}_c$ compatible with $\mathcal{R}_c$\;
Construct a $4$-leaf basic g-network root of $\mathcal{B}$\;
}
$N \leftarrow$ Combine the $4$-leaf basic g-network root of each block\;
\Return $N$
\end{algorithm}

\begin{corollary}
\label{coro:polynomial}
Recognizing a $4$-leaf power of a basic g-network and constructing one of its roots, if there is any, can be done in polynomial time.
\end{corollary}

\begin{pf}
We will prove that Algorithm \ref{algorithm2} is polynomial.

Calculating the blocks of $G$ is done in linear time. 

For each block $\mathcal{B}$ of $G$, let $V_\mathcal{B}, E_\mathcal{B}$ be the number of vertices and edges of $\mathcal{B}$. We know that $MC(\mathcal{B})$ is calculated in $|V_\mathcal{B}|.|E_\mathcal{B}|^2$ (Theorem \ref{theo:mc}) and $BC(\mathcal{B})$ is constructible in linear time $O(|V_\mathcal{B}|+|E_\mathcal{B}|)$. 

We consider now the complexity to calculate the subgraphs of $\mathcal{S}(B)$.
It can be deduced from Lemmas \ref{lem:type1}, \ref{lem:type1_small} and the lists of roots of $C_6$, $C_8$ that the subgraphs of $\mathcal{C}'(\mathcal{B})$, $\mathcal{C}_1(\mathcal{B})$, $\mathcal{C}_3(\mathcal{B})$, $\mathcal{C}_6(\mathcal{B})$, $\mathcal{C}_8(\mathcal{B})$  must be pairwise node disjoint, otherwise the cycles in their roots intersect. Moreover, they are the only subgraphs of $\mathcal{S}(\mathcal{B})$ that contain cycles. It implies that if $\mathcal{B}$ is a $4$-leaf basic g-network power, then each non trivial block of $MC(\mathcal{B})$ is a one of these subgraphs. 

Remark that:

- Each subgraph of $\mathcal{C}'(\mathcal{B})$, $\mathcal{C}_1(\mathcal{B})$, $\mathcal{C}_6(\mathcal{B})$, $\mathcal{C}_8(\mathcal{B})$ is a cycle, 

- Each subgraph of $\mathcal{C}_3(\mathcal{B})$ contains $8$ nodes, $12$ edges, 

- The number of nodes of $MC(\mathcal{B})$ is bounded by $3.|E_\mathcal{B}|$ (Corollary \ref{co:3e}). 

So if $\mathcal{B}$ is a $4$-leaf basic g-network power, then the number of edges of $MC(\mathcal{B})$ is bounded by $c.|E_\mathcal{B}|$, where $c$ is a constant. 
Hence, calculating the blocks of $MC(\mathcal{B})$ and checking the form of each one can be done in $O(|E_\mathcal{B}|)$. 

Calculating $\mathcal{C}_2(\mathcal{B})$ is also done in $O(|E_\mathcal{B}|)$.

Calculating the other subgraphs of $\mathcal{C}_o(\mathcal{B})$, $\mathcal{C}_c(\mathcal{B})$ is done in $O(|E_\mathcal{B}|^4)$ because each subgraph has at most $4$ nodes. So, its takes in total $O(|E_\mathcal{B}|^4)$ to construct $\mathcal{S}(\mathcal{B})$.

Next, constructing the root set of each subgraph is done in at most $O(|E_\mathcal{B}|)$ following the constructible Lemmas \ref{lem:type1}, \ref{lem:type1_small} for the subgraphs in $\mathcal{C}_e(B)$, and following the provided finite list of roots, the construction of $r(S)$ in Definition \ref{def:r_S} for the subgraphs in $\mathcal{C}_o(\mathcal{B}) \cup \mathcal{C}_c(\mathcal{B})$. There are at most $O(|E_\mathcal{B}|)$ subgraphs, then in total it takes $O(|E_\mathcal{B}|^2)$ to find the roots set for each subgraph of $\mathcal{S}(\mathcal{B})$.

By Theorem \ref{theo:condition}, applying the condition between two non-disjoint blocks in order to calculate a cycle-root set $\mathcal{R}_c$ is done in linear time $O(|V_\mathcal{B}|+|E_\mathcal{B}|)$.

Calculating $\mathcal{B} \setminus \mathcal{R}_c$, determining if it is the square of a forest \cite{KC98,BLS08} and constructing a root of it by Theorem \ref{theo:4power}  are done in linear time $O(|V_\mathcal{B}|+|E_\mathcal{B}|)$.

Finally, combining the root of the blocks of $G$ is done in linear time $O(|V|+|E|)$.

Then, we have the total complexity is $O(|V|.|E|^2+|E|^4) = O(|E|^4)$ (because $G$ is connected).
\end{pf}

\begin{corollary}
\label{coro:square}
Recognizing a square of a basic g-network and constructing one of its root, if there is any, can be done in polynomial time.
\end{corollary}

\begin{pf} It is easily to see that $G$ is the square of a basic g-network iff it is the $4$-leaf power of a basic g-network without invisible vertex. So, it can be deduced from Theorem \ref{theo:4power} that $G$ is the square of a basic g-network iff $G$ is biconnected, has a cycle-root set $\mathcal{R}_c$ such that the cycles in $\mathcal{R}_c$ are without invisible vertex, and $G \setminus \mathcal{R}_c$ is the square of a forest. So, any cycle of $\mathcal{C}'(G)$ must have all edges of weight $2$. We do not need to calculate $\mathcal{C}_i(G)$ for $i=4 , \dots , 8$ because all roots of these subgraphs contain at least one invisible vertex. We do not neither have to calculate $\mathcal{C}_c(G)$ because $G$ is biconnected. So, the complexity of this problem is equal to the complexity of calculating the maximal cliques graph, which is $O(|V|.|E|^2)$.
\end{pf}

\section{Conclusions and Perspectives}

Of course the complexity of our algorithm could be improved, but we hope that these partial structural results on $k$-leaf $\mathcal N$ network power   could help to answer the biological question described in the introduction.
We have proven here that the recognition of $4$-leaf g-network power graphs is polynomial. It would be natural to enlarge this result to the
 class of extended Cactus graphs  (with leaves but for which two cycles share at most one vertex).
 
 But we propose a stronger conjecture in the same flavour that \cite{GW07}:
 
 \textbf{Conjecture:}
 The recognition of $4$-leaf $\mathcal N$ power graphs is polynomial for every bounded treewidth  $\mathcal N$ class of phylogenetic networks.

\bibliography{k-leaf}

\begin{thebibliography}{10}

\bibitem{Bap09}
E.~Bapteste, MA~O'Malley, RG~Beiko, M~Ereshefsky, JP~Gogarten, L~Franklin-Hall,
  FJ~Lapointe, J~Dupr{\'e}, T~Dagan, and Martin~W. Boucher~Y.
\newblock Prokaryotic evolution and the tree of life are two different things.
\newblock {\em Biology Direc.}, pages 4--34, 2009.

\bibitem{Bap07}
Eric Bapteste and Richard~M. Burian.
\newblock On the need for integrative phylogenomics, and some steps toward its
  creation.
\newblock {\em Biology and Philosophy}, 25(4):711--736.

\bibitem{B08}
Andreas Brandst{\"a}dt.
\newblock On leaf powers.
\newblock Technical report, University of Rostock, 2010.

\bibitem{BL06}
Andreas Brandst{\"a}dt and Van~Bang Le.
\newblock Structure and linear time recognition of 3-leaf powers.
\newblock {\em Information Processing Letters}, 98:133--138, 2006.

\bibitem{BLS08}
Andreas Brandst{\"a}dt, Van~Bang Le, and R.~Sritharan.
\newblock Structure and linear-time recognition of 4-leaf powers.
\newblock {\em ACM Transactions on Algorithms}, 5(1), 2008.

\bibitem{DGHN06}
Michael Dom, Jiong Guo, Falk H{\"u}ffner, and Rolf Niedermeier.
\newblock Error compensation in leaf power problems.
\newblock {\em Algorithmica}, 44(4):363--381, 2006.

\bibitem{LVT08}
Lima-Mendez G, Van~Helden J, Toussaint A, and Leplae R.
\newblock Reticulate representation of evolutionary and functional
  relationships between phage genomes.
\newblock {\em Molecular Biology and Evolution}, 25(4):762--777, 2008.

\bibitem{GNS09}
Alain G\'{e}ly, Lhouari Nourine, and Bachir Sadi.
\newblock Enumeration aspects of maximal cliques and bicliques.
\newblock {\em Discrete Applied Mathematics}, 157(7):1447--1459, april 2009.

\bibitem{GW07}
Frank Gurski and Egon Wanke.
\newblock The clique-width of tree-power and leaf-power graphs.
\newblock In {\em Proceedings of the 33rd international conference on
  Graph-theoretic concepts in computer science}, WG'07, pages 76--85, Berlin,
  Heidelberg, 2007. Springer-Verlag.

\bibitem{HPV98}
Michel Habib, Christophe Paul, and Laurent Viennot.
\newblock A synthesis on partition refinement: A useful routine for strings,
  graphs, boolean matrices and automata.
\newblock In {\em STACS}, pages 25--38, 1998.

\bibitem{KC98}
Paul~E. Kearney and Derek~G. Corneil.
\newblock Tree powers.
\newblock {\em J. Algorithms}, 29(1):111--131, 1998.

\bibitem{L09}
FJ~Lapointe, P~Lopez, Y~Boucher, J~Koenig, and E.~Bapteste.
\newblock Clanistics: a multi-level perspective for harvesting unrooted gene
  trees.
\newblock {\em Trends Microbiology}, 18(8):341--347, 2010.

\bibitem{NRT02}
N.~Nishimura, P.~Ragde, and D.M. Thilikos.
\newblock On graph powers for leaf-labeled trees.
\newblock {\em J. of Algorithms}, 42:69--108, 2002.

\bibitem{R06}
Dieter Rautenbach.
\newblock Some remarks about leaf roots.
\newblock {\em Discrete Mathematics}, 306:1456--1461, 2006.

\end{thebibliography}
\bibliographystyle{plain}

\end{document}